\documentclass[pra,twocolumn,showpacs,amsmath,amssymb,superscriptaddress,floatfix,nofootinbib]{revtex4-1}
\usepackage{graphicx,amsmath,amssymb,amsfonts,dsfont,natbib,physics}
\usepackage{multirow}
\usepackage{lipsum,xcolor}
\usepackage{verbatim}
\usepackage{commath}
\usepackage{bm}
\usepackage{float}
\usepackage{pifont}

\newcommand{\mpsgraphics}[1]{\raisebox{-0.5\height}{\includegraphics[scale=0.15]{#1}}}

\newcommand{\sumal}[2]{\underset{#1}{\overset{#2}{\sum}}}

\newcommand{\nn}{\nonumber}

\newcommand{\wH}{\widehat{H}}
\newcommand{\wh}{\widehat{h}}
\newcommand{\wO}{\widehat{O}}
\newcommand{\wOtwo}{\widehat{O}^{(2)}}
\newcommand{\wB}{\widetilde{BB}}
\newcommand{\Btl}{\widetilde{B}_l}
\newcommand{\Btr}{\widetilde{B}_r}

\newcommand{\bZ}{Z}
\newcommand{\bzero}{0}

\newcommand{\mA}{\mathcal{A}}
\newcommand{\mB}{\mathcal{B}}

\newcommand{\mBt}{\widetilde{\mathcal{B}}}
\newcommand{\mC}{\mathcal{C}}
\newcommand{\mE}{\mathcal{E}}
\newcommand{\mH}{\mathcal{H}}
\newcommand{\mO}{\mathcal{O}}

\newcommand{\mP}{\mathcal{P}}

\newcommand{\fourpartdef}[7]
{
	\left\{
		\begin{array}{ll}
			#1 & \mbox{if } #2 \\
			#3 & \mbox{if } #4 \\
			#5 & \mbox{if } #6 \\
			#7 & \mbox{otherwise}
		\end{array}
	\right.
}

\newcommand{\wmO}{\widehat{\mO}}

\begin{document}

\tolerance 10000

\newcommand{\vk}{{\bf k}}   
\title{Large Classes of Quantum Scarred Hamiltonians from Matrix Product States}
\author{Sanjay Moudgalya}
\affiliation{Department of Physics, Princeton University, NJ 08544, USA}

\author{Edward O'Brien}
\affiliation{Rudolf Peierls Centre for Theoretical Physics, Clarendon Laboratory, Parks Road, Oxford, OX1 3PU, United Kingdom}

\author{B. Andrei Bernevig}
\affiliation{Department of Physics, Princeton University, NJ 08544, USA}

\author{Paul Fendley}
\affiliation{Rudolf Peierls Centre for Theoretical Physics, Clarendon Laboratory, Parks Road, Oxford, OX1 3PU, United Kingdom}
\affiliation{All Souls College, Oxford, OX1 4AL, United Kingdom}

\author{Nicolas Regnault}
\affiliation{Department of Physics, Princeton University, NJ 08544, USA}
\affiliation{Laboratoire de Physique de l'Ecole normale sup\'{e}rieure, ENS, Universit\'{e} PSL, CNRS, Sorbonne Universit\'{e}, Universit\'{e} Paris-Diderot, Sorbonne Paris Cit\'{e}, Paris, France}

\begin{abstract}
Motivated by the existence of exact many-body quantum scars in the AKLT chain, we explore the connection between Matrix Product State (MPS) wavefunctions and many-body quantum scarred Hamiltonians. 
We provide a method to systematically search for and construct parent Hamiltonians with towers of exact eigenstates composed of quasiparticles on top of an MPS wavefunction. 
These exact eigenstates have low entanglement in spite of being in the middle of the spectrum, thus violating the strong Eigenstate Thermalization Hypothesis (ETH).  
Using our approach, we recover the AKLT chain starting from the MPS of its ground state, and we derive the most general nearest-neighbor Hamiltonian that shares the AKLT quasiparticle tower of exact eigenstates. 
We further apply this formalism to other simple MPS wavefunctions, and derive new families of Hamiltonians that exhibit AKLT-like quantum scars. 
As a consequence, we also construct a scar-preserving deformation that connects the AKLT chain to the integrable spin-1 pure biquadratic model.
Finally, we also derive other families of Hamiltonians that exhibit new types of exact quantum scars, including a $U(1)$-invariant perturbed Potts model.
\end{abstract}
\maketitle
\date{\today}


\section{Introduction}\label{sec:intro}
The study of ergodicity and its breaking in isolated many-body quantum systems has been a growing area of research.
A central principle that governs the thermalization of initial states under time-evolution by a Hamiltonian is the Eigenstate Thermalization Hypothesis (ETH)~\cite{deutsch1991quantum, srednicki1994chaos}, which states in its strong form that all eigenstates of an ergodic system display thermal behavior. Most Hamiltonians are believed to satisfy ETH, but two mechanisms of ETH-violation are widely known: integrability and many-body localization~\cite{rahul2015review}, where \textit{all} eigenstates violate ETH.  
Quantum many-body systems that exhibit so-called quantum many-body scars have been recently added to the list of ETH-violating phenomena.
In these systems \textit{some}, but not all, eigenstates of a Hamiltonian violate ETH. 
The first exact examples of quantum scars include a systematic embedding of non-thermal eigenstates in a thermal spectrum~\cite{mori2017eth}, and an equally spaced tower of ETH-violating eigenstates discovered in the spin-1 Affleck-Kennedy-Lieb-Tasaki (AKLT) chain~\cite{Affleck1988, Moudgalya2018a, Moudgalya2018b}.
The interest in quantum scars primarily originates from an experimental observation of anomalous dynamics in a Rydberg atom experiment~\cite{bernien2017probing}, where quenches from a specially prepared initial state showed strong revivals and slow thermalization.  
Such anomalous dynamics were traced numerically to the initial state having a high overlap with an equally spaced tower of apparently ETH-violating eigenstates in the so-called PXP model \cite{Fendley2004}, a Rydberg-blockade Hamiltonian modelling this experiment~\cite{turner2017quantum, turner2018quantum}.
Various attempts to explain the anomalous dynamics phenomenon include connections to classical scars on an emergent classical manifold~\cite{ho2018periodic, michailidis2019slow}, proximity to integrability~\cite{khemani2019signatures}, existence of momentum-$\pi$ quasiparticles on top of an exact~\cite{lin2019exact} or approximate~\cite{iadecola2019quantum} eigenstate, and construction of parent Hamiltonians with almost-perfect revivals~\cite{choi2018emergent}, including by using ideas from Lie algebras~\cite{bull2020quantum}.
Furthermore, the presence of approximate revivals has also been demonstrated numerically in a variety of other models resembling the PXP model~\cite{ho2018periodic, 2bull2019scar, moudgalya2019quantum, hudomal2019}.
In addition, several works have explored interacting systems that show anomalous dynamics and the phenomenology of quantum scars.
These include kinetically constrained Hamiltonians~\cite{znidaric2019coexistence, sala2019ergodicity, moudgalya2019thermalization, pancotti2019quantum, yang2019hilbertspace, zhao2020quantum, robinson2019signatures, james2019nonthermal, lerose2019quasilocalized,  surace2019lattice, lin2019pxp, alhambra2019revivals} as well as Floquet systems~\cite{pai2019robust, khemani2019localization}, where quantum scars without Hamiltonian analogues can arise due to periodic driving~\cite{mukherjee2019collapse, haldar2019scars}.

The exact eigenstates in the AKLT chain also are composed of multiple momentum-$\pi$ quasiparticles on top of the ground state~\cite{Affleck1988, Moudgalya2018a, Moudgalya2018b}.
Subsequently, many similar examples of exact ETH-violating eigenstates were discovered.
Some non-integrable systems exhibit some solvable eigenstates~\cite{lin2019exact, ok2019topological, lee2020exact}, whereas exact towers of states embedded in a thermal spectrum were discovered in a variety of models, for example in the spin-1 XY models~\cite{schecter2019weak,chattopadhyay2019quantum}, a spin-1/2 domain-wall conserving model~\cite{iadecola2019quantum2, mark2020unified}, and systems with Onsager symmetries~\cite{Vernier2019, shibata2019onsagers}. %
These towers of equally spaced eigenstates also resemble the $\eta$-pairing states that long have been known to exist in the Hubbard and related models~\cite{Yang1987, vafek2017entanglement, yu2018beyond}. 
In some of these models, the quantum scars can be understood using a  formalism developed by Ref.~\cite{mori2017eth}, where ETH-violating eigenstates can be embedded systematically in the middle of an ETH-satisfying spectrum.
However, it has not been clear if some other models -  for example the AKLT chain - are isolated scarred points in the space of Hamiltonians or if they are part of a much larger family of quantum scarred Hamiltonians, although recent work in Ref.~\cite{mark2020unified} has shed light on this question for the AKLT chain. 
Given that the ground state of the AKLT chain~\cite{Affleck1987} is also a paradigmatic example of a Matrix Product State (MPS)~\cite{Klumper1993}, it is natural to wonder if the powerful tools developed in the context of MPS~\cite{perezgarcia2007matrix, Schollwock2011, orus2014practical} can be used to understand the exact excited states in the AKLT chain.
The exact excited states for the AKLT chain are known to also have simple MPS descriptions, which motivates the search for a general connection between MPS wavefunctions and quantum scarred Hamiltonians.  
In this work, we provide a general formalism for constructing quantum scarred Hamiltonians starting from an MPS wavefunction.
Given an MPS wavefunction, the so-called parent Hamiltonian construction provides a family of Hamiltonians for which  said MPS is an eigenstate.
In the large family of such parent Hamiltonians, we illustrate a method to look systematically for the subfamilies of Hamiltonians with quantum scars. 
Using this approach, we recover the analytical examples of quantum scars of the AKLT chain~\cite{Moudgalya2018a, Moudgalya2018b} and generalize them in three directions.
First, we obtain a 6-parameter family of nearest-neighbor Hamiltonians that all have the AKLT tower of states as exact eigenstates.  
Second, we start with a generalization of the AKLT MPS and obtain a class of Hamiltonians with new towers of exact eigenstates.  
Using this generalization, we also show that the AKLT chain can be continuously deformed to the (integrable) spin-1 biquadratic model, while preserving the quantum scars. 
Finally, we use our formalism to show examples of new types of quantum scars in a Potts model perturbed to have a $U(1)$ symmetry and exact ground states~\cite{OBrien2019}, and we discuss generalizations therein. 
This paper is organized as follows.
In Sec.~\ref{sec:MPS}, we review the basic concepts of MPS and quasiparticle excitations in the MPS language used in the rest of the paper.
In Sec.~\ref{sec:parent}, we review the construction of parent Hamiltonian of an MPS ground state using the AKLT chain as an example. 
In Sec.~\ref{sec:scars}, we give the main result of the paper, the extension of the parent Hamiltonian construction to include a tower of states composed of single-site quasiparticles. 
%
%
We illustrate this method by obtaining a family of Hamiltonians for which the AKLT tower of states remain eigenstates.  
In Sec.~\ref{sec:newfamilyakltlike}, we use our formalism to obtain a new family of quantum scarred models starting from a generalized AKLT MPS. We construct a continuous scar-preserving path from the AKLT chain to the integrable spin-1 biquadratic model.
Further, in Sec.~\ref{sec:newtype}, we discuss the extension of our formalism to a tower of two-site quasiparticles, and we show that the $U(1)$-invariant perturbed Potts model of Ref.~\cite{OBrien2019} exhibits such a tower of states.
We present our conclusions in Sec.~\ref{sec:conclusions}.
\section{Review of Matrix Product States (MPS)}\label{sec:MPS}
\subsection{Ground State}\label{sec:GS}
Consider a one-dimensional quantum chain with a $d$-dimensional Hilbert space on each of the $L$ sites. 
The many-body basis of the system is labelled by $\ket{m_1, m_2, \cdots, m_L}$, where $m_j$ runs over a basis of the single-site Hilbert space.
A wavefunction $\ket{\psi}$ on such a system is a Matrix Product State (MPS) if its decomposition in this basis reads~\cite{perezgarcia2007matrix} 
\begin{equation}
    \ket{\psi} = \sum_{\{m_j\}}{}{\ \textrm{Tr}\left[A^{[m_1]}_1 A^{[m_2]}_2 \cdots A^{[m_L]}_L\right]\ket{m_1, m_2, \cdots m_L}}.
\label{eq:MPSdefn}
\end{equation}
Here $A^{[m_j]}_j$ is a $\chi \times \chi$ matrix, $\chi$ being the bond dimension of the MPS and thus the $\{A_j\}$'s are $d \times \chi \times \chi$ tensors. The trace arises from the periodic boundary conditions we impose.
In this work, we use the following graphical and shorthand notations to represent such a wavefunction 
\begin{eqnarray}
    \ket{\psi} &=& \mpsgraphics{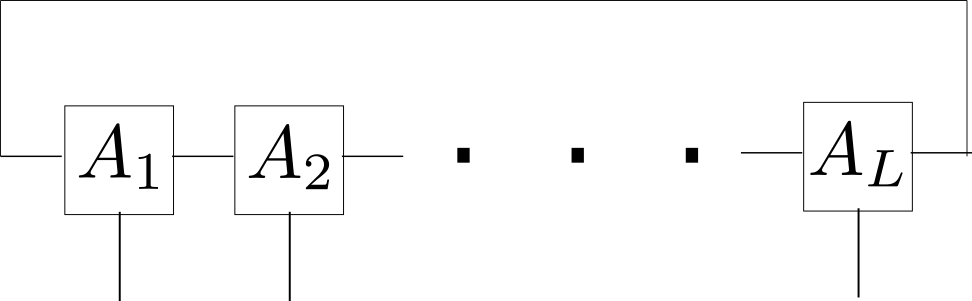} \nn \\
    &=& \ket{\left[A_1 A_2\cdots A_{L-1} A_L\right]}.
\label{eq:shorthands}
\end{eqnarray}
In Eq.~(\ref{eq:shorthands}), we use the brackets $[\ ]$ to indicate that the auxiliary indices at the ends have been contracted.
It is also sometimes useful to address segments of the wavefunction $\ket{\psi}$ of Eq.~(\ref{eq:shorthands}), for which we use the shorthand notation without the brackets $[\ ]$, for example 
\begin{equation}
    \ket{A_1 A_2} = \mpsgraphics{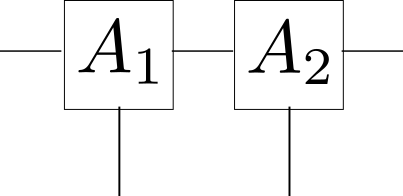} \equiv \sumal{m_1, m_2}{}{A^{[m_1]}_1 A^{[m_2]}_2\ket{m_1, m_2}}.
\label{eq:twositeshorthand}
\end{equation}
In Eq.~(\ref{eq:twositeshorthand}), $A^{[m_1]}_1 A^{[m_2]}_2$ is a $\chi \times \chi$ matrix for given values of $m_1$ and $m_2$.
Although the tensors $\{A_j\}$ in Eq.~(\ref{eq:MPSdefn}) can be site-dependent, a translation-invariant wavefunction can always be represented by an MPS with a site-independent tensor $A$~\cite{perezgarcia2007matrix}.
That is, any translation invariant MPS wavefunction $\ket{\psi_A}$ can be represented as
\begin{equation}
    \ket{\psi_A} = \ket{[AA\cdots A]}, 
\label{eq:MPSTI}
\end{equation}
where we have used the shorthand notation of Eq.~(\ref{eq:shorthands}).
A well-known class of wavefunctions with MPS forms are Valence Bond States (VBS)~\cite{Klumper1993, totsuka1995matrix, karimipour2008matrix}. Among these, the Affleck-Lieb-Kennedy-Tasaki (AKLT) states~\cite{Affleck1987} have been used to prove rigorously several important results, such as the existence of the Haldane gap in integer-spin chains~\cite{Affleck1988}. 
For the AKLT state, the MPS tensors $\{A^{[m]}\}$ are given by~\cite{Schollwock2011, Moudgalya2018b}
\begin{eqnarray}
    &A^{[+]} = \sqrt{\frac{2}{3}}
    \begin{pmatrix}
        0 & 1 \\
        0 & 0
    \end{pmatrix} = \sqrt{\frac{2}{3}}\sigma^+\nn \\
    \;\;
    &A^{[0]} = \frac{1}{\sqrt{3}}
    \begin{pmatrix}
        -1 & 0 \\
        0 & 1
    \end{pmatrix} = -\frac{1}{\sqrt{3}}\sigma^z \nn \\
    &A^{[-]} = \sqrt{\frac{2}{3}}
    \begin{pmatrix}
        0 & 0 \\
        -1 & 0
    \end{pmatrix} = -\sqrt{\frac{2}{3}}\sigma^-.
\label{eq:AKLTMPS}
\end{eqnarray}
Here we use the labels $+$, $0$, and $-$ to label the $S_z = +1$, $0$, and $-1$ spin-1 basis states respectively.
Thus, the physical dimension $d = 3$ and the bond dimension $\chi = 2$. 

In addition to such exact examples of MPSs, ground-state wavefunctions of gapped local Hamiltonians can be \textit{approximated} by an MPS with a small bond dimension $\chi$~\cite{verstraete2006matrix}.
Such a property has led to major developments in numerical simulations of one-dimensional systems~\cite{vidal2004efficient, Schollwock2011, orus2014practical}.
\subsection{Quasiparticle Excitations}\label{sec:QPeigenstate}
In addition to efficiently describing ground states of gapped one-dimensional Hamiltonians,  MPSs can also be used to efficiently describe quasiparticle excitations above the ground state. These techniques were pioneered by works on so-called tangent space methods~\cite{haegeman2013post, vanderstraeten2015excitations, vanderstraeten2019tangent}.
A single-site quasiparticle excitation with momentum $k$ on top of the MPS state $\ket{\psi_A}$ is given by
\begin{eqnarray}
    &\ket{\psi_A\left(B, k\right)} = \sumal{j = 1}{L}{e^{i k j}\sumal{\{m_j\}}{}{\left(\ \ket{m_1 m_2 \cdots m_L}\times \right.}}\nn \\
    &\left.\textrm{Tr}\left[\cdots A^{[m_{j-1}]} B^{[m_j]} A^{[m_{j+1}]} \cdots \right]\right),
\label{eq:QPansatz}
\end{eqnarray}
where $B^{[m_j]}$ is a $\chi \times \chi$ matrix with physical dimension $d$. 
Using the shorthand notation of Eq.~(\ref{eq:shorthands}),
we denote Eq.~(\ref{eq:QPansatz}) as
\begin{equation}
    \ket{\psi_A\left(B, k\right)} = \sumal{j = 1}{L}{e^{ikj}\overset{j}{\ket{\left[A\cdots A B A\cdots A\right]}}}. 
\end{equation}
where the $j$ on top of the $B$ operator tags its position on the lattice. In the context of the Single-Mode Approximation, the quasiparticles are usually described by in terms of a single-site ``quasiparticle creation operator" $\wO$, such that 
\begin{equation}
    B^{[m]} = \sumal{m, n}{}{\wO_{m, n} A^{[n]}},
\label{eq:QPop}
\end{equation}
which we denote in shorthand as
\begin{equation}
    \mpsgraphics{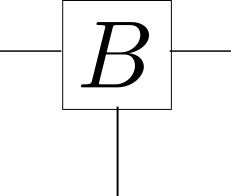} = \mpsgraphics{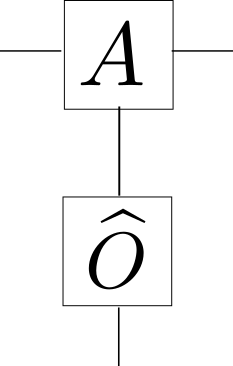}\;\;\;\textrm{or}\;\;\;\ket{B} = \wO\ket{A}.
\end{equation} 
Note that we could have a quasiparticle tensor $B$ which does not have the form of Eq.~(\ref{eq:QPop}) (for example $\wO$ could act on several neighboring sites).\footnote{$B$ has $d \chi^2$ entries while $\mO$ has $d^2$ entries. Thus, if $\chi^2 > d$ ($d\chi^2 > d^2$), not all choices of $B$ can be expressed in the form of Eq.~(\ref{eq:QPop}).}
However, for pedagogical reasons, in this work we restrict ourselves to $B$ of the form of Eq.~(\ref{eq:QPop}).
For example, in Ref.~\cite{Moudgalya2018a}, the AKLT chain was shown to have an exact low-energy eigenstate given by
\begin{equation}
    \ket{\psi_A\left(B, \pi\right)} = \sumal{j = 1}{L}{(-1)^j (S^+_j)^2}\ket{\psi_A}, 
\end{equation}
where $\{A^{[m]}\}$'s in $\ket{\psi_A}$ are the ground state AKLT MPS tensors of Eq.~(\ref{eq:AKLTMPS}) and $\{B^{[m]}\}$ are then given by (following Eq.~(\ref{eq:QPop}))
\begin{eqnarray}
    &B^{[m]} = \sumal{n \in \{+, 0, -\}}{}{(S^+)^2_{m,n} A^{[n]}},\nn \\
    &\implies B^{[+]} = A^{[-]} = -\sqrt{\frac{2}{3}}\sigma^-,\;\; B^{[0]} = B^{[-]} = 0.
\label{eq:BAKLTMPS}
\end{eqnarray}
Here $B^{[+]}$ is the only non-trivial matrix, a direct consequence of the $(S^+)^2$ operator acting on spin-1.
\subsection{Tower of Quasiparticle States}\label{sec:QPtower}
In addition to single quasiparticles, multiple identical quasiparticle states can be described in the MPS formalism using multiple tensors.
For example, the expression for a state with two quasiparticles described by tensor $B$ with momenta $k$ reads
\begin{equation}
    \ket{\psi_A\left(B^2, k\right)} \equiv \left(\sumal{j}{}{e^{ikj} \wO_j}\right)^2\ket{\psi_A}. 
\end{equation}
Such a state can also be expressed in the MPS language as
\begin{eqnarray}
    &\ket{\psi_A\left(B, k\right)} = \sumal{j_1\neq j_2}{}{e^{i k( j_1 + j_2)}\overset{j_1\hspace{10mm}j_2}{\ket{[A\cdots A B A \cdots A B A \cdots A]}}} \nn \\
    &+ 2\sumal{j}{}{e^{2 i k j} \overset{j}{\ket{[A\cdots A B^2 A \cdots A]}}},
\label{eq:twoQPMPS}
\end{eqnarray}
where we have used the shorthand notation of Eq.~(\ref{eq:shorthands}) and defined 
\begin{equation}
    \ket{B^m} \equiv \wO^m\ket{A},\;\; m \geq 1.
\end{equation}
For example, in the AKLT chain, $\wO = (S^+)^2$ and hence $\ket{B^2} = 0$. %
Similarly, a state with a number $n$ of $B$ quasiparticles reads
\begin{eqnarray}
    &\ket{\psi_A\left(B^n, k\right)} = \left(\sumal{j}{}{e^{ikj}\wO_j}\right)^n\ket{\psi_A} \nn \\
    &= \sumal{\{j_l\}}{}{e^{i k \sumal{l = 1}{n}{j_l}}\overset{j_1\hspace{11mm}j_l\hspace{11mm}j_n}{\ket{[A \cdots A B A \cdots A B A \cdots A B A \cdots A]}}},\;\;
\label{eq:towereig}
\end{eqnarray}
where $B$ is replaced by $B^m$ if $m$ of the $j_l$'s are equal. 
If these states $\{\ket{\psi_A\left(B^n, k\right)}\}$ are eigenstates of the Hamiltonian, they form a tower of (quasiparticle) states corresponding to the quantum many-body scars. 
\section{Parent Hamiltonian}\label{sec:parent}
\subsection{General Construction}
Given an MPS wavefunction $\ket{\psi_A}$ of the form of Eq.~(\ref{eq:MPSTI}) with a finite bond dimension $\chi$, we can construct the most general Hamiltonian for which $\ket{\psi_A}$ is a \textit{frustration-free} eigenstate.\footnote{Note that parent Hamiltonian constructions are typically restricted to constructing Hamiltonians with $\ket{\psi_A}$ as the ground state. However, such constructions straightforwardly work for highly excited eigenstates.}  
%
That is, we can construct a Hamiltonian $H$ that is a sum of local terms acting on a finite number of consecutive physical sites such that each of the local terms vanishes on $\ket{\psi_A}$.
Thus, we are looking for Hamiltonians that satisfy the property
\begin{equation}
    \wH = \sumal{j = 1}{L}{\wh_{j}},\;\;\; \wh_{j} \ket{\psi_A} = 0\;\;\forall j,
\label{eq:FFeigenstate}
\end{equation}
where $\wh_j$ is a local operator with a finite support, $j$ denoting the leftmost site of this finite support. 
In general, $\wh_j$ in Eq.~(\ref{eq:FFeigenstate}) is a local operator that acts on several consecutive sites.
However, in this work, we always restrict ourselves to the case where $\wh_j$ is a two-site operator, and the generalization of our formalism to multisite $\wh_j$ is straightforward. 
%
Denoting the two-site $\wh_j$ diagrammatically as 
\begin{equation}
    \wh_j = \mpsgraphics{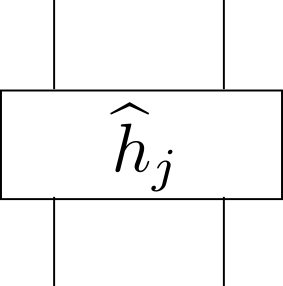}
\end{equation}
a sufficient condition for Eq.~(\ref{eq:FFeigenstate}) is if the operator $\wh_j$ satisfies
\begin{equation}
\mpsgraphics{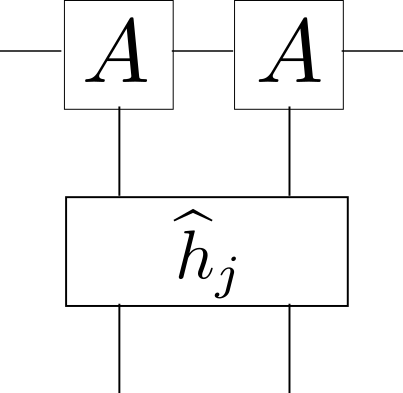} = 0\;\;\textrm{or}\;\; \wh_j \ket{AA} = 0.
\label{eq:cond1}
\end{equation}
To obtain such an operator $\wh_j$, we consider the MPS on two consecutive sites $\ket{AA}$.
$\ket{AA}$ can be interpreted as a map from the space of $\chi \times \chi$ matrices $\mH_{\chi^2}$ to vectors on the physical Hilbert space of two sites $\mH_{d^2}$ as follows:
\begin{eqnarray}
	&\ket{AA} : \mH_{\chi^2} \longrightarrow \mH_{d^2}\nn \\
	&X \longmapsto \sumal{m, n}{}{\ \textrm{Tr}\left[X A^{[m]} A^{[n]}\right]}\ket{m, n}.
\label{eq:Anmap}
\end{eqnarray}
Diagrammatically, this map reads
\begin{equation}
    X \;\;\longmapsto \;\; \raisebox{-0.4\height}{\includegraphics[scale=0.15]{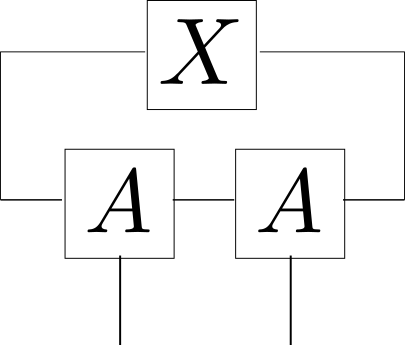}}
\end{equation}
To construct the local operator $\wh_j$ that satisfies Eq.~(\ref{eq:cond1}), consider the subspace $\mA$ in the physical Hilbert space of two sites ($\mA \subseteq \mH_{d^2}$) defined as 
\begin{eqnarray}
    &\mA \equiv \textrm{span}_X \left\{\mpsgraphics{AAX.png}\right\}\nn \\
    &= \textrm{span}_X\left\{\sumal{m, n}{}{\textrm{Tr}\left[X A^{[m]} A^{[n]}\right]\ket{m, n}}\right\},
\label{eq:mAdefn}
\end{eqnarray}
where $X$ runs over a complete basis of $\chi \times \chi$ matrices. 
For example, a complete basis is the set of matrices $\{X^{(m,n)}\}$, each of which has a single non-zero element given by $X^{(m,n)}_{ij} = \delta_{m,i} \delta_{n,j}$.
Such a choice of basis obviously is not unique.
For $\chi = 2$, a convenient choice is $\{\frac{1}{\sqrt{2}}\mathds{1}, \sigma^+, \sigma^-, \frac{1}{\sqrt{2}}\sigma^z\}$, which we use in App.~\ref{app:examples}.
Since $\mathcal{H}_{\chi^2}$ is a $\chi^2$-dimensional space, the dimension of $\mA$ is at most $\chi^2$. 
Provided $d^2 > \chi^2$, $\mA$ has a smaller dimension than the $\mH_{d^2}$ and thus $\mA$ is strictly contained within (but not equal to) $\mH_{d^2}$.
That is, $\mA$ is a proper subspace of $\mH_{d^2}$ ($\mA \subset \mH_{d^2}$). 
We then can define $\mA^c$, the complement of $\mA$ in $\mH_{d^2}$, as
\begin{equation}
    \mA^c \equiv \mH_{d^2}/\mA. 
\label{eq:Acdefn}
\end{equation}
For the local operator $\wh_j$ to vanish on $\ket{\psi_A}$, it is then sufficient to choose any operator that is supported in $\mA^c$.
That is the $d^2 \times d^2$ matrix of $\wh_j$ that satisfies Eq.~(\ref{eq:FFeigenstate}) has the following block-diagonal form in the basis of $\{\mA^c, \mA\}$:
\begin{equation}
    \wh_j = 
    \overset{\mA^c\;\; \mA}{
    \left(\begin{array}{c|c}
        \bZ^{(\mA^c)}_j & \bzero \\
        \hline
        \bzero & \bzero 
    \end{array}\right)}\hspace{-1mm}
    \begin{array}{c}
    {\scriptstyle \mA^c} \\
    {\scriptstyle \mA}\end{array},
\label{eq:hmatrixform}
\end{equation}
where $Z^{(\mA^c)}_j$ is an arbitrary Hermitian matrix with dimension that of $\mA^c$. 
Thus, the Hamiltonian $\wH$ of Eq.~(\ref{eq:FFeigenstate}) with $\wh_j$ of the form of Eq.~(\ref{eq:hmatrixform}) is a ``parent Hamiltonian" of the the MPS wavefunction $\ket{\psi_A}$. 
If we also require that $\ket{\psi_A}$ be the ground state of the Hamiltonian $\wH$, we then require $Z^{(\mA^c)}_j$ to be a positive definite matrix.
Note that when $Z^{(\mA^c)}_j$ is not positive definite, $\ket{\psi_A}$ is still an eigenstate of $\wH$ with area-law entanglement but it is generically located in the middle of the spectrum.
Indeed, it is then an typical example of a quantum scar of $\wH$ captured by the Shiraishi-Mori embedding formalism~\cite{mori2017eth}.
\subsection{AKLT State Example}	
We now illustrate the parent Hamiltonian construction for the AKLT ground state, with MPS tensors given in Eq.~(\ref{eq:AKLTMPS}).
Since $d^2 = 9 > \chi^2 = 4$ for the AKLT MPS, we are guaranteed that $\mA \subset \mH_{d^2}$, allowing the construction of nearest-neighbor terms $\wh_j$ that vanish on the MPS state $\ket{\psi_A}$.
As shown in App.~\ref{app:examples}, the subspace $\mA$ defined in Eq.~(\ref{eq:mAdefn}) can be explicitly computed using the AKLT tensors of Eq.~(\ref{eq:AKLTMPS}). 
As shown there in Eq.~(\ref{eq:mA}), we obtain
\begin{equation}
    \mA = \textrm{span}\{\ket{J_{1,1}}, \ket{J_{1,0}}, \ket{J_{1,-1}}, \ket{J_{0, 0}}\}
\label{eq:AKLTmA}
\end{equation}
where $\ket{J_{j,m}}$ is the total angular momentum eigenstate of two spin-1's with total spin $j$ and its $z$-projection $m$; they are listed in App.~\ref{sec:totalspin}.   
That is, the Hilbert space of two spin-1's decomposes into total angular momentum sectors with total spin $2$, $1$, or $0$ as
\begin{equation}
    1 \otimes 1 = 2 \oplus 1 \oplus 0, 
\end{equation}
and $\mA$ spans the total spin 1 and 0 subspaces. In the spin-$1/2$ Schwinger boson language, this is evident as there is a spin $1/2$ singlet between any two adjacent sites. The remaining spin $1/2$'s, one on each adjacent site, cannot clearly sum to spin-$2$.
Hence its orthogonal subspace $\mA^c$ spans the total spin-$2$ subspace, i.e.\ 
\begin{equation}
    \mA^c = \textrm{span}\{\ket{J_{2,2}}, \ket{J_{2,1}}, \ket{J_{2,0}}, \ket{J_{2, -1}}, \ket{J_{2, -2}}\}.
\label{eq:AKLTmAc}
\end{equation}
Thus, following Eq.~(\ref{eq:hmatrixform}), with the elements of $\bZ^{(\mA^c)}_j$ defined as
\begin{equation}
    (\bZ^{(\mA^c)}_j)_{m,n} = z^{(m,n)}_j,
\end{equation}
the most general nearest-neighbor Hamiltonian with $\ket{\psi_A}$ as a frustration-free eigenstate reads
\begin{equation}
    \wH = \sumal{j}{}{\wh_j},\;\; \wh_j = \sumal{m, n }{}{z^{(m, n)}_j \ket{J_{2,m}}\bra{J_{2,n}}}
\label{eq:mostgenAKLT}
\end{equation}
with Hermiticity imposing $z^{(n, m)}_j = (z^{(m, n)}_j)^\ast$.
However, imposing symmetries on the Hamiltonians restricts the form of $\bZ_j$. 
For example, translation invariance requires that $\bZ_j$ be independent of $j$.
$S_z$-spin conservation $U(1)$ symmetry requires that $\bZ_j$ be diagonal, since the operators $\ket{J_{2,m}}\bra{J_{2,n}}$ do not preserve the spin $S_z$ for $m \neq n$. 
Furthermore, imposing $SU(2)$ symmetry on the parent Hamiltonian requires that all the operators $\ket{J_{2,m}}\bra{J_{2,m}}$ appear with the same coefficient in the Hamiltonian.
Thus, with translation invariance and $SU(2)$ symmetry, the local term of the parent Hamiltonian is uniquely determined to be
\begin{equation}
    \wh_j = c \sumal{m = -2}{2}{\ket{J_{2,m}}\bra{J_{2,m}}} = c\ P^{(2, 1)}, 
\label{eq:SU2translationlocal}
\end{equation}
where $c$ is an arbitrary constant $P^{(2, 1)}$ is the projector of two spin-1's onto total spin 2, which is nothing but the local term of the AKLT Hamiltonian. 
\section{Quantum Scarred Hamiltonians}\label{sec:scars}
Having constructed the most general nearest-neighbor Hamiltonian for which $\ket{\psi_A}$ is a frustration-free eigenstate, we would like to determine the set of conditions on $\bZ^{(\mA^c)}_j$ in Eq.~(\ref{eq:hmatrixform}) such that the Hamiltonian $\wH$ exhibits a quasiparticle tower of states.
For example, in the case of the AKLT MPS, we know that the AKLT Hamiltonian ($\bZ^{(\mA^c)}_j = \mathds{1}$) exhibits a tower of states~\cite{Moudgalya2018a, Moudgalya2018b}.
Here we show that there are other choices of $\bZ^{(\mA^c)}_j$ for which the states in the AKLT tower are eigenstates. 
\subsection{One Quasiparticle Eigenstate}
We now illustrate the formalism to construct a Hamiltonian with a single quasiparticle eigenstate in addition to a frustration free MPS eigenstate, similar to the case discussed in Sec.~\ref{sec:QPeigenstate}.
That is, given an MPS wavefunction $\ket{\psi_A}$, we want to obtain a Hamiltonian, with  $\ket{\psi_A}$ as an eigenstate, that also  has a quasiparticle eigenstate of the form $\ket{\psi_A\left(B, k\right)}$ of Eq.~(\ref{eq:QPansatz}) with energy $\mE$.
As we show in App.~\ref{app:QPeigenstateMPS}, a sufficient local condition is (using the shorthand notation of Eq.~(\ref{eq:twositeshorthand}))
\begin{eqnarray}
    &\mpsgraphics{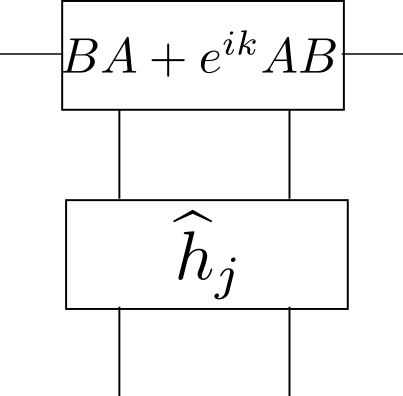} = \mE\; \mpsgraphics{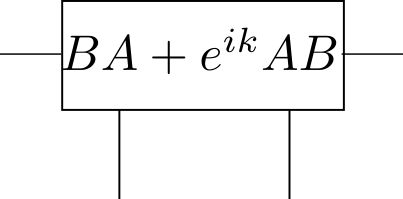}\nn \\
    &\textrm{or}\;\; \wh_j \left(\ket{BA} + e^{ik} \ket{AB}\right) = \mE \left(\ket{BA} + e^{ik} \ket{AB}\right),
\label{eq:cond2}
\end{eqnarray}
where
\begin{equation}
    \mpsgraphics{BAkAB.png} \equiv \mpsgraphics{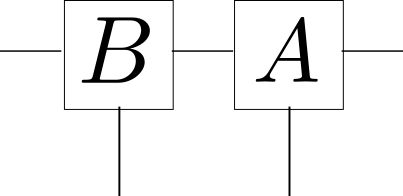} + e^{ik} \mpsgraphics{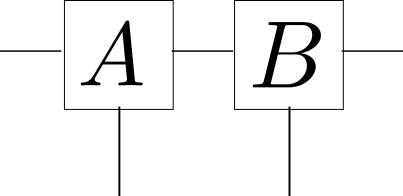} 
\end{equation}
To find operators that satisfy Eq.~(\ref{eq:cond2}), similar to Eq.~(\ref{eq:Anmap}), we view $\left(\ket{BA} + e^{ik} \ket{AB}\right)$ as a map from the space of $\chi \times \chi$ matrices $\mH_{\chi^2}$ to the physical Hilbert space of two sites $\mH_{d^2}$:
\begin{eqnarray}
    &\left(\ket{BA} + e^{ik} \ket{AB}\right): \mH_{\chi^2} \longrightarrow \mH_{d^2} \nn \\
    & X \longmapsto \mpsgraphics{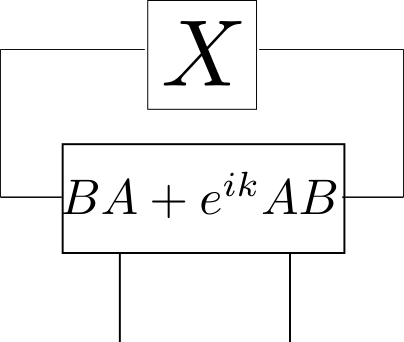}
\end{eqnarray}
We define the subspace $\mB$ ($\subseteq \mH_{d^2}$) as
\begin{eqnarray}
    &\mB \equiv \textrm{span}_X\left\{\mpsgraphics{BAkABX.png}\right\} \nn \\
    &= \textrm{span}_X \left\{\sumal{m,n}{}{\textrm{Tr}\left[X \left(B^{[m]} A^{[n]} + e^{ik} A^{[m]} B^{[n]}\right)\right]\ket{m, n}}\right\},\nn \\
\label{eq:mBdefn}
\end{eqnarray}
where $X$ runs over a complete basis of $\chi \times \chi$ matrices.
In terms of a single-site quasiparticle creation operator $\wO$ for which the tensors $B$ and $A$ satisfy Eq.~(\ref{eq:QPop}), the subspace $\mB$ reads
\begin{equation}
    \mB = \underbrace{\left(\wO \otimes \mathds{1} + e^{ik} \mathds{1} \otimes \wO\right)}_{\wmO} \mA \equiv \textrm{span}\{ \wmO\ket{\psi} : \ket{\psi} \in \mA\}.
\label{eq:mBopdefn}
\end{equation}
Since $\mB$ has a dimension of at most $\chi^2$, if $d^2 > \chi^2$, $\mB$ is a proper subspace of $\mH_{d^2}$ ($\mB \subset \mH_{d^2}$).
Defining the complement as
\begin{equation}
    \mB^c \equiv \mH_{d^2}/\mB, 
\end{equation}
the $\wh_j$ satisfying Eq.~(\ref{eq:cond2}) reads
\begin{equation}
    \wh_j = 
    \overset{\mB^c\;\; \mB}{
    \left(\begin{array}{c|c}
        Z^{(\mB^c)}_j & \bzero \\
        \hline
        \bzero & \mE\mathds{1} 
    \end{array}\right)}\hspace{-1mm}
    \begin{array}{c}
    {\scriptstyle \mB^c} \\
    {\scriptstyle \mB}
    \end{array},
\label{eq:hmatrixformcond2}
\end{equation}
where $Z^{(\mB)}_j$ is an arbitrary matrix with the same dimension as $\mB^c$. 
However, to obtain $\wh_j$ that satisfies both Eq.~(\ref{eq:cond1}) and Eq.~(\ref{eq:cond2}) with $\mE \neq 0$, it is essential that $\mA$ lies within the subspace $\mB^c$ in Eq.~(\ref{eq:hmatrixformcond2}).
In other words, we require
\begin{equation}
    \mA \subseteq \mB^c\;\;\implies\;\; \mB \subseteq \mA^c. 
\label{eq:impcond}
\end{equation}
In fact, operators $\wO$ and momentum $e^{ik}$ for which $\mB$ satisfies Eq.~(\ref{eq:impcond}) can be found by ensuring the orthogonality of states in $\mA$ and $\mB$ by solving the linear equation
\begin{equation}
    \mpsgraphics{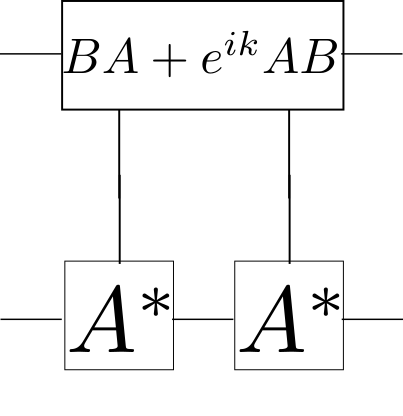} = 0\;\;\iff\;\;\raisebox{-0.5\height}{\includegraphics[scale = 0.15]{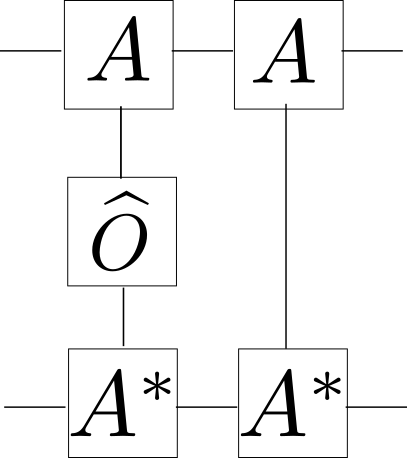}} = -e^{ik} \raisebox{-0.5\height}{\includegraphics[scale = 0.15]{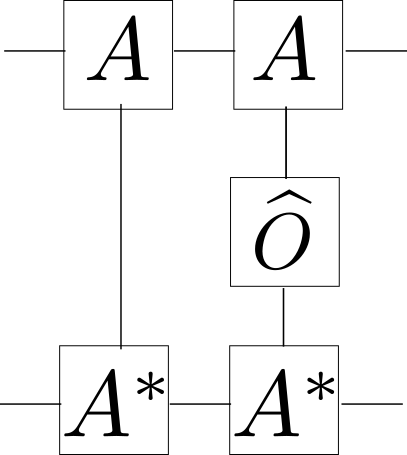}}.
\label{eq:impcondopequation}
\end{equation}
Assuming Eq.~(\ref{eq:impcond}) is satisfied, the term $\wh_j$ has the structure
\begin{equation}
    \wh_j = 
    \overset{\hspace{2mm}\mA^c/\mB\hspace{3mm}\mB\hspace{3mm}\mA\hspace{4mm}}{
    \left(\begin{array}{c|c|c}
        \bZ^{(\mA^c/\mB)}_j & \bzero & \bzero \\
        \hline
        \bzero & \mE\mathds{1} & \bzero \\
        \hline
        \bzero & \bzero & \bzero
    \end{array}\right)}\hspace{-2mm}\begin{array}{c}
        {\scriptstyle\mA^c/\mB}  \\
        {\scriptstyle \mB} \\
        {\scriptstyle \mA}
    \end{array},
\label{eq:hmatrixformgen}
\end{equation}
where $Z^{(\mA^c/\mB)}_j$ here is an arbitrary Hermitian matrix with the same dimension as $\mA^c/\mB$.  
\subsection{Tower of Quasiparticle Eigenstates}
Given the most general Hamiltonian of the form of Eq.~(\ref{eq:hmatrixformgen}) that has a single-quasiparticle eigenstate, we now wish to construct a Hamiltonian with a tower of quasiparticle eigenstates of the form of Eq.~(\ref{eq:towereig}) in Sec.~\ref{sec:QPtower}. 
One way to do so is to impose emergent constraints on the quasiparticles, similar to the tower of states in the AKLT chain~\cite{Moudgalya2018a} (as we show in Sec.~\ref{sec:AKLTtowergen}) and the ones in Ref.~\cite{iadecola2019quantum}.
For example if the quasiparticles are naturally constrained to be at least one site away from each other, the quasiparticles do not interact with each other under a \textit{nearest-neighbor} Hamiltonian and, as we show in this section, we can construct eigenstates composed of multiple identical quasiparticles. 
In terms of the MPS, such a condition reads
\begin{eqnarray}
    	&\ket{BB} = \raisebox{-0.5\height}{\includegraphics[scale = 0.15]{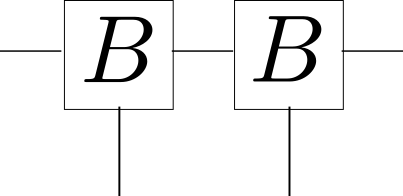}} = 0 \iff \left(\wO \otimes \wO\right)\mA = 0, \label{eq:constraint1} \\
    	&\textrm{and}\;\;\;\wO \ket{B} = \wO^2 \ket{A} = 0, \label{eq:constraint2}
\end{eqnarray}
$\mA$ is defined in Eq.~(\ref{eq:mAdefn}). 
Eq.~(\ref{eq:constraint1}) prohibits two quasiparticles on neighboring sites, and Eq.~(\ref{eq:constraint2}) prohibits two quasiparticles on the same site.
As shown in App.~\ref{app:QPeigenstateMPS}, these conditions result in a tower of exact eigenstates of the Hamiltonian $\wH$ with local terms of the form of Eq.~(\ref{eq:hmatrixformgen}).
The states of the tower have the form $\{\ket{S_{2n}}\}$
\begin{equation}
     \ket{S_{2n}} = \mP^n \ket{\psi_A},\;\; \wH \ket{S_{2n}} = 2 \mE n \ket{S_{2n}},\;\; n \leq {L/2}
\label{eq:towerform}
\end{equation}
where
\begin{equation}
    \mP = \sumal{j = 1}{L}{e^{ikj} \wO_j}.
\label{eq:mPdefn}
\end{equation}
Due to Eq.~(\ref{eq:constraint1}), the tower is guaranteed to end on the state after $\left(L/2 + 1\right)$ applications of $\mP$ on the state $\ket{\psi_A}$ since 
\begin{eqnarray}
    &\mP^{\frac{L}{2}}\ket{\psi_A} \propto \left(\ket{BABA\cdots} \pm \ket{ABABA\cdots}\right)\nn \\
    &\implies \mP^{\frac{L}{2} + 1}\ket{\psi_A} = 0.  
\end{eqnarray}
\subsection{Models with AKLT Tower of States}\label{sec:AKLTtowergen}
We now show that the scars in the AKLT chain~\cite{Moudgalya2018a, Moudgalya2018b} can be explained in this formalism, and we construct a family of nearest-neighbor Hamiltonians for which all the scars of the AKLT chain are eigenstates. 
We start with the spin-1 AKLT ground state MPS of Eq.~(\ref{eq:AKLTMPS}), and take the operator $\wO$ and momentum $k$ to be
\begin{equation}
    \wO = (S^+)^2,\;\; k = \pi.
\label{eq:OAKLT}
\end{equation}
The subspace $\mB$ defined in Eq.~(\ref{eq:mBopdefn}) then reads
\begin{equation}
    \mB = \underbrace{\left((S^+)^2 \otimes \mathds{1} - \mathds{1} \otimes (S^+)^2\right)}_{\equiv \widetilde{(S^+)^2}}\mA, 
\label{eq:Stdefn}
\end{equation}
where $\mA$ for the AKLT MPS is shown in Eq.~(\ref{eq:mAdefn}). 
We can compute the subspace $\mB$ by noting two important properties of the operator $\widetilde{(S^+)^2}$: (i) it is a spin-2 operator, (ii) it is antisymmetric (i.e. $\widetilde{(S^+)^2} \longrightarrow -\widetilde{(S^+)^2}$) under exchange of the two sites involved in Eq.~(\ref{eq:Stdefn}).  
We then can deduce the following (see Eqs.~(\ref{eq:mBAKLT})-(\ref{eq:genBsubspacenorm}) in App.~\ref{app:examples} for an explicit derivation):
\begin{enumerate}
    \item The vector $\ket{J_{1,1}}$ in $\mA$ with spin $1$ vanishes under the action of $\widetilde{(S^+)^2}$ since a vector with spin $3$ cannot be formed from two spin-1's.
    \item Since the vector $\ket{J_{0, 0}}$ in $\mA$ is symmetric under exchange, it vanishes under the action of $\widetilde{(S^+)^2}$ since an antisymmetric vector with spin 2 cannot be formed from two spin-1's.
    \item The remaining vectors in $\mA$: $\ket{J_{1,0}}$ and $\ket{J_{1,-1}}$ are antisymmetric under exchange, and thus under the action of $\widetilde{(S^+)^2}$ result in symmetric states with spins $2$ and $1$ respectively (i.e. $\ket{J_{2,2}}$ and $\ket{J_{2,1}}$ respectively). 
\end{enumerate}
Thus,
\begin{eqnarray}
    \mA &=& \textrm{span}\{ \ket{J_{0,0}}, \ket{J_{1,-1}}, \ket{J_{1,0}}, \ket{J_{1,1}} \} \nn \\
    \implies \mB &=& \textrm{span}\{ \ket{J_{2,1}}, \ket{J_{2, 2}}\}, \nn \\
    \implies \mA^c/\mB &=& \textrm{span}\{ \ket{J_{2,0}}, \ket{J_{2,-1}}, \ket{J_{2,-2}}\}.
\label{eq:mAmB}     
\end{eqnarray}
Clearly, $\mA$ and $\mB$ in Eq.~(\ref{eq:mAmB}) are orthogonal subspaces, and Eq.~(\ref{eq:impcond}) is satisfied.
Furthermore, Eq.~(\ref{eq:constraint1}) is satisfied since $\left((S^+)^2 \otimes (S^+)^2\right)$, by virtue of being a spin-4 operator, vanishes on all states of $\mA$.
Eq.~(\ref{eq:constraint2}) is also satisfied since $\left((S^+)^2\right)^2 = 0$. 
Thus, the most general nearest-neighbor Hamiltonian with the AKLT tower of states as eigenstates has the form of Eq.~(\ref{eq:hmatrixformgen}). It reads
\begin{eqnarray}
    &\wH = \sumal{j}{}{\wh_{j}},\nn \\
    &\wh_j = \mE\left(\ket{J_{2,1}}\bra{J_{2,1}} + \ket{J_{2,2}}\bra{J_{2,2}}\right) \nn \\
    &+ \sumal{m,n =-2}{0}{z^{(m, n)}_{j} \ket{J_{2, m}}\bra{J_{2,n}}}.
\label{eq:genAKLT}
\end{eqnarray}
with
\begin{equation}
    z^{(n,m)}_j = (z^{(m,n)}_j)^\ast. 
\end{equation}
Note that $\mE$ in Eq.~(\ref{eq:genAKLT})) is only an overall scale. This 6-parameter family of nearest-neighbor Hamiltonians was also obtained very recently in Ref.~\cite{mark2020unified}.
If we demand conservation of $S_z$, we need to set
\begin{equation}
    z^{(m, n)}_{j} \propto \delta_{m,n}, 
\label{eq:Szcondition}
\end{equation}
yielding a three-dimensional family of Hamiltonians. 
The AKLT Hamiltonian is recovered by setting
\begin{equation}
    z^{(m,n)} = \mE \delta_{m,n}. 
\end{equation} 
%
%

%
%
Instead of assuming $\wO$ and $k$ in Eq.~(\ref{eq:OAKLT}), we can also arrive at that choice by brute force solving Eq.~(\ref{eq:impcondopequation}) for $e^{ik}$ and $\wO$ given the AKLT MPS of Eq.~(\ref{eq:AKLTMPS}).
Solving for the 10 variables ($k$ and $9$ parameters in $\mO$) using symbolic computation software, we obtain the solutions
\begin{eqnarray}
    &k = \pi\;\;\; \textrm{and}\nn \\
    &\wO \in \textrm{span}\{ (S^+)^2, \{S^+, S^z\}, 2 (S^z)^2 - \{S^+, S^-\},\nn \\
    &\{S^-, S^z\}, (S^-)^2 \}
\label{eq:AKLTsolns}
\end{eqnarray}
where $\{\cdot, \cdot\}$ represents the anticommutator. 
This guarantees the existence of Hamiltonians with local terms of the form Eq.~(\ref{eq:hmatrixformgen}) having one-quasiparticle eigenstates with energy $\mE$ of the form
\begin{equation}
    \ket{\psi_A\left(B, \pi\right)} = \sumal{j}{}{(-1)^j \wO_j}\ket{\psi_A}.
\label{eq:akltqp}
\end{equation}
where $\wO_j$ is chosen from Eq.~(\ref{eq:AKLTsolns}).
In general, the subspace $\mB$ in Eq.~(\ref{eq:hmatrixformgen}) depends on the choice of $\wO$ in Eq.~(\ref{eq:AKLTsolns}), and generically we obtain distinct families of Hamiltonians for distinct $\wO$'s.
The families of Hamiltonians obtained for different choices of $\wO$ intersect at the AKLT Hamiltonian (i.e. when $Z_j = \mE \mathds{1}$ in Eq.~(\ref{eq:hmatrixformgen})). The five independent excited states there span the entire multiplet of spin-2 magnon state, as shown in App.~\ref{sec:spin2magnon}.\footnote{The AKLT Hamiltonian is $SU(2)$ symmetric, and hence an eigenstate with spin $s$ is $(2s + 1)$-fold degenerate, e.g.\ the spin-2 magnon state is $5$-fold degenerate.} 
We further impose Eqs.~(\ref{eq:constraint1}) and (\ref{eq:constraint2}) on $\wO$, and by numerical brute force we obtain precisely two choices for $\wO$: 
\begin{equation}
    \wO = (S^+)^2\;\;\textrm{or}\;\; \wO = (S^-)^2.
\label{eq:TowerSolns}
\end{equation}
These are the only choices of single-site operators that generate the tower of states starting from the AKLT MPS. 
The fact that the AKLT state satisfies Eq.~(\ref{eq:constraint1}) is a consequence of ``string order" in the AKLT ground state~\cite{denNijs1989string, kennedy1992hidden, oshikawa1992hidden, perezgarcia2008string}.
That is, when decomposed in the spin-1 product state basis, the AKLT ground state has zero weight on configurations that have the form $\ket{\cdots + 0^n + \cdots}$ or $\ket{\cdots - 0^n - \cdots}$ for $n \geq 0$, where $0^n$ represents a string of $n$ $0$'s.
In particular, nearest neighbor configurations of $\ket{++}$ and $\ket{--}$ do not appear in the AKLT ground state. 
Since the operators $(S^+)^2 \otimes (S^+)^2$ and $(S^-)^2 \otimes (S^-)^2$) are non-vanishing only on the configurations $\ket{--}$ and $\ket{++}$ respectively, they vanish on the AKLT ground state, thus satisfying Eq.~(\ref{eq:constraint1}).
These two towers are actually equivalent in the AKLT chain since they correspond to highest and lowest states of the same multiplet of the $SU(2)$ symmetry.  
However, since the $\mB$ subspaces depend on the choice of $\wO$ in Eq.~(\ref{eq:TowerSolns}), we can deform away from the AKLT Hamiltonian (by breaking the $SU(2)$ symmetry) and preserve only one tower of states: \textit{either} the highest weight state \textit{or} the lowest weight states of the SU(2) multiplet of the AKLT tower of states. 
\section{New Families of AKLT-like Quantum Scarred Hamiltonians}\label{sec:newfamilyakltlike}
\subsection{Quantum Scarred Hamiltonians from Generalized AKLT MPS}
Having established the formalism to construct quantum scarred models starting from an MPS, we now deform away from the AKLT MPS and obtain new families of quantum scarred Hamiltonians.
Since we start with a different MPS, the tower of states presented in this section is \textit{distinct} from the AKLT tower of states. 
In particular, we consider the following generalization of the AKLT MPS of Eq.~(\ref{eq:AKLTMPS})
\begin{equation}
    A^{[+]} = c_+ \sigma^+,\;\;\;A^{[0]} = c_0 \sigma^z,\;\;\;A^{[-]} =  c_- \sigma^-,
\label{eq:genAKLTMPSmain}
\end{equation}
where one of $c_+$, $c_0$ and $c_-$ is fixed by the normalization of the MPS wavefunction $\ket{\psi_A}$.
Note that for $(c_+, c_0, c_-) = \frac{1}{\sqrt{\alpha + 1}}(\sqrt{\alpha}, -1, -\sqrt{\alpha})$, the MPS of Eq.~(\ref{eq:genAKLTMPSmain}) coincides with some of the ones considered in Ref.~\cite{Schutz1993}.
By numerical brute force,  we find that Eqs.~(\ref{eq:impcondopequation}), (\ref{eq:constraint1}) and (\ref{eq:constraint2}) are satisfied for for the MPS $A$ if $\wO = (S^+)^2$ and $k = \pi$.  
Thus, there exist Hamiltonians for which the MPS of Eq.~(\ref{eq:genAKLTMPSmain}) are frustration-free eigenstates and a tower of eigenstates can be built from them with the same raising operator $\mP$ as that of the AKLT tower of states.
As we show in App.~\ref{app:examples}, the subspaces $\mA$ and $\mB$ for the MPS of Eq.~(\ref{eq:genAKLTMPSmain}) read
\begin{eqnarray}
    \mA &=& \textrm{span}\{\ket{K_{0,0}}, \ket{K_{1,-1}}, \ket{K_{1,0}}, \ket{K_{1,1}}\} \nn \\
    \implies \mB &=& \textrm{span}\{\ket{K_{2,1}}, \ket{K_{2,2}}\} \nn \\ 
    \implies \mA^c/\mB &=& \textrm{span}\{\ket{K_{2,0}}, \ket{K_{2,-1}}, \ket{K_{2,-2}}\},
\label{eq:gensubspaces}
\end{eqnarray}
where we have defined
\begin{eqnarray}
    \ket{K_{0,0}} &\equiv& \frac{c_+ c_- \left(\ket{+-} + \ket{-+}\right) + 2 c_{0}^2 \ket{00}}{\sqrt{2\left(|c_+ c_-|^2 + 2|c_0|^4\right)}} \nn \\
    \ket{K_{2,0}} &\equiv& \frac{- c_0^2 \left(\ket{+-} + \ket{-+}\right) + c_+ c_- \ket{00}}{\sqrt{|c_+ c_-|^2 + 2|c_0|^4}} \nn \\
    \ket{K_{j,m}} &\equiv& \ket{J_{j,m}}\;\;\textrm{if}\;\;(j, m) \notin \{(0, 0), (2,0)\}.
\label{eq:Kdefn}
\end{eqnarray}
Note that the subspaces $\mA$ and $\mB$ in Eq.~(\ref{eq:gensubspaces}) are orthogonal irrespective of the values of $(c_+, c_0, c_-)$, and Eq.~(\ref{eq:impcond}) is satisfied. 
Furthermore Eqs.~(\ref{eq:constraint1}) and (\ref{eq:constraint2}) are satisfied for the same reasons as those for the AKLT MPS (see Sec.~\ref{sec:AKLTtowergen}).
Since the dimensions of the subspaces $\mA$ and $\mB$ are the same as that in the AKLT case (see Eq.~(\ref{eq:mAmB})), we can similarly derive the 6 parameter family of Hermitian nearest-neighbor Hamiltonian that has a tower of states generated from the MPS eigenstate of Eq.~(\ref{eq:genAKLTMPSmain}), composed of the local term $\wh_j$ of the form of Eq.~(\ref{eq:hmatrixform}).
Thus, the most general Hamiltonian with such a tower reads
\begin{eqnarray}
    &\wH = \sumal{j}{}{\wh_{j}},\nn \\
    &\wh_j = \mE\left(\ket{K_{2,1}}\bra{K_{2,1}} + \ket{K_{2,2}}\bra{K_{2,2}}\right) \nn \\
    &+ \sumal{m,n =-2}{0}{z^{(m, n)}_{j} \ket{K_{2, m}}\bra{K_{2,n}}}.
\label{eq:genMPSHamil}
\end{eqnarray}
\subsection{Deformation to Integrability}\label{sec:deformation}
Continuous deformations maintaining exact ground states connect the AKLT and the spin-1 biquadratic chains \cite{Klumper1993}. 
Here we give a continuous deformation that preserves a tower of exact quasiparticle states as well. The biquadratic chain is integrable~\cite{Parkinson1987integrability, Barber1989, Ercolessi2014Analysis},\footnote{Note that only a part of the spectrum is known to be integrable with periodic boundary conditions~\cite{Parkinson1987integrability, Parkinson1988spin}, whereas the full spectrum is integrable with open boundary conditions~\cite{Barber1989}} in contrast to the other Hamiltonians considered in this paper. 
We start with the observation that (see App.~\ref{app:SS2})
%
%
 \begin{align}
 \left(\vec{S} \cdot \vec{S}\right)^2 =&\left(\ket{+ -} + \ket{- +} - \ket{0 0}\right)\left(\bra{+ -} + \bra{- +} - \bra{0 0}\right)\nn\\
 &+1\ ,\cr
\label{eq:SS2simp}
\end{align}
where $\vec{S} \cdot \vec{S}$ is the usual two-site Heisenberg interaction:
\begin{equation}
    \vec{S} \cdot \vec{S} \equiv S^+ \otimes S^- + S^- \otimes S^+ + S^z \otimes S^z. 
\label{eq:circdefn}
\end{equation}
Thus, if we set
\begin{equation}
    c_0^2 = c_+ c_-, 
\end{equation}
using Eq.~(\ref{eq:Kdefn}), Eq.~(\ref{eq:SS2simp}) can be written as
\begin{equation}
    \frac{1}{3}\left(\left(\vec{S} \cdot \vec{S}\right)^2 - 1\right) =  \ket{K_{2,0}}\bra{K_{2,0}}.
\label{eq:SS2proj}
\end{equation}
The Hamiltonian built from Eq.~(\ref{eq:SS2proj}) is thus of the form of Eq.~(\ref{eq:genMPSHamil}) with the parameters
\begin{equation}
    \mE = 0,\;\; z^{(m,n)}_j = \delta_{m, 0}\delta_{n,0}.
\end{equation}
Thus, in the space of quantum scarred Hamiltonians considered here, the AKLT and pure biquadratic models are located at the following points
\begin{eqnarray*}
    \textrm{AKLT}:\;\left(\mE, z^{(m,n)}_j, \frac{c_+}{c_0}, \frac{c_-}{c_0}\right) &=& \left(1, \delta_{m,n}, -\sqrt{2}, \sqrt{2}\right) \nn \\
    \textrm{Biquadratic}:\;\left(\mE, z^{(m,n)}_j, \frac{c_+}{c_0}, \frac{c_-}{c_0}\right) &=& \left(0, \delta_{m,0}\delta_{n,0}, 1, 1\right).
\end{eqnarray*}

We consider a path between the two given by
\begin{align}
    (\mE, z^{(m,n)}_j, &\frac{c_+}{c_0}, \frac{c_-}{c_0}) = (2\cos\theta\,,\nn \\
    &2\cos\theta\ \delta_{m,n} + (\csc\theta + 2\sin\theta - 4\cos\theta)\delta_{m,0}\delta_{n,0}\,,\nn \\
    & -\sqrt{\cot\theta - 1}, \sqrt{\cot\theta - 1}).\nn \\
\end{align}
We recover the AKLT chain (up to a constant factor and constant shift) by setting $\theta = \cot^{-1}3$, and the pure biquadratic model by setting $\theta = \pi/2$. The Hamiltonian parametrized by $\theta$ reads
\begin{eqnarray}
    \wH &= \sumal{j}{}{\left[\cos\theta\left(\vec{S}_j\cdot\vec{S}_{j+1}\right) + \sin\theta \left(\vec{S}_j\cdot\vec{S}_{j+1}\right)^2\right.} \nn \\
    &\quad\left.+ \cos\theta\left(\cot\theta - 3\right)\ P^{00}_{j,j+1}\right],
\label{eq:scarfamily}
\end{eqnarray}
up to an overall factor and constant shift. We thus recover the usual bilinear-biquadratic chain plus an additional term proportional to
\begin{equation}
    P^{00} = \ket{0 0}\bra{0 0}. 
\end{equation}
The model of Eq.~(\ref{eq:scarfamily}) thus has a tower of exact eigenstates with spacing $\mE = 4\cos\theta$ starting from the ground state.

We note that at the purely biquadratic point ($\theta = \frac{\pi}{2}$), all the states of the tower are degenerate ($\mE = 0$). 
Indeed, we can verify that 
\begin{eqnarray}
    &\left[(S^+_j)^2 - (S^+_{j+1})^2, \left(\vec{S}_j\cdot \vec{S}_{j+1}\right)^2\right] = 0 \nn \\
    &\implies\left[\sumal{j}{}{(-1)^j (S^+_j)^2}, \sumal{j}{}{\left(\vec{S}_j \cdot \vec{S}_{j+1}\right)^2}\right] = 0.
\label{eq:towersymmetry}
\end{eqnarray}
The operator generating the tower of states is thus a symmetry of this integrable Hamiltonian.
\section{New Type of Quantum Scars: Two-Site Quasiparticle Operators}\label{sec:newtype}
In the previous sections, we assumed that the quasiparticle that constitutes the tower is a one-site operator, and we found a family of quantum scarred Hamiltonians where the quasiparticle creation operator is the same as the one in the AKLT chain~\cite{Moudgalya2018a}. 
Here we relax the constraint that the quasiparticle be a single-site operator, and find examples of  Hamiltonians that contain a tower of states of a different type. In this way, we clearly show that our construction gives rise to many Hamiltonians which contain scar states. 
\subsection{Scars with two-site quasiparticles}\label{sec:twosite}
We first set up the general formalism for obtaining Hamiltonians that have a two-site quasiparticle tower of states.
Similarly to the case of single-site quasiparticles, we focus on Hamiltonians that satisfy Eq.~(\ref{eq:FFeigenstate}), i.e. those which have a frustration-free MPS eigenstate $\ket{\psi_A}$ of the form Eq.~(\ref{eq:MPSTI}).  
As illustrated in Sec.~\ref{sec:parent}, the most general nearest-neighbor Hamiltonian with such a property has a local term of the form of Eq.~(\ref{eq:hmatrixform}). 
A two-site quasiparticle $\wB$ has the form
\begin{equation}
    \wB^{[m,n]} = \sumal{l, r}{}{\wOtwo_{ml,nr} A^{[l]} A^{[r]}}.
\label{eq:QPnondiag}
\end{equation}
In shorthand, we write
\begin{equation}
    \mpsgraphics{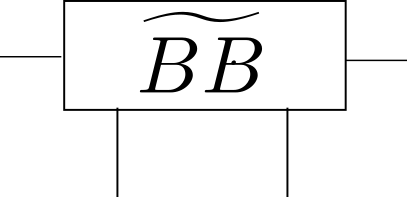} = \mpsgraphics{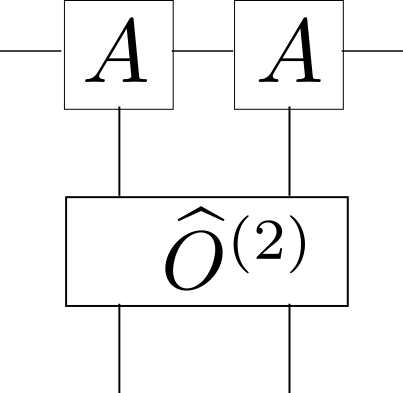}\;\;\textrm{or}\;\;\ket{\wB} = \wOtwo\ket{AA},
\end{equation}
where $\wOtwo$ is a nearest-neighbor two-site operator. 
The wavefunction with the quasiparticle dispersing with momentum $k$ then reads
\begin{equation}
    \ket{\psi_A(\wB, k)} = \sumal{j}{}{e^{ikj} \overset{j,j+1}{\ket{[A\cdots A\wB A \cdots A]}}}.
\label{eq:twositeQPexpr}
\end{equation}
In App.~\ref{app:twositeQPMPS} we show that a set of sufficient conditions for the existence of an eigenstate of the form of Eq.~(\ref{eq:twositeQPexpr}) with energy $(2\mE_1 + \mE_2)$ reads
\begin{eqnarray}
    &\mpsgraphics{hAA.png} = 0\;\;\textrm{or}\;\;\wh_j\ket{AA} = 0,\label{eq:twositesuff1} \\
    &\hspace{-8mm}\mpsgraphics{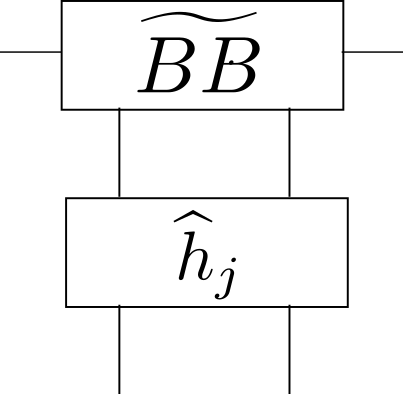} = \mE_2\mpsgraphics{BBtilde.png}\;\;\textrm{or}\;\;(\wh_j - \mE_2)\ket{\wB} = 0, \label{eq:twositesuff2} \\ 
    &\mpsgraphics{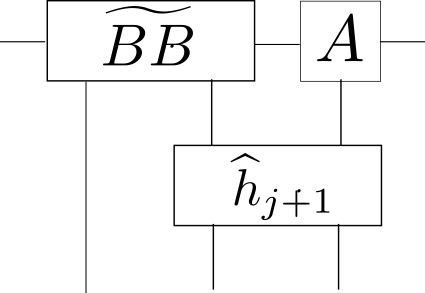} + e^{ik} \mpsgraphics{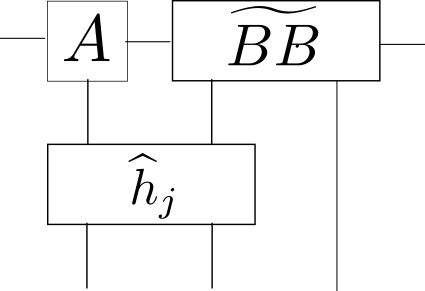} \nn \\
    &=\mE_1\left(\mpsgraphics{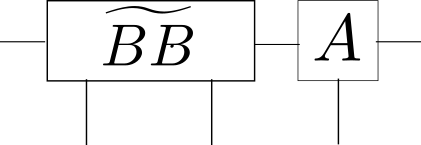} + e^{ik}\mpsgraphics{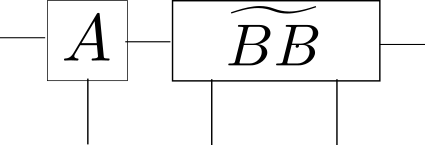}\right) \nn \\
    &\textrm{or}\nn \\
    &(\wh_{j+1} - \mE_1) \overset{j+1}{\ket{\wB A}} + e^{ik} (\wh_{j} - \mE_1) \overset{j+1}{\ket{A \wB}} = 0 \label{eq:twositesuff3}
\end{eqnarray}
where we have used the shorthand notations of the form of Eq.~(\ref{eq:twositeshorthand}).
We now proceed to construct local terms $\{\wh_j\}$ that satisfy Eqs.~(\ref{eq:twositesuff1})-(\ref{eq:twositesuff3}). 
Hamiltonian terms satisfying Eqs.~(\ref{eq:twositesuff1}) and (\ref{eq:twositesuff2}) can be built similarly to the single-site quasiparticle case. 
That is, we first construct the subspaces $\mA$ and $\mBt$, where $\mA$ is defined in Eq.~(\ref{eq:mAdefn}) and $\mBt$ is now defined as
\begin{eqnarray}
    &\mBt \equiv \textrm{span}_X\left\{\mpsgraphics{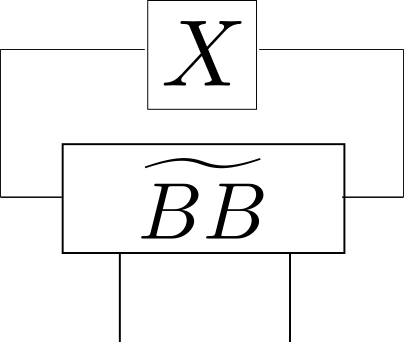}\right\} \nn \\
    &=\textrm{span}_X\{\sumal{m,n}{}{\textrm{Tr}[\wB^{[m,n]}]\ket{m,n}}\} = \wOtwo \mA.
\label{eq:mBtdefn}
\end{eqnarray}
The Hamiltonian term that satisfies Eqs.~(\ref{eq:twositesuff1}) and (\ref{eq:twositesuff2}) then reads
\begin{equation}
    \wh_j = 
    \overset{\hspace{5mm}\mA^c/\mBt \hspace{6mm}\mBt\hspace{4mm}\mA\hspace{4mm}}{
    \left(\begin{array}{c|c|c}
        \bZ^{(\mA^c/\mBt)}_j & \bzero & \bzero \\
        \hline
        \bzero & \mE_2\mathds{1} & \bzero \\
        \hline
        \bzero & \bzero & \bzero
    \end{array}\right)}\hspace{-2mm}\begin{array}{c}
        {\scriptstyle \mA^c/\mBt}  \\
        {\scriptstyle \mBt} \\
        {\scriptstyle \mA}
    \end{array},
\label{eq:hmatrixformgentwo}
\end{equation}
where $Z^{(\mA^c/\mB)}_j$ here is an arbitrary Hermitian matrix with the same dimension as $\mA^c/\mB$, $\mA^c$ being the complement of $\mA$ (see Eq.~(\ref{eq:Acdefn}))  .
To aid in finding a solution to Eq.~(\ref{eq:twositesuff3}), we decompose the two-site quasiparticle $\wB$ as 
\begin{equation}
    \mpsgraphics{BBtilde.png} = \mpsgraphics{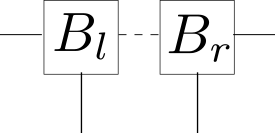}\;\;\textrm{or}\;\;\ket{\wB} = \ket{\Btl \Btr},
\label{eq:QPdecomposition}
\end{equation}
where $\Btl$ and $\Btr$ are one-site MPSs.
Note that the decomposition of Eq.~(\ref{eq:QPdecomposition}) is not unique, and $\Btl^{[\alpha]}$ and $\Btr^{[\alpha]}$ (where $\alpha$ is the physical index) need not be square matrices.
That is, the contracted auxiliary index in Eq.~(\ref{eq:QPdecomposition}) (denoted by a dashed line) can have a different dimension than that of contracted auxiliary indices (denoted by solid lines).
As we show in App.~\ref{app:sufficiency}, it is sufficient to find a two-site operator $\wh_j$ and a one-site MPS $C$ such that 
\begin{eqnarray}
    &\mpsgraphics{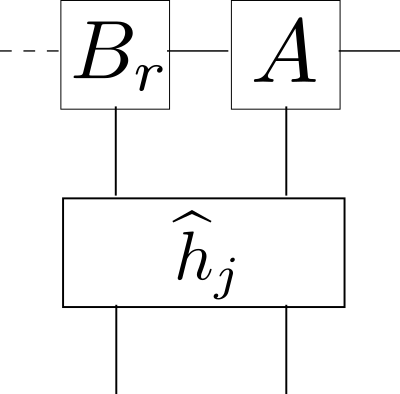} = \mE_1\mpsgraphics{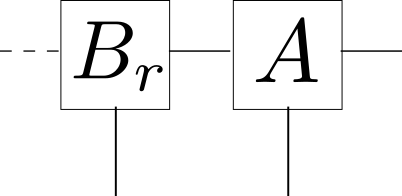} + \mpsgraphics{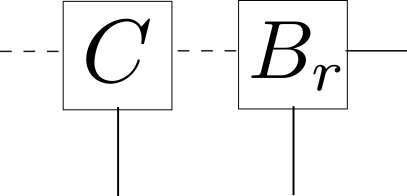} \nn \\
    &\textrm{or}\nn \\
    &\left(\wh_j - \mE_1\right)\overset{j\hspace{5mm}}{\ket{\Btr A}} = \overset{j\hspace{5mm}}{\ket{C \Btr}},\label{eq:rightcond} \\
    &\mpsgraphics{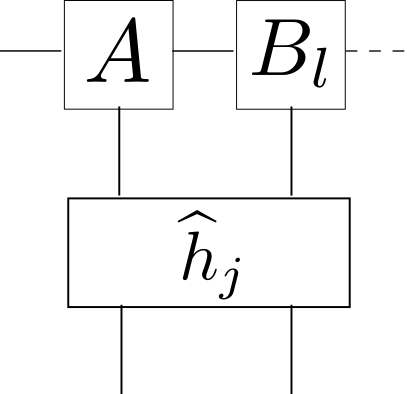} = \mE_1\mpsgraphics{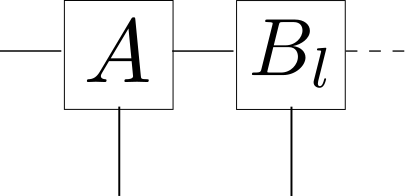} - e^{-ik}\mpsgraphics{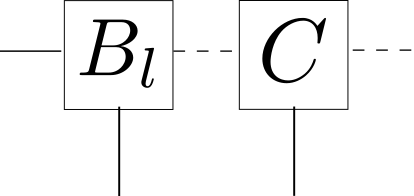} \nn \\
    &\textrm{or}\nn \\
    &\left(\wh_j - \mE_1\right)\overset{j\hspace{3mm}}{\ket{A \Btl}} = -e^{-ik}\overset{j\hspace{5mm}}{\ket{\Btl C}}\label{eq:leftcond}.
\end{eqnarray}
Generically it is not clear we can find terms $\wh_j$ of the form of Eq.~(\ref{eq:hmatrixformgentwo}) (for some $Z^{(\mA^c/\mB)}$) that also satisfy Eqs.~(\ref{eq:rightcond}) and (\ref{eq:leftcond}).
However, in the next subsection, we show that when we restrict ourselves to a particular form of the MPS matrices, we can fix the form of $Z^{(\mA^c/\mB)}$ such that Eq.~(\ref{eq:twositesuff3}) is satisfied.
Similar to the one-site quasiparticle, we can obtain a tower of equally spaced eigenstates composed of the $\wB$ quasiparticles if $\wB$ obeys the additional constraints that generalize Eq.~(\ref{eq:constraint2}).
As we show in App.~\ref{app:twositeQPMPStower}, a sufficient condition is to constrain the quasiparticles to be at least one site away from each other. 
That is, we require
\begin{eqnarray}
    &\wOtwo_j\ket{\wB} = 0,\label{eq:twoconstraint1} \\
    &\wOtwo_j\overset{j+1}{\ket{A\wB}} = \wOtwo_{j+1} \overset{j+1}{\ket{\wB A}} = 0,\label{eq:twoconstraint2} \\
    & \ket{\wB \wB} = \mpsgraphics{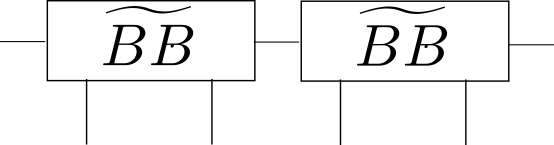} = 0,\nn \\
\label{eq:twoconstraint3}
\end{eqnarray}
where $\wOtwo_j$ is the two-site quasiparticle creation operator.
Similar to Eq.~(\ref{eq:towerform}), we then obtain a tower of equally spaced eigenstates $\{\ket{S^{(2)}_{3n}}\}$, where
\begin{eqnarray}
    &\ket{S^{(2)}_{3n}} = (\mP^{(2)})^n \ket{\psi_A},\;\; \wH\ket{S^{(2)}_{3n}} = n(2\mE_1 + \mE_2)\ket{S^{(2)}_{3n}},\nn \\
    &n \leq L/3, 
\end{eqnarray}
and
\begin{equation}
    \mP^{(2)} = \sumal{j=1}{L}{e^{ikj} \wOtwo_j}. 
\end{equation}
Note that the tower is guaranteed to end on the state after $\left(L/3 + 1\right)$ applications of $\mP^{(2)}$ on the state $\ket{\psi_A}$ since we can have at most $L/3$ $\wB$ quasiparticles on the chain that satisfy the constraints of Eqs.~(\ref{eq:twoconstraint1})-(\ref{eq:twoconstraint3}).
\subsection{Concrete example: Perturbed Potts MPS}\label{sec:pertpotts}
\begin{figure*}[t!]
\begin{tabular}{cc}
\includegraphics[scale =0.45]{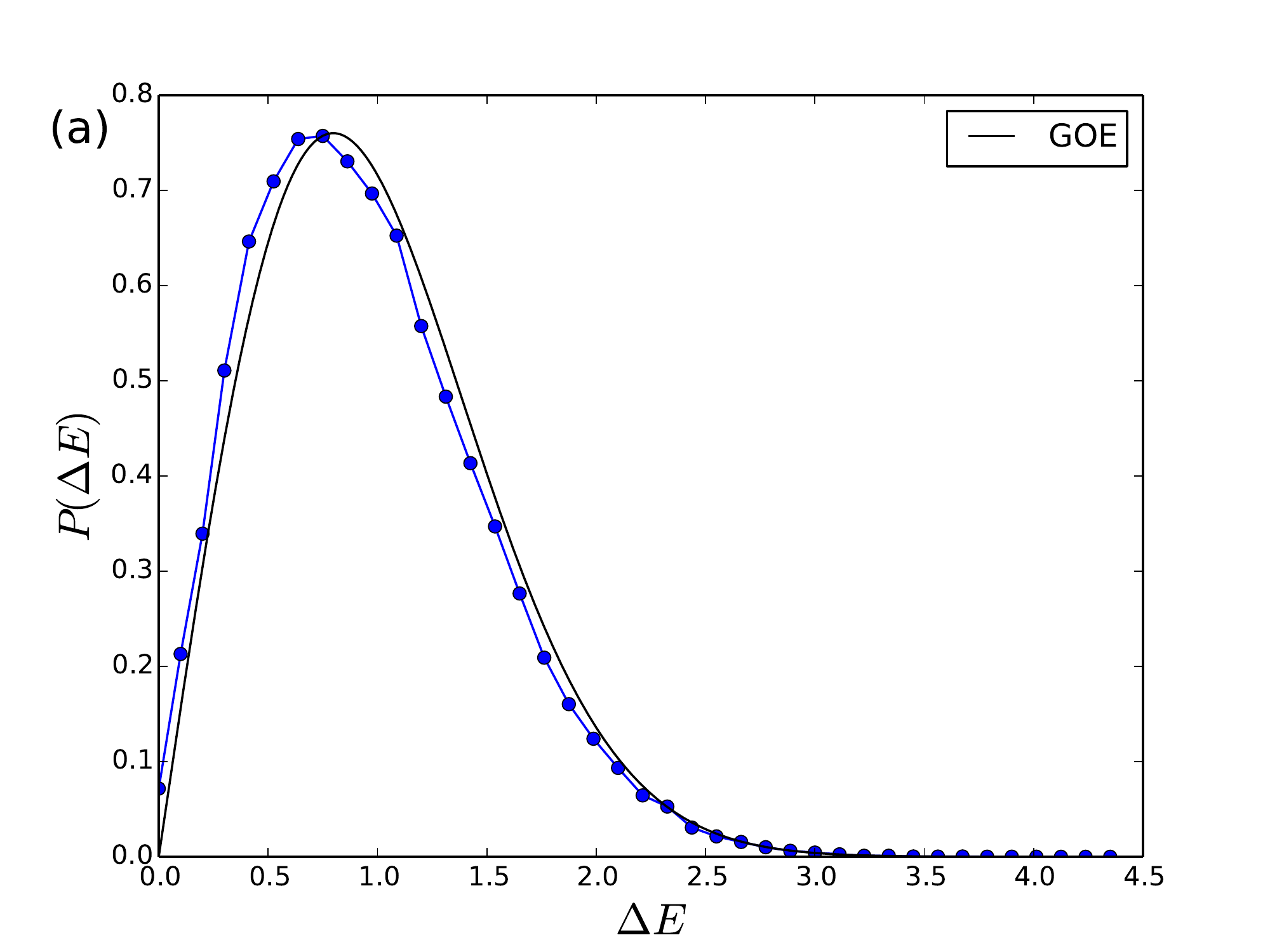}
&\includegraphics[scale = 0.45]{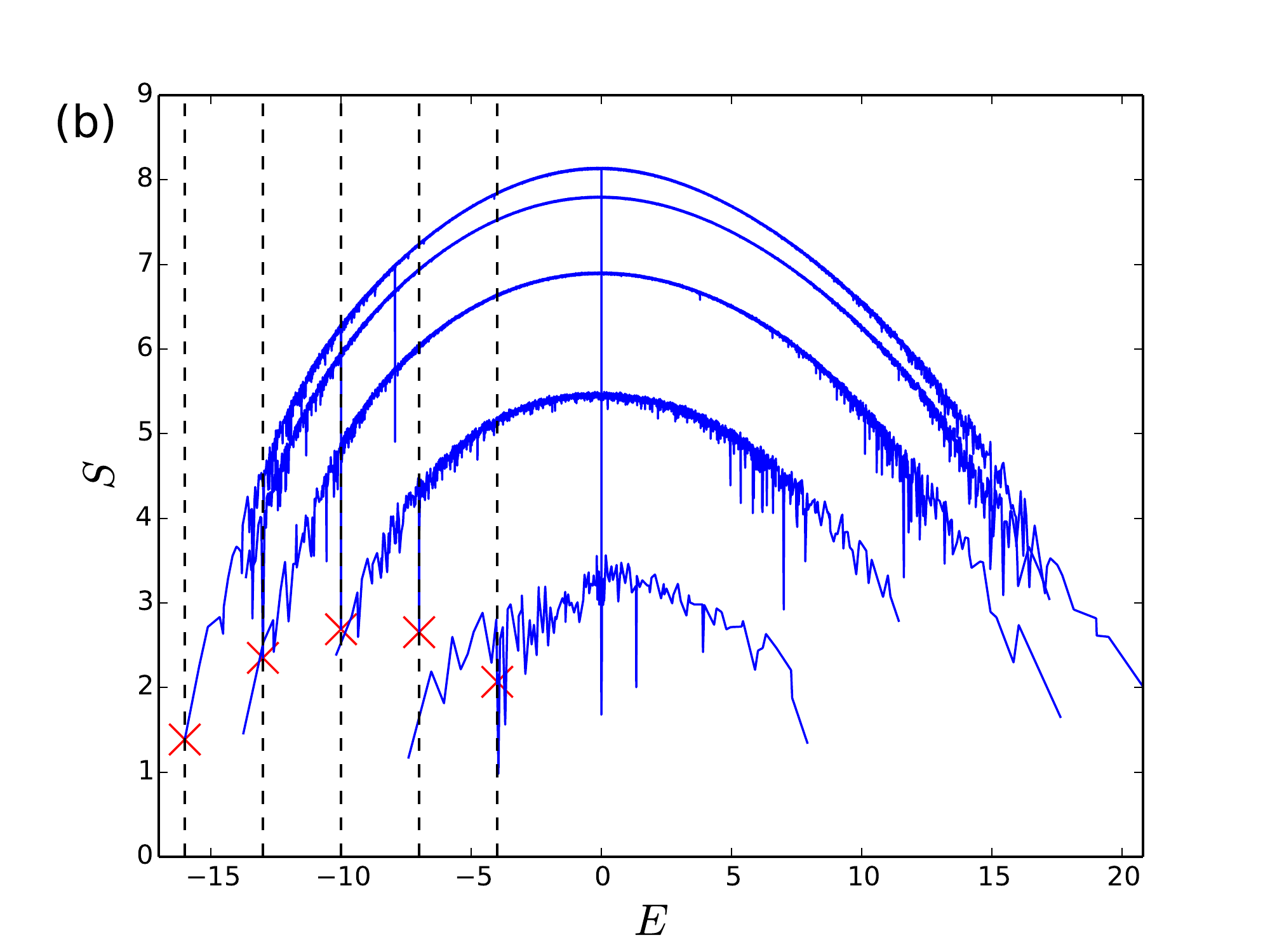}
\end{tabular}
\caption{(Color online) (a) Level statistics for the spectrum of the perturbed Potts model Eq.~(\ref{eq:PP}). The data shows a good fit to a Gaussian Orthogonal Ensemble (GOE) statistics, indicating the non-integrability of the model. Data is shown for system size $L = 16$ and the quantum number sector $(S_z, e^{ik}, I, P_z) = (0, 1, 1, 1)$, where $k$ is the momentum, and $I$ and $P_z$ are quantum numbers defined in Sec.~\ref{sec:pertpotts}.
(b) Entanglement entropies $S_{EE}$, of the eigenstates in each of the sectors $(S_z, e^{ik}, I) = (3n, (-1)^n, (-1)^n)$, $n\leq 4$.
The crosses indicates the two-site quasiparticle tower discussed in the main text and the vertical lines indicate their energies.
}
\label{level_spacing}
\end{figure*}
We now provide a concrete example of a model where we find a two-site quasiparticle tower of states of the form discussed in Sec.~\ref{sec:twosite}.
Throughout this section, we use as an example the MPS
\begin{equation}
    A^{[+]} = \frac{1}{\sqrt{2}}\sigma^+,\;\;A^{[0]} = \frac{1}{\sqrt{2}}\mathds{1}_{2\times 2},\;\; A^{[-]} = \frac{1}{\sqrt{2}}\sigma^-.
\label{eq:PottsMPS}
\end{equation}
This MPS is the ground state of the Hamiltonian \cite{Klumper1993,OBrien2019} 
\begin{align}
H_{\text{PP}} = \sum\limits_j \left({S_j^+}^2{S_{j+1}^-}^{\!\!\!\!2} - S_j^+S_{j+1}^-  + \text{h.c.}\right)\ .
\label{eq:PP}
\end{align}  
As apparent, $H_{\rm PP}$ has a $U(1)$ symmetry corresponding to the total spin $\sum_j S_j^z$. 
In addition, it has a spin-flip symmetry ($P_z$) given by $S_j^+ \leftrightarrow S_j^-$ and $S_j^z \leftrightarrow -S_j^z$, and inversion symmetry ($I$), defined by taking all operators at site $j$ to site $L+1-j$ for a chain of length $L$.  
The Hamiltonian of Eq.~(\ref{eq:PP}) arises from perturbing the $S_3$-invariant three-state Potts chain by shortest-range $U(1)$-invariant interaction \cite{OBrien2019}.  %

We show that $H_{\rm PP}$ has a two-site quasiparticle tower of exact eigenstates, and derive a family of Hamiltonians with a similar tower of states.
In order to do so,
we construct the subspace $\mA$ defined in Eq.~(\ref{eq:mAdefn}).
As shown in Eq.~(\ref{eq:APotts}) in App.~\ref{app:examplespotts}, the subspace reads
\begin{equation}
    \mA = \textrm{span}\{\ket{J_{2,0}}, \ket{J_{2,-1}}, \ket{J_{1,0}}, \ket{J_{2,1}}\}. 
\label{eq:mAOFModel}
\end{equation}
We now consider the quasiparticle creation operator
\begin{equation}
    \mP^{(2)} = \sumal{j}{}{(-1)^j \underbrace{\left((S^+_j)^2 S^+_{j+1} + S^+_j (S^+_{j+1})^2\right)}_{=\wOtwo_j}}.
\label{eq:OFoperator}
\end{equation}
As shown in Eq.~(\ref{eq:genBtsubspacenorm}) in App.~\ref{app:examplespotts}, the subspace $\mBt$ defined in Eq.~(\ref{eq:mBtdefn}) reads
\begin{equation}
    \mBt = \{\ket{J_{2,2}}\}.
\label{eq:mBtOFModel}
\end{equation}
Thus, a nearest-neighbor Hamiltonian term $\wh_j$ that satisfies Eqs.~(\ref{eq:twositesuff1}) and (\ref{eq:twositesuff2}) has the form of Eq.~(\ref{eq:hmatrixformgentwo}). 
We now obtain the most general Hamiltonian of the form of Eq.~(\ref{eq:hmatrixformgentwo}) that also satisfies Eq.~(\ref{eq:twositesuff3}) (i.e. Eqs.~(\ref{eq:rightcond}) and (\ref{eq:leftcond})).
%
%
Following Eq.~(\ref{eq:QPnondiag}), we obtain that the only non-vanishing element of the quasiparticle tensor $\wB$ obtained with the operator $\wOtwo$ in Eq.~(\ref{eq:OFoperator}) reads (see Eq.~(\ref{eq:Btexpr}))
\begin{equation}
    \wB^{[+, +]} = \sigma^- = \begin{pmatrix}0 \\ 1\end{pmatrix} \begin{pmatrix}1 & 0\end{pmatrix}. 
\label{eq:BBt}
\end{equation}
Using Eqs.~(\ref{eq:QPdecomposition}) and (\ref{eq:BBt}), we obtain
\begin{equation}
    {\Btl}^{[+]} = \begin{pmatrix} 0 \\ 1\end{pmatrix},\;\;{\Btr}^{[+]} = \begin{pmatrix}1 & 0\end{pmatrix}.
\label{eq:BtlBtr}
\end{equation}
In App.~\ref{app:examplespotts} we find that Eqs.~(\ref{eq:rightcond}) and (\ref{eq:leftcond}) are satisfied if (see Eqs.~(\ref{eq:E1E2}), (\ref{eq:C0set}), and (\ref{eq:hsolnC}))
\begin{eqnarray}
    &\mE_1 = \mE_2 \equiv \mE,\;\; C^{[\alpha]} = - c_0\mE \delta_{\alpha, 0} \nn \\
    &\wh_j \ket{J_{1,1}} = 2\mE \ket{J_{1,1}}.
\end{eqnarray}
The Hamiltonian term that satisfies Eqs.~(\ref{eq:twositesuff1})-(\ref{eq:twositesuff3}) then reads
\begin{equation}
    \wh_j = 
    \overset{\hspace{5mm}(\mA^c/\mBt)/\mC \hspace{6mm} \mC \hspace{6mm}\mBt\hspace{4mm}\mA\hspace{4mm}}{
    \left(\begin{array}{c|c|c|c}
        \bZ^{(\mA^c/\mBt)/\mC}_j & \bzero & \bzero & \bzero \\
        \hline
        \bzero & 2\mE\mathds{1} & \bzero & \bzero \\
        \hline
        \bzero & \bzero & \mE\mathds{1} & \bzero \\
        \hline
        \bzero & \bzero & \bzero & \bzero
    \end{array}\right)}\hspace{-2mm}\begin{array}{c}
        {\scriptstyle (\mA^c/\mBt)/\mC}  \\
        {\scriptstyle \mC} \\
        {\scriptstyle \mBt} \\
        {\scriptstyle \mA}
    \end{array},
\label{eq:hmatrixformgenPotts}
\end{equation}
where $\bZ^{(\mA^c/\mBt)/\mC}_j$ is an arbitrary matrix with the same dimension as $(\mA^c/\mBt)/\mC$ and we have defined the subspace $\mC$ as
\begin{equation}
    \mC = \{ \ket{J_{1,1}}\},
\label{eq:Cdefn}
\end{equation}
Any Hamiltonian of the form of Eq.~(\ref{eq:hmatrixformgenPotts}) hosts a quasiparticle excited state of the form of Eq.~(\ref{eq:twositeQPexpr}) created by the operator $\mP^{(2)}$ of Eq.~(\ref{eq:OFoperator}).
In addition, by brute force we have verified that the operator $\wOtwo$ in Eq.~(\ref{eq:OFoperator}) satisfies Eqs.~(\ref{eq:twoconstraint1})-(\ref{eq:twoconstraint3}) with the MPS of Eq.~(\ref{eq:PottsMPS}).
Thus we find a 6-parameter family of Hamiltonians hosting a tower of two-site quasiparticles with energies $E = 3 n \mE$. The Hamiltonians are of the form $\wH=\sum_h \wh_j$, where
\begin{eqnarray}
    \wh_j &= \mE\ket{J_{2,2}}\bra{J_{2,2}} + 2\mE\ket{J_{1,1}}\bra{J_{1,1}} \nn \\
    &\quad+ \sumal{m, n = 0}{2}{z^{(m, n)}_j \ket{J_{m,-m}}\bra{J_{n,-n}}}\ .
\label{eq:mostgenPotts}
\end{eqnarray}
As before, $\mE$ in Eq.~(\ref{eq:mostgenPotts}) is merely an overall scale. 
Indeed, the terms for the perturbed Potts model of Eq.~(\ref{eq:PP}) read (up to overall constant factors and energy shifts)
\begin{align}
    \wh_j =& \ket{J_{2,2}}\bra{J_{2,2}} + \ket{J_{2,-2}}\bra{J_{2,-2}} + 3\ket{J_{0,0}}\bra{J_{0,0}}\nn \\
    &+ 2\left(\ket{J_{1,1}}\bra{J_{1,1}} + \ket{J_{1,-1}}\bra{J_{1,-1}}\right)\ ,
\label{eq:OFProjectors}
\end{align}
which can be obtained from Eq.~(\ref{eq:mostgenPotts}) by setting
\begin{equation}
    \mE = 1,\;\; z^{(m,n)}_j = (3 - m)\delta_{m,n}. 
\end{equation}
As discussed in App.~\ref{app:Pottsfamily}, we can repeat this exercise for the generalized perturbed Potts MPS that reads
\begin{equation}
    A^{[+]} = c_+ \sigma^+,\;\; A^{[0]} = c_0 \mathds{1}_{2\times 2},\;\; A^{[-]} = c_-\sigma^-,
\label{eq:genPottsMPSarb}
\end{equation}
where one of $c_+$, $c_0$, and $c_-$ is fixed by normalization.
We note that we can build a scar-preserving deformation from the perturbed Potts model of Eq.~(\ref{eq:PP}) to the spin-1 pure biquadratic Hamiltonian similar to the scar-preserving deformation of the AKLT chain illustrated in Sec.~\ref{sec:deformation}. 
However, unlike in the AKLT case, the quasiparticle operator $\mP^{(2)}$ of Eq.~(\ref{eq:OFoperator}) is \textit{not} a symmetry of the pure biquadratic Hamiltonian. 
\subsection{Numerical evidence of Quantum Scars}\label{sec:numevidence}
As a check on our calculations, we here present numerical evidence for the two-site quasiparticle tower of states in the perturbed Potts model of Eq.~(\ref{eq:PP}). In Fig.~\ref{level_spacing}(a) we also give the level-spacing distribution for one of the quantum-number sectors for periodic boundary conditions.
The fact that it fits to the Gaussian Orthogonal Ensemble gives strong evidence for the non-integrability of the model.
%
%
The presence of these excited states can be seen by dips in the von Neumann entanglement entropy $S_{EE}$, as shown in Fig.~\ref{level_spacing}(b). The states of the two-site quasiparticle tower are denoted by the red crosses.
While we only analyzed above one family of exact excited states in the perturbed Potts model, there are others analogous to the Arovas states found previously in the AKLT chain~\cite{Arovas1989, Moudgalya2018a, Moudgalya2018b} as well as the family of excitations found in Ref.~\cite{chattopadhyay2019quantum} (all of which have energy $E = 0$ here).
\section{Conclusions}\label{sec:conclusions}
We have provided a formalism to search and construct quantum scarred models starting from a Matrix Product State wavefunction.
The scarred Hamiltonians we construct have a quasiparticle tower of exact eigenstates in their spectra. 
We have illustrated our method thoroughly for single-site quasiparticles by constructing a 6-parameter family of nearest-neighbor Hamiltonians that have the exact quantum scars of the AKLT chain as eigenstates. 
Applying our construction to a more general class of MPS wavefunctions, we showed that the scars of AKLT chain~\cite{Moudgalya2018a, Moudgalya2018b} can be continuously deformed to a symmetry of the pure biquadratic spin-1 model, an integrable model. 
Further, we generalized our construction to the case of two-site quasiparticles and we obtain new types of quantum scarred models.
We illustrated these results with the help of a concrete example of the perturbed Potts model~\cite{OBrien2019}, which we show that hosts a tower of exact eigenstates composed of two-site quasiparticles.
We believe that our formalism can be generalized to include a wide variety of known models with quantum scars, including the spin-$S$ AKLT chains.
We also expect that many more models with quantum scars can be obtained by relaxing several assumptions introduced for pedagogical reasons in this work, such as periodic boundary conditions, single-site or two-site quasiparticle creation operators or nearest-neighbor Hamiltonians. 
It would also be interesting to work out the exact relation between the MPS construction of scars and the unified formalisms recently proposed in Refs.~\cite{mark2020unified} and \cite{bull2020quantum}, and formulate a dimension independent understanding of scars. 
It should also be possible to extend our formalism to higher dimensions using Projected Entangled Pair States (PEPS)~\cite{schuch2010peps} and search for higher dimensional quantum scarred models, a question we defer for future work. 
On a different note, given that the PXP model has exact MPS eigenstates~\cite{lin2019exact, shiraishi2019connection} as well as an approximate MPS ground state~\cite{mark2019new, lesanovsky2011many}, it is natural to ask if the scars exhibited there have any connections to the formalism developed here.
Furthermore, the deformation to integrability raises questions of whether quantum scarred Hamiltonians are always connected to integrable ones, as suggested by numerical explorations around the PXP model~\cite{khemani2019signatures}. 
\textit{Note added}: Recently Ref.~\cite{mark2020unified} derived a general nearest-neighbor Hamiltonian that exhibits the scars of the AKLT chain as eigenstates using a different approach. Our results agree where they overlap. 
\section*{Acknowledgements}
We thank Huan He, Tom Iadecola, Olexei Motrunich, Arijeet Pal, Laurens Vanderstraeten, and Frank Verstraete for useful discussions. 
B.A.B. and N.R. were supported by the Department of Energy Grant No. de-sc0016239, the Schmidt Fund for Innovative Research, Simons Investigator Grant No. 404513  the Packard Foundation. Further support was provided by the National Science Foundation EAGER Grant No. DMR 1643312, and NSF-MRSEC DMR-1420541. E.O.B. and P.F.  were supported by the Engineering and Physical Sciences Research Council, UK, through grant EP/N509711/1 1734484 (EOB) along with grants EP/S020527/1 and EP/N01930X (PF).

\appendix
\section{Examples of $\mA$ and $\mB$ subspaces for the AKLT-like MPS}\label{app:examples}
Here we compute the subspaces $\mA$ and $\mB$ of Eqs.~(\ref{eq:mAdefn}) and (\ref{eq:mBdefn}) for an MPS of the form 
\begin{equation}
    A^{[+]} = c_+ \sigma^+,\;\;\;A^{[0]} = c_0 \sigma^0,\;\;\;A^{[-]} =  c_- \sigma^-,
\label{eq:genAKLTMPS}
\end{equation}
where $\sigma^+$, $\sigma^-$, and $\sigma^0 \equiv \sigma^z$ are the Pauli matrices.
Note that the AKLT MPS of Eq.~(\ref{eq:AKLTMPS}) is recovered by setting
\begin{equation}
    \left(c_+, c_0, c_-\right) = \frac{1}{\sqrt{3}}\left(\sqrt{2}, -1, -\sqrt{2}\right).
\label{eq:cAKLT}
\end{equation}
Compactly, we can write
\begin{equation}
    A^{[m]} = c_{m} \sigma^m. 
\label{eq:Amcompact}
\end{equation}
We first compute the subspace $\mA$ for the two-site MPS defined as
\begin{equation}
    \mA \equiv \textrm{span}_X \left\{\sumal{m, n \in \{+, 0, -\}}{}{\textrm{Tr}\left[X A^{[m]} A^{[n]}\right]\ket{m,n}}\right\},
\label{eq:mAdefnapp}
\end{equation}
where $X$ runs over a basis of $2 \times 2$ matrices.  
Using Eqs.~(\ref{eq:Amcompact}) and (\ref{eq:mAdefnapp}), we obtain 
\begin{equation}
    \mA = \textrm{span}_X \left\{\sumal{m, n \in \{+, 0, -\}}{}{c_m c_n \textrm{Tr}\left[X \sigma^{m} \sigma^{n}\right]\ket{m,n}}\right\}.
\label{eq:Acomp}
\end{equation}
Choosing $X$ from the convenient basis $\{\frac{1}{\sqrt{2}}\mathds{1}, \sigma^+, \frac{1}{\sqrt{2}}\sigma^z, \sigma^-\}$ of $2 \times 2$ matrices, the subspace $\mA$ in Eq.~(\ref{eq:Acomp}) can be straightforwardly computed to be
\begin{eqnarray}
    &\mA = \textrm{span}\{\frac{c_+ c_-}{\sqrt{2}} \left(\ket{+-} + \ket{-+}\right) + \sqrt{2} c_{0}^2 \ket{00},\nn \\
    &c_- c_0 \left(\ket{-0} - \ket{0-}\right), \nn \\
    &\frac{c_+ c_-}{\sqrt{2}} \left(\ket{+-} - \ket{-+}\right), \nn \\
    &c_{0} c_+ \left(-\ket{+0} + \ket{0+}\right)\}.
\label{eq:genAsubspace}
\end{eqnarray}
After normalization, it reads
\begin{eqnarray}
    &\hspace{-5mm}\mA = \textrm{span}\{\frac{1}{\sqrt{2\left(|c_+ c_-|^2 + 2|c_0|^4\right)}}\left(c_+ c_- \left(\ket{+-} + \ket{-+}\right) + 2 c_{0}^2 \ket{00}\right),\nn \\
    &\frac{1}{\sqrt{2}}\left(\ket{-0} - \ket{0-}\right), \nn \\
    &\frac{1}{\sqrt{2}}\left(\ket{+-} - \ket{-+}\right), \nn \\
    &\frac{1}{\sqrt{2}}\left(\ket{+0} - \ket{0+}\right)\},
\label{eq:genAsubspacenorm}
\end{eqnarray}
Thus, for the AKLT MPS, using Eq.~(\ref{eq:cAKLT}) we obtain 
\begin{eqnarray}
    &\mA = \textrm{span}\{ \sqrt{\frac{1}{3}}\left(\ket{+-} + \ket{-+} - \ket{00}\right), \nn \\
    &\frac{1}{\sqrt{2}}\left(\ket{-0} - \ket{0-}\right), \nn \\
    &\frac{1}{\sqrt{2}}\left(\ket{+-} - \ket{-+}\right) \nn \\
    &\frac{1}{\sqrt{2}}\left(\ket{+0} - \ket{0+}\right)\}.
\label{eq:AAKLT}
\end{eqnarray}
The states in Eq.~(\ref{eq:AAKLT}) are indeed proportional the total angular momentum eigenstates obtained from two spin-1's: 
\begin{equation}
    1 \otimes 1 = 2 \oplus 1 \oplus 0.
\end{equation}
Thus, $\mA$ for the AKLT MPS reads
\begin{equation}
    \mA = \textrm{span}\{ \ket{J_{0, 0}}, \ket{J_{1, -1}}, \ket{J_{1, 0}}, \ket{J_{1, 1}} \},
\label{eq:mA}
\end{equation} 
where the total angular momentum eigenstates $\ket{J_{j,m}}$ are enumerated in App.~\ref{sec:totalspin}.
Using the operator $\wO = (S^+)^2$, the $\mB$ subspace defined in Eq.~(\ref{eq:mBopdefn}) reads
\begin{equation}
    \mB = \underbrace{\left((S^+)^2\otimes \mathds{1} - \mathds{1} \otimes (S^+)^2\right)}_{=\widetilde{(S^+)^2}}\mA,
\label{eq:mBAKLT}
\end{equation}
Note that the action of $\widetilde{(S^+)^2}$ on $\ket{J_{1,1}}$ in $\mA$ vanishes since $\widetilde{(S^+)^2}$ is a spin-2 operator:
\begin{equation}
    ((S^+)^2 \otimes \mathds{1} - \mathds{1} \otimes (S^+)^2)\frac{\ket{+ 0} - \ket{0 +}}{\sqrt{2}} = 0.
\end{equation}
Furthermore, we obtain
\begin{equation}
    ((S^+)^2 \otimes \mathds{1} - \mathds{1} \otimes (S^+)^2)\frac{\left(c_+ c_- \left(\ket{+-} + \ket{-+}\right) + 2 c_{0}^2 \ket{00}\right)}{\sqrt{2\left(|c_+ c_-|^2 + 2|c_0|^4\right)}} = 0.
\end{equation} 
On the remaining vectors in $\mA$, $\ket{J_{1,0}}$ and $\ket{J_{1,-1}}$, we find that
\begin{eqnarray}
    &((S^+)^2 \otimes \mathds{1} - \mathds{1} \otimes (S^+)^2) \frac{\ket{+-} - \ket{-+}}{\sqrt{2}} = -\sqrt{2}\ket{++} \propto \ket{J_{2,2}} \nn \\
    &((S^+)^2 \otimes \mathds{1} - \mathds{1} \otimes (S^+)^2) \frac{\ket{-0} - \ket{0-}}{\sqrt{2}} = \frac{\ket{+0}+\ket{0+}}{\sqrt{2}} \propto \ket{J_{2,1}},\nn \\
\end{eqnarray}
which have been shown heuristically in the main text. 
Thus, we obtain
\begin{eqnarray}
    &\mB = \textrm{span}\{\frac{1}{\sqrt{2}}\left(\ket{+0} + \ket{0+}\right), \ket{++}\} = \textrm{span}\{\ket{J_{2,1}}, \ket{J_{2,2}}\},\nn \\
\label{eq:genBsubspacenorm}
\end{eqnarray}
which is independent of the $c_m$'s.
Thus, using Eqs.~(\ref{eq:AAKLT}) and (\ref{eq:genBsubspacenorm}), we obtain
\begin{eqnarray}
    &\mA^c/\mB = \textrm{span}\{ \ket{J_{2,-1}}, \ket{J_{2,-2}},\nn \\
    &\frac{1}{\sqrt{|c_+ c_-|^2 + 2|c_0|^4}}\left(-c_0^2\left(\ket{+-} + \ket{-+}\right) + c_{+} c_- \ket{00}\right) \}.\nn \\
\label{eq:mAc/mB}
\end{eqnarray}
\section{Total Angular Momentum Eigenstates}\label{sec:totalspin}
In this Appendix, we list the various total angular momentum eigenstates of two spin-1's.
We denote the single site spin-1 basis vectors with $S_z = +1, 0, -1$ by $\ket{+}, \ket{0}, \ket{-}$ respectively. 
Labelling the state with total spin $j$, $j \in \{0, 1, 2\}$ and its $z$-projection $m$, $m \in \{-j, -j+1, \cdots, j\}$ as $\ket{J_{j,m}}$, they read
\begin{eqnarray}
    &\ket{J_{2,\pm 2}} = \ket{\pm\ \pm}\nn \\
    &\ket{J_{2,\pm 1}} = \frac{1}{\sqrt{2}}\left(\ket{\pm \ 0} + \ket{0\ \pm}\right) \nn \\
    &\ket{J_{2,0}} = \frac{1}{\sqrt{6}}\left(\ket{+\ -} + 2\ket{0\ 0} + \ket{-\ +}\right) \nn \\
    &\ket{J_{1, \pm 1}} = \frac{1}{\sqrt{2}}\left(\ket{\pm\ 0} - \ket{0\ \pm}\right) \nn \\
    &\ket{J_{1, 0}} = \frac{1}{\sqrt{2}}\left(\ket{+\ -} - \ket{-\ +}\right) \nn \\
    &\ket{J_{0, 0}} = \frac{1}{\sqrt{3}}\left(\ket{+\ -} - \ket{0\ 0} + \ket{-\ +}\right).
\label{eq:totalJ}
\end{eqnarray}
\section{Single-Site Quasiparticle Exact Eigenstates in the MPS Language}\label{app:QPeigenstateMPS}
\subsection{Single quasiparticle}
In this section we show that the conditions of Eqs.~(\ref{eq:cond1}) and (\ref{eq:cond2}) imply the existence of a quasiparticle eigenstate of the Hamiltonian Eq.~(\ref{eq:FFeigenstate}).
Rewriting the conditions here for convenience, they read 
\begin{eqnarray}
    &\wh_j \ket{AA} = 0\label{eq:QPcondition1} \\
    &\wh_j\left(\ket{BA} + e^{ik} \ket{AB}\right) = \mE\left(\ket{BA} +e^{ik} \ket{AB}\right)\nn \\
    &\implies \left(\wh_j - \mE\right)\left(\ket{BA} + e^{ik}\ket{AB}\right) = 0.\label{eq:QPcondition2}
\end{eqnarray}
Note that in Eqs.~(\ref{eq:QPcondition1}) and (\ref{eq:QPcondition2}), $\wh_j$ is a two-site operator, with $j$ being the left site. 
Using these two expressions, the action of the Hamiltonian $\wH$ on the state with one quasiparticle reads
\begin{eqnarray}
    &\wH \overset{j}{\ket{[A \cdots A B A \cdots A]}} = \left(\wh_{j-1} + \wh_j\right)\overset{j}{\ket{[A \cdots A B A \cdots A]}} \nn \\
    &= \left(\wh_{j-1} + \mE\right)\overset{j}{\ket{[A \cdots A B A \cdots A]}} \nn \\
    &- e^{ik} \left(\wh_j - \mE\right)\overset{j}{\ket{[A \cdots A A B \cdots A]}},\nn \\
\label{eq:locQPcond}
\end{eqnarray}
where the first line follows from Eq.~(\ref{eq:cond1}) and the second line from Eq.~(\ref{eq:cond2}) with $\wh_j$.
Defining the shorthand notation
\begin{equation}
    \ket{B_j} \equiv \overset{j}{\ket{[A\cdots A B A\cdots A]}}, 
\end{equation}
Eq.~(\ref{eq:locQPcond}) can be rewritten as
\begin{equation}
    \wH\ket{B_j} = \left(\wh_{j-1} + \mE\right)\ket{B_j} - e^{ik} \left(\wh_j - \mE\right)\ket{B_{j+1}}.
\label{eq:Hamilaction}
\end{equation}
Thus,
\begin{eqnarray}
    &\wH\ket{\psi_A\left(B, k\right)} = \wH\sumal{j=1}{L}{e^{ikj}\ket{B_j}} \nn \\
    &= \sumal{j=1}{L}{e^{ikj}\left[\left(\wh_{j-1}\ket{B_j} - e^{ik} \wh_j\ket{B_{j+1}}\right) \right.}\nn \\
    &{\left.+ \mE\left(\ket{B_j} + e^{ik} \ket{B_{j+1}}\right)\right]} \nn \\
    &= \sumal{j=1}{L}{e^{ikj}\left[\left(\wh_{j-1}\ket{B_{j}} - \wh_{j-1}\ket{B_{j}}\right) + \mE\left(\ket{B_j} + \ket{B_{j}}\right)\right]} \nn \\
    &=2\mE \sumal{j=1}{L}{e^{ikj}\ket{B_j}} = 2\mE\ket{\psi_A\left(B, k\right)}.
\end{eqnarray}
Thus, the conditions of Eqs.~(\ref{eq:cond1}) and (\ref{eq:cond2}) guarantee a quasiparticle eigenstate of $\wH$ with energy $E = 2\mE$. 
\subsection{Tower of states}\label{sec:singlesitetower}
Here we show that in addition to Eqs.~(\ref{eq:cond1}) and (\ref{eq:cond2}), Eqs.~(\ref{eq:constraint1}) and (\ref{eq:constraint2}) guarantee the existence of a tower of quasiparticle exact eigenstates (rewriting here for convenience)
\begin{eqnarray}
    \wO^2\ket{B} = 0\label{eq:QPconstraints1} \\
    \ket{BB} = 0.
\label{eq:QPconstraints2}
\end{eqnarray}
We first illustrate the exactness for two quasiparticles dispersing in the ground state background, by defining the configuration of two quasiparticles as
\begin{equation}
    \ket{B_{j_1}, B_{j_2}} = \overset{\hspace{1mm}j_1\hspace{11mm}j_2}{\ket{[A \cdots A B A \cdots A B A \cdots A]}}.
\label{eq:twoQPconfig}
\end{equation}
Note that
\begin{equation}
    \ket{B_j, B_{j+1}} = 0\;\;\textrm{and}\;\;\ket{B_j, B_j} = 0.
\label{eq:Bvanish}
\end{equation}
As a consequence of Eqs.~(\ref{eq:QPconstraints1}) and (\ref{eq:QPconstraints2}), we are guaranteed to have at least one $A$ in between the $B$'s in the configuration of Eq.~(\ref{eq:twoQPconfig}).
Thus, the Hamiltonian $\wH$ acts independently on each of the quasiparticles. 
That is, similar to Eq.~(\ref{eq:Hamilaction}), we obtain (with subscripts taken modulo $L$) 
\begin{widetext}
\begin{equation}
    \hspace{-10mm}\wH\ket{B_{j}, B_{j + n}} = \fourpartdef{\left(\wh_{j - 1} + \mE + \wh_{j + n - 1} + \mE\right)\ket{B_{j}, B_{j + n}} - e^{ik} \left(\wh_{j} - \mE\right)\ket{B_{j+1}, B_{j + n}} - e^{ik} \left(\wh_{j + n} - \mE\right)\ket{B_{j}, B_{j + n +1}}}{3 \leq n \leq L - 3}{\left(\wh_{j - 1} + \mE + \wh_{j + 1} + \mE\right)\ket{B_{j}, B_{j + 2}} - e^{ik} \left(\wh_{j + 2} - \mE\right)\ket{B_{j}, B_{j + 3}}}{n = 2}{\left(\wh_{j - 1} + \mE + \wh_{j + L - 3} + \mE\right)\ket{B_{j}, B_{j + L - 2}} - e^{ik} \left(\wh_{j} - \mE\right)\ket{B_{j + 1}, B_{j + L - 2}}}{n = L - 2}{0},
\label{eq:twoQPscatteringlong}
\end{equation}
where we have used Eq.~(\ref{eq:Bvanish}).
To write Eq.~(\ref{eq:twoQPscatteringlong}) compactly, we first obtain a useful identity by applying Eqs.~(\ref{eq:QPcondition2}) and (\ref{eq:Bvanish}) 
\begin{equation}
    \left(\wh_{j+1} - \mE\right)\ket{B_j, B_{j+2}} = \left(\wh_{j+1} - \mE\right)\overset{\;\;\;j\;\;\;j+2}{\ket{[A\cdots A B A B A\cdots A]}}=-e^{-ik}\left(\wh_{j+1} - \mE\right)\overset{\;\;\;j\;\;\;j+2}{\ket{[A\cdots A B B A A\cdots A]}} = 0.
\label{eq:usefulidentity}
\end{equation}
Note that Eq.~(\ref{eq:twoQPscattering}) can be written as
\begin{equation}
    \wH\ket{B_{j_1}, B_{j_2}} = \left(\wh_{j_1 - 1} + \mE + \wh_{j_2 - 1} + \mE\right)\ket{B_{j_1}, B_{j_2}} - e^{ik} \left(\wh_{j_1} - \mE\right)\ket{B_{j_1+1}, B_{j_2}} - e^{ik} \left(\wh_{j_2} - \mE\right)\ket{B_{j_1}, B_{j_2+1}}\;\;\forall j_1, j_2,
\label{eq:twoQPscattering}
\end{equation}
where the conditions of Eq.~(\ref{eq:Bvanish}) are implicitly assumed, and we have used Eq.~(\ref{eq:usefulidentity}).
Using Eq.~(\ref{eq:twoQPscattering}), we obtain that
\begin{eqnarray}
    &\wH\sumal{j_1, j_2 = 1}{L}{e^{ik(j_1 + j_2)}\ket{B_{j_1}, B_{j_2}}} = \sumal{j_1, j_2 = 1}{L}{e^{ik(j_1 + j_2)}\left[\left(\wh_{j_1 - 1} + \mE\right)\ket{B_{j_1}, B_{j_2}} - e^{ik} \left(\wh_{j_1} - \mE\right)\ket{B_{j_1+1}, B_{j_2}}\right.} \nn \\
    &\left.+ \left(\wh_{j_2 - 1} + \mE\right)\ket{B_{j_1}, B_{j_2}}  - e^{ik} \left(\wh_{j_2} - \mE\right)\ket{B_{j_1}, B_{j_2+1}}\right]\nn \\
    &= \sumal{j_1, j_2 = 1}{L}{e^{ik(j_1 + j_2)}\left[\left(\wh_{j_1 - 1} + \mE\right)\ket{B_{j_1}, B_{j_2}} - e^{ik} \left(\wh_{j_1} - \mE\right)\ket{B_{j_1+1}, B_{j_2}}\right]}\nn \\
    &+ \sumal{j_1, j_2 = 1}{L}{\left[e^{ik(j_1 + j_2)}\left(\wh_{j_2 - 1} + \mE\right)\ket{B_{j_1}, B_{j_2}}  - e^{ik} \left(\wh_{j_2} - \mE\right)\ket{B_{j_1}, B_{j_2+1}}\right]}\nn \\
    &= 2\sumal{j_1, j_2 = 1}{L}{e^{ik(j_1 + j_2)}\left(\wh_{j_1 - 1} + \mE\right)\ket{B_{j_1}, B_{j_2}}} - 2\sumal{j_1, j_2 = 1}{L}{e^{ik(j_1 + 1 + j_2)} \left(\wh_{j_1} - \mE\right)\ket{B_{j_1+1}, B_{j_2}}}\nn \\
    &= 4\mE\sumal{j_1, j_2 = 1}{L}{e^{ik(j_1 + j_2)}\ket{B_{j_1}, B_{j_2}}},
\label{eq:twoQPexplicit}
\end{eqnarray}
where in the third step we have interchanged $j_1$ and $j_2$ in the second sum. 
\end{widetext}
Similarly, we obtain an exact eigenstate with $n$ quasiparticles of momentum $k$, provided the quasiparticles are constrained to be separated by at least one site. 
Thus, we obtain a quasiparticle tower of exact eigenstates $\{\ket{S_{2n}}\}$ with energies $\{2n\mE\}$.
Note that this tower of eigenstates can consist of identical quasiparticles of any momentum $k$ provided there exists a tensor $B$ for which Eqs.~(\ref{eq:QPcondition1}), (\ref{eq:QPcondition2}), (\ref{eq:QPconstraints1}), and (\ref{eq:QPconstraints2}) are satisfied.
However, in the main text we only discuss examples where $k = \pi$. 
\section{SU(2) Multiplet of the Spin-2 Magnon for the AKLT chain}\label{sec:spin2magnon}
We obtain the quasiparticle creation operators for the spin-2 magnon multiplet of the AKLT chain.
Representing the AKLT ground state as $\ket{G}$, the highest weight state of the spin-2 magnon exact eigenstate (up to a normalization constant) reads~\cite{Moudgalya2018a}
\begin{equation}
    \ket{S_2} \equiv \mP\ket{G} = \sumal{j = 1}{L}{(-1)^j (S^+_j)^2}\ket{G}. 
\end{equation}
Since the AKLT Hamiltonian is $SU(2)$ symmetric~\cite{Moudgalya2018a} and $\ket{S_2}$ has a total spin $2$, we can obtain $5$ linearly independent eigenstates with the same energy. 
They are
\begin{equation}
    \{\ket{S_2}, S^-\ket{S_2}, (S^-)^2\ket{S_2}, (S^-)^3\ket{S_2}, (S^-)^4\ket{S_2}\}, 
\label{eq:multiplet}
\end{equation}
where $S^-$ is the lowering operator
\begin{equation}
    S^- \equiv \sumal{j = 1}{L}{S^-_j}. 
\label{eq:loweringop}
\end{equation}
We now express the rest of the states in the multiplet of Eq.~(\ref{eq:multiplet}) as quasiparticles states of the form of Eq.~(\ref{eq:akltqp}).
Note that with the onsite spin-1 operators defined as in Eq.~(\ref{eq:spin1ops}), their commutation relations read
\begin{equation}
    [S^+_j, S^-_k] = S^z_j\delta_{j,k},\;\;[S^z_j, S^\pm_k] = \pm S^\pm_j\delta_{j,k}.
\label{eq:spincomm}
\end{equation}
We start with the $S_z = +1$ state and using the fact that $S^-\ket{G} = 0$, we express it as
\begin{eqnarray}
    S^-\ket{S_2} &=& [S^-, \mP]\ket{G} = \sumal{j = 1}{L}{(-1)^j [S^-_j, (S^+_j)^2]}\ket{G} \nn \\
    &\propto& \sumal{j = 1}{L}{(-1)^j\{S^z_j, S^+_j\}}\ket{G},
\label{eq:onelowqp}
\end{eqnarray}
where we have used Eq.~(\ref{eq:spincomm}) and omitted an overall normalization factor. 
Similarly, we can apply the lowering operator on Eq.~(\ref{eq:onelowqp}) and write
\begin{eqnarray}
    (S^-)^2\ket{S_2} &=& \sumal{j = 1}{L}{(-1)^j[S^-_j, \{S^z_j, S^+_j\}]}\ket{G} \nn \\
    &\propto& \sumal{j = 1}{L}{(-1)^j(2(S^z_j)^2 - \{S^+_j, S^-_j\}))}\ket{G},
\end{eqnarray}
Repeating the same steps again, we also obtain
\begin{eqnarray}
    (S^-)^3\ket{S_2} &=& \sumal{j = 1}{L}{(-1)^j\{S^-_j, S^z_j\}}\ket{G} \nn \\
    (S^-)^4\ket{S_2} &=& \sumal{j = 1}{L}{(-1)^j(S^-_j)^2}\ket{G}.
\end{eqnarray}
Hence, an arbitrary eigenstate in the multiplet of the spin-2 magnon exact eigenstate is given by
\begin{eqnarray}
    &\ket{\psi} = \sumal{j = 1}{L}{(-1)^j \wO_j}\ket{G},\nn \\
    &\wO\in \textrm{span}\{(S^+)^2, \{S^z, S^+\}, 2(S^z)^2 - \{S^+, S^-\},\nn \\
    &\{S^-, S^z\}, (S^-)^2\}. 
\end{eqnarray}
\section{Derivation of Eq.~(\ref{eq:SS2simp})}\label{app:SS2}
We consider the spin-1 operators
\begin{eqnarray}
    &S^+ = 
    \begin{pmatrix}
    0 & 1 & 0 \\
    0 & 0 & 1 \\
    0 & 0 & 0
    \end{pmatrix},\;S^- = \left(S^+\right)^\dagger,\;S^z = 
    \begin{pmatrix}
    1 & 0 & 0 \\
    0 & 0 & 0 \\
    0 & 0 & -1
    \end{pmatrix}\;\;\;\;
\label{eq:spin1ops}
\end{eqnarray}
where the basis is in the order $\{\ket{+}, \ket{0}, \ket{-}\}$. We further denote the $3 \times 3$ identity matrix by $\mathds{1}$.
Since $\vec{S} \cdot \vec{S}$ in Eq.~(\ref{eq:circdefn}) commutes with the total spin operator on two sites $S^z_2 \equiv S^z \otimes \mathds{1} + \mathds{1} \otimes S^z$, 
it is straightforward to compute the matrix elements of the restriction of $(\vec{S} \cdot \vec{S})^2 - \mathds{1} \otimes \mathds{1}$ onto the spin $S^z_2 = 0$ sector as~\cite{Parkinson1988spin, Ercolessi2014Analysis}
\begin{eqnarray}
    &\left.\left((\vec{S}\cdot\vec{S})^2 - \mathds{1}\otimes \mathds{1}\right)\right|_{S^z_2 = 0} = 
    \begin{pmatrix}
        1 & -1 & 1 \\
        -1 & 1 & -1 \\
        1 & -1 & 1
    \end{pmatrix} \nn \\
    &= \begin{pmatrix}
    1 \\
    -1 \\
    1
\end{pmatrix}\begin{pmatrix}1 & -1 & 1\end{pmatrix}, 
\label{eq:ss2form}
\end{eqnarray}
where the two spin-1 basis is in the order $\{\ket{+-}, \ket{00}, \ket{-+}\}$.
Further, it is also straightforward to compute that $(\vec{S} \cdot \vec{S})^2 - \mathds{1} \otimes \mathds{1}$ is the zero matrix when restricted to the $S^z_2 \neq 0$ sectors.
Hence we directly deduce Eq.~(\ref{eq:SS2simp}).

\section{Two-Site Quasiparticle Exact Eigenstates in the MPS Language}\label{app:twositeQPMPS}
\subsection{Single quasiparticle}
In this section we show that the conditions of Eq.~(\ref{eq:twositeQPexpr}) imply the existence of a quasiparticle eigenstate of the Hamiltonian Eq.~(\ref{eq:FFeigenstate}).
Rewriting the conditions here for convenience, they read
\begin{eqnarray}
    &\wh_j\ket{AA} = 0,\label{eq:twositesuff1app} \\
    &(\wh_{j+1} - \mE_1) \overset{j,j+1\hspace{2mm}}{\ket{\wB A}} + e^{ik} (\wh_{j} - \mE_1) \overset{\hspace{2mm}j+1,j+2}{\ket{A\wB}}= 0 \label{eq:twositesuff2app} \\
    &(\wh_j - \mE_2) \ket{\wB} = 0, \label{eq:twositesuff3app}
\end{eqnarray}
The action of the Hamiltonian $\wH$ on the state with one quasiparticle reads
\begin{widetext}
\begin{eqnarray}
    &\wH \overset{j, j+1}{\ket{[A \cdots A \wB A \cdots A]}} = \left(\wh_{j-1} + \wh_j + \wh_{j+1}\right)\overset{j,j+1}{\ket{[A \cdots A \wB A \cdots A]}} \nn \\
    &=\wh_{j-1}\overset{j,j+1}{\ket{[A \cdots A \wB A \cdots A]}} + \wh_{j+1}\overset{j,j+1}{\ket{[A \cdots A \wB A \cdots A]}} + \mE_2\overset{j,j+1}{\ket{[A \cdots A \wB A \cdots A]}} \nn \\
    &=\wh_{j-1}\overset{j,j+1}{\ket{[A \cdots A \wB A \cdots A]}} + \left(\mE_1 \overset{j,j+1}{\ket{[A \cdots A \wB A \cdots A]}} -e^{ik}\left(\wh_j - \mE_1\right)\overset{j+1,j+2}{\ket{[A \cdots A \wB \cdots A]}}\right) + \mE_2\overset{j,j+1}{\ket{[A \cdots A \wB A \cdots A]}} \nn \\
    &=\left(\wh_{j-1} + \mE_1 + \mE_2\right)\overset{j,j+1}{\ket{[A \cdots A \wB A \cdots A]}}- e^{ik} \left(\wh_j - \mE_1\right)\overset{j+1,j+2}{\ket{[A \cdots A \wB \cdots A]}}.
\label{eq:loctwositeQPcond}
\end{eqnarray}
\end{widetext}
where we have used Eq.~(\ref{eq:twositesuff1app}) in the first line, Eq.~(\ref{eq:twositesuff3app}) in the second line, and Eq.~(\ref{eq:twositesuff2app}) in the third. 
Defining the shorthand notation
\begin{equation}
    \ket{\wB_{j}} \equiv \overset{j,j+1}{\ket{[A\cdots A \wB A\cdots A]}}, 
\end{equation}
Eq.~(\ref{eq:loctwositeQPcond}) can be rewritten as
\begin{eqnarray}
    &\wH\ket{\wB_{j}} = \left(\wh_{j-1} + \mE_1 + \mE_2\right)\ket{\wB_{j}} \nn \\
    &- e^{ik} \left(\wh_j - \mE_1\right)\ket{\wB_{j+1}}.
\label{eq:Hamilactiontwo}
\end{eqnarray}
Thus, we obtain
\begin{widetext}
\begin{eqnarray}
    &\wH\ket{\psi_A\left(\wB, k\right)} = \wH\sumal{j=1}{L}{e^{ikj}\ket{\wB_{j}}} = \sumal{j=1}{L}{e^{ikj}\left[\left(\wh_{j-1}\ket{\wB_{j}} - e^{ik} \wh_j\ket{\wB_{j+1}}\right) + \mE_1\left(\ket{\wB_{j}} + e^{ik} \ket{\wB_{j+1}}\right) + \mE_2 \ket{\wB_{j}}\right]} \nn \\
    &= \sumal{j=1}{L}{e^{ikj}\left[\left(\wh_{j-1}\ket{\wB_{j}} - \wh_{j-1}\ket{\wB_{j}}\right) + \mE_1\left(\ket{\wB_{j}} + \ket{\wB_{j}}\right) + \mE_2 \ket{\wB_j}\right]} \nn \\
    &=\left(2\mE_1 + \mE_2\right) \sumal{j=1}{L}{e^{ikj}\ket{\wB_{j}}} = \left(2\mE_1 + \mE_2\right)\ket{\psi_A\left(\wB, k\right)}.
\end{eqnarray}
\end{widetext}
The conditions of Eqs.~(\ref{eq:cond1}) and (\ref{eq:cond2}) guarantee a quasiparticle eigenstate of $\wH$ with energy $E = 2\mE_1 + \mE_2$. 
\subsection{Tower of states}\label{app:twositeQPMPStower}
Here we show that in addition to Eqs.~(\ref{eq:twositesuff1app})-(\ref{eq:twositesuff3app}), the quasiparticle constraints of  Eqs.~(\ref{eq:twoconstraint1})-(\ref{eq:twoconstraint3}) guarantees the existence of a quasiparticle tower of exact eigenstates.
We first illustrate the exactness for two quasiparticles dispersing in the ground state background, by defining the configuration of two quasiparticles as
\begin{equation}
    \ket{\wB_{j_1}, \wB_{j_2}} = \overset{\hspace{1mm}j_1,j_1+1\hspace{11mm}j_2,j_2+1}{\ket{[A \cdots A \wB A \cdots A \wB A \cdots A]}}.
\label{eq:twoQPconfigtwo}
\end{equation}
Note that as a consequence of Eq.~(\ref{eq:twoconstraint3}), we are guaranteed to have at least one $A$ in between the $\wB$'s in the configuration of Eq.~(\ref{eq:twoQPconfigtwo}).
That is,
\begin{equation}
    \ket{\wB_{j}, \wB_{l}} = 0\;\;\textrm{if}\;\; |j - l| \leq 2.
\label{eq:Bvanishtwo}
\end{equation}
Thus, the Hamiltonian $\wH$ acts independently on each of the quasiparticles. 
The proof proceeds straightforwardly following the single-site quasiparticle case (see App.~\ref{sec:singlesitetower}).
Indeed, after simplification using Eqs.~(\ref{eq:twositesuff2app}) and (\ref{eq:Bvanishtwo}), we obtain 
\begin{eqnarray}
    &\wH\ket{\wB_{j_1}, \wB_{j_2}} = \left(\wh_{j_1 - 1} + \mE_1 + \mE_2\right)\ket{\wB_{j_1}, \wB_{j_2}} \nn \\
    &+ \left(\wh_{j_2 - 1} + \mE_1 + \mE_2\right)\ket{\wB_{j_1}, \wB_{j_2}} \nn \\
    &- e^{ik} \left(\wh_{j_1} - \mE_1\right)\ket{\wB_{j_1+1}, \wB_{j_2}} \nn \\
    &- e^{ik} \left(\wh_{j_2} - \mE_1\right)\ket{\wB_{j_1}, \wB_{j_2+1}}\;\;\forall j_1, j_2,
\label{eq:twoQPscatteringtwo}
\end{eqnarray}
where the conditions of Eq.~(\ref{eq:Bvanishtwo}) are implicit.
As for Eq.~(\ref{eq:twoQPexplicit}), we obtain that
\begin{eqnarray}
    &\wH\sumal{j_1, j_2 = 1}{L}{e^{ik(j_1 + j_2)}\ket{\wB_{j_1}, \wB_{j_2}}} \nn \\
    &= 2(2\mE_1 + \mE_2)\sumal{j_1, j_2 = 1}{L}{e^{ik(j_1 + j_2)}\ket{\wB_{j_1}, \wB_{j_2}}},
\end{eqnarray}
Similarly, we obtain an exact eigenstate with $n$ quasiparticles of momentum $k$, provided the quasiparticles are constrained to be separated by at least one site. 
Thus, we obtain a quasiparticle tower of exact eigenstates $\{\ket{S_{2n}}\}$ with energies $\{n(2\mE_1 + \mE_2)\}$.
\section{Sufficiency of Eqs.~(\ref{eq:rightcond}) and (\ref{eq:leftcond})}\label{app:sufficiency}
Here we show that Eqs.~(\ref{eq:rightcond}) and (\ref{eq:leftcond}) are sufficient for Eq.~(\ref{eq:twositesuff3}) to be satisfied.  
By left-multiplying Eq.~(\ref{eq:rightcond}) by $\ket{\Btr}$ and replacing $j$ by $j+1$, we obtain 
\begin{equation}
    \left(\wh_{j+1} - \mE_1\right)\overset{j+1}{\ket{\Btl \Btr A}} = \left(\wh_{j+1} - \mE_1\right)\overset{j+1}{\ket{\wB A}} = \overset{j+1}{\ket{\Btl C \Btr}},
\label{eq:suff1}
\end{equation}
where we have used Eq.~(\ref{eq:QPdecomposition}).
Similarly, by right-multiplying Eq.~(\ref{eq:leftcond}) by $\ket{\Btl}$, we obtain 
\begin{equation}
    \left(\wh_j - \mE_1\right)\overset{j+1\hspace{2mm}}{\ket{A \Btl \Btr}} = \left(\wh_j - \mE_1\right)\overset{j+1}{\ket{A \wB}} = -e^{-ik}\overset{j+1}{\ket{\Btl C \Btr}},
\label{eq:suff2}
\end{equation}
Using Eqs.~(\ref{eq:suff1}) and (\ref{eq:suff2}), we immediately see that Eq.~(\ref{eq:twositesuff3}) follows from Eqs.~(\ref{eq:rightcond}) and (\ref{eq:leftcond}).
\section{Examples of $\mA$ and $\mBt$ subspaces for the Potts-like MPS}\label{app:examplespotts}
Here we present the derivation of the subspaces $\mA$ and $\mBt$ of Eqs.~(\ref{eq:mAdefn}) and (\ref{eq:mBtdefn}) for an MPS of the form 
\begin{equation}
    A^{[+]} = c_+ \sigma^+,\;\;\;A^{[0]} = c_0 \mathds{1}_{2\times 2}\equiv c_0 \sigma^0,\;\;\;A^{[-]} =  c_- \sigma^-,
\label{eq:genPottsMPS}
\end{equation}
where $\sigma^+$, and $\sigma^-$ are the Pauli matrices.
Note that the Potts MPS of Eq.~(\ref{eq:PottsMPS}) is recovered by setting
\begin{equation}
    \left(c_+, c_0, c_-\right) = \frac{1}{\sqrt{2}}\left(1, 1, 1\right).
\label{eq:cPotts}
\end{equation}
The computation of the $\mA$ proceeds similarly to that illustrated for the generalized AKLT MPS in Eqs.~(\ref{eq:Amcompact})-(\ref{eq:genAsubspacenorm}) in App.~\ref{app:examples}.
Using the MPS matrices of Eq.~(\ref{eq:genPottsMPS}) instead, we obtain
\begin{eqnarray}
    &\hspace{-5mm}\mA = \textrm{span}\{\frac{1}{\sqrt{2\left(|c_+ c_-|^2 + 2|c_0|^4\right)}}\left(c_+ c_- \left(\ket{+-} + \ket{-+}\right) + 2 c_{0}^2 \ket{00}\right),\nn \\
    &\frac{1}{\sqrt{2}}\left(\ket{-0} + \ket{0-}\right), \nn \\
    &\frac{1}{\sqrt{2}}\left(\ket{+-} - \ket{-+}\right), \nn \\
    &\frac{1}{\sqrt{2}}\left(\ket{+0} + \ket{0+}\right)\},
\label{eq:genAsubspacenormPotts}
\end{eqnarray}
Thus, for the perturbed Potts MPS, using Eq.~(\ref{eq:cPotts}) we obtain 
\begin{eqnarray}
    &\mA = \textrm{span}\{ \sqrt{\frac{1}{6}}\left(\ket{+-} + \ket{-+} + 2\ket{00}\right), \nn \\
    &\frac{1}{\sqrt{2}}\left(\ket{-0} + \ket{0-}\right), \nn \\
    &\frac{1}{\sqrt{2}}\left(\ket{+-} - \ket{-+}\right) \nn \\
    &\frac{1}{\sqrt{2}}\left(\ket{+0} + \ket{0+}\right)\}.
\label{eq:APotts}
\end{eqnarray}
In terms of total angular momentum eigenvectors, $\mA$ of Eq.~(\ref{eq:APotts}) reads
\begin{equation}
    \mA = \textrm{span}\{ \ket{J_{2, 0}}, \ket{J_{2, -1}}, \ket{J_{1, 0}}, \ket{J_{2, 1}} \},
\label{eq:mAPotts}
\end{equation} 
where the total angular momentum eigenstates $\ket{J_{j,m}}$ are enumerated in App.~\ref{sec:totalspin}.
We then compute the $\mBt$ subspace defined in Eq.~(\ref{eq:mBtdefn}), which reads
\begin{equation}
    \mBt = \left((S^+)^2\otimes S^+ + S^+ \otimes (S^+)^2\right)\mA,
\label{eq:mBtPotts}
\end{equation}
Note that the action of $\wOtwo$ on the basis states with $S_z \geq 0$ in $\mA$ of Eq.~(\ref{eq:mAPotts}) vanishes since $\wOtwo$ is a spin-3 operator.
Further, we obtain
\begin{equation}
    ((S^+)^2 \otimes S^+ + S^+ \otimes (S^+)^2)\frac{\ket{- 0} + \ket{0 -}}{\sqrt{2}} = \ket{++} = \ket{J_{2,2}}.
\end{equation}
Thus,
\begin{equation}
    \mBt = \{\ket{J_{2,2}}\}
\label{eq:genBtsubspacenorm}
\end{equation}
which is independent of the $c_m$'s.
Using Eqs.~(\ref{eq:genAsubspacenormPotts}) and (\ref{eq:genBtsubspacenorm}), we obtain
\begin{eqnarray}
    &(\mA^c/\mBt) = \textrm{span}\{ \ket{J_{2,-2}}, \ket{J_{1,-1}}, \ket{J_{1,1}}\nn \\
    &\frac{1}{\sqrt{|c_+ c_-|^2 + 2|c_0|^4}}\left(-c_0^2\left(\ket{+-} + \ket{-+}\right) + c_{+} c_- \ket{00}\right) \}.\nn \\
\label{eq:mAcmB}
\end{eqnarray}
\section{Solution of Eqs.~(\ref{eq:rightcond}) and (\ref{eq:leftcond}) for the generalized perturbed Potts MPS}\label{sec:SolnPottsMPS}
Here, we obtain a solution to Eqs.~(\ref{eq:rightcond}) and (\ref{eq:leftcond}) using the MPS of Eq.~(\ref{eq:genPottsMPS}) and $\wh_j$ of the form of Eq.~(\ref{eq:hmatrixform}), where the subspaces $\mA$ and $\mBt$ are obtained in Eqs.~(\ref{eq:genAsubspacenormPotts}) and (\ref{eq:genBtsubspacenorm}) respectively.
In particular, we use the following properties of $\wh_j$:
\begin{eqnarray}
    &\wh_j \mBt = \mE_2 \mBt \implies \wh_j\ket{++} = \mE_2\ket{++},\label{eq:hproperty1} \\
    &\wh_j \mA = 0 \implies \wh_j\frac{1}{\sqrt{2}}\left(\ket{+0} + \ket{0+}\right) = 0.
\label{eq:hproperty2}
\end{eqnarray}
In this section, it is convenient to represent MPS tensors as vectors on the physical indices with matrix coefficients.  
For example, we express the MPS $A$ of Eq.~(\ref{eq:genPottsMPS}) as
\begin{eqnarray}
    \ket{A} &=& c_+ \sigma^+ \ket{+}\ +\ c_0 \sigma^0 \ket{0}\ +\ c_- \sigma^-\ket{-}\nn \\
    &=& \sumal{\alpha \in \{+, 0, -\}}{}{c_\alpha \sigma^\alpha\ket{\alpha}}. 
\label{eq:AMPSrep}
\end{eqnarray}
Consequently multisite MPS can be obtained with a matrix multiplication of the coefficients and a tensor product over the physical indices.
Using the operator $\wOtwo = (S^+)^2 \otimes S^+ + S^+ \otimes (S^+)^2$, using Eq.~(\ref{eq:AMPSrep}), we straightforwardly obtain the expression for the $\wB$ quasiparticle tensor defined in Eq.~(\ref{eq:QPnondiag}):
\begin{equation}
    \ket{\wB} = 2 c_0 c_- \sigma^- \ket{++} = 2 c_0 c_- \begin{pmatrix} 0 & 0 \\ 1 & 0\end{pmatrix}\ket{++}, 
\label{eq:Btexpr}
\end{equation}
where the $2 \times 2$ matrix is over the auxiliary indices and the $\ket{\cdot}$ is over the physical index of the $\wB$ tensor.  
As shown in Eq.~(\ref{eq:QPdecomposition}), we can always decompose (in a non-unique way) $\ket{\wB} = \ket{\Btl \Btr} = \ket{B_l} \otimes \ket{B_r}$.
Applied to Eq.~(\ref{eq:Btexpr}), we get
\begin{equation}
    \ket{\wB} = \left(\sqrt{2 c_0 c_-} \begin{pmatrix}0\\1\end{pmatrix}\ket{+}\right) \otimes \left(\sqrt{2 c_0 c_-} \begin{pmatrix}1 & 0\end{pmatrix}\ket{+}\right),
\label{eq:Btapp}
\end{equation}
leading to
\begin{equation}
    \ket{\Btl} = \sqrt{2 c_0 c_-}\begin{pmatrix}0 \\ 1\end{pmatrix}\ket{+},\;\;\ket{\Btr} = \sqrt{2 c_0 c_-}\begin{pmatrix}1 & 0\end{pmatrix}\ket{+}.
\label{eq:BtlBtrapp}
\end{equation}
Using Eqs.~(\ref{eq:AMPSrep}) and (\ref{eq:BtlBtrapp}), $\ket{A\Btl}$ and $\ket{\Btr A}$ read
\begin{eqnarray}
    \ket{A\Btl} = \sqrt{2 c_0 c_-}\left(\begin{pmatrix}c_+ \\ 0\end{pmatrix}\ket{++} + \begin{pmatrix}0 \\ c_0\end{pmatrix}\ket{0 +}\right) \label{eq:ABlapp} \\
    \ket{\Btr A} = \sqrt{2 c_0 c_-}\left(\begin{pmatrix} 0 & c_+ \end{pmatrix}\ket{++} +  \begin{pmatrix}c_0 & 0\end{pmatrix}\ket{+ 0}\right).\label{eq:BrAapp}
\end{eqnarray}
The most general $\ket{C}$ in this case has the form
\begin{equation}
    \ket{C} = \sumal{\alpha \in \{+, 0, -\}}{}{C^{[\alpha]}\ket{\alpha}}, 
\end{equation}
where $\{C^{[\alpha]}\}$ are numbers. 
Consequently, using Eq.~(\ref{eq:BtlBtrapp}), $\ket{\Btl C}$ and $\ket{C \Btr}$ read
\begin{eqnarray}
    \ket{\Btl C} = \sqrt{2 c_0 c_-}\sumal{\alpha \in \{+, 0, -\}}{}{\begin{pmatrix}0 \\ C^{[\alpha]} \end{pmatrix}\ket{+ \alpha}} \label{eq:BlCapp} \\
    \ket{C \Btr} = \sqrt{2 c_0 c_-}\sumal{\alpha \in \{+, 0, -\}}{}{\begin{pmatrix} C^{[\alpha]} & 0 \end{pmatrix}\ket{\alpha  +}}. \label{eq:CBrapp}
\end{eqnarray}
Using Eqs.~(\ref{eq:ABlapp}) and (\ref{eq:BlCapp}), Eq.~(\ref{eq:rightcond}) reads
\begin{widetext}
\begin{eqnarray}
    &\begin{pmatrix} c_+ \\ 0\end{pmatrix}\left(\wh_j - \mE_1\right)\ket{++} + \begin{pmatrix}0 \\ c_0\end{pmatrix}\left(\wh_j - \mE_1\right)\ket{0 +}  = \sumal{\alpha \in \{+, 0, -\}}{}{\begin{pmatrix}0 \\ C^{[\alpha]}\end{pmatrix}\ket{+ \alpha}} \nn \\
    &\implies \begin{pmatrix} c_+ \\ 0\end{pmatrix}\left(\mE_2 - \mE_1\right)\ket{++} + \begin{pmatrix}0 \\ c_0\end{pmatrix}\left(\wh_j - \mE_1\right)\ket{0 +}  = \sumal{\alpha \in \{+, 0, -\}}{}{\begin{pmatrix}0 \\ C^{[\alpha]}\end{pmatrix}\ket{+ \alpha}},
\label{eq:int}
\end{eqnarray}
where we have used Eq.~(\ref{eq:hproperty1}). 
\end{widetext}
Equating the two components of the row vector in Eq.~(\ref{eq:int}), we obtain
\begin{eqnarray}
    &c_+\left(\mE_2 - \mE_1\right)\ket{++} = 0 \implies \mE_2 = \mE_1 \equiv \mE\label{eq:E1E2} \\
    &\left(\wh_j - \mE_1\right)\ket{0+} = \sumal{\alpha\in\{+,0,-\}}{}{\frac{C^{[\alpha]}}{c_0}\ket{+\alpha}}.\label{eq:rightcondsoln}
\end{eqnarray}
Similarly, solving Eq.~(\ref{eq:leftcond}), we obtain
\begin{eqnarray}
    \left(\wh_j - \mE\right)\ket{0 +} = \sumal{\alpha \in \{+, 0, -\}}{}{\frac{C^{[\alpha]}}{c_0}\ket{+ \alpha}},
\label{eq:leftcondsoln}
\end{eqnarray}
where $\mE \equiv \mE_1 = \mE_2$.
Adding Eqs.~(\ref{eq:rightcondsoln}) and (\ref{eq:leftcondsoln}), we obtain
\begin{eqnarray}
    &\left(\wh_j - \mE\right)\frac{\left(\ket{+ 0} + \ket{0 +}\right)}{\sqrt{2}} = \sumal{\alpha\in \{+,0,-\}}{}{\frac{C^{[\alpha]}}{c_0}\frac{\left(\ket{+ \alpha} + \ket{\alpha +}\right)}{\sqrt{2}}} \nn \\
    &\hspace{-10mm}\implies \wh_j \ket{J_{2,1}} = \left(\mE + \frac{C^{[0]}}{c_0}\right)\ket{J_{2,1}} + \sqrt{2}\frac{C^{[+]}}{c_0}\ket{J_{2,2}} + \frac{C^{[-]}}{c_0}\ket{J_{2,0}}.\nn \\
\end{eqnarray}
However, using Eq.~(\ref{eq:hproperty2}), we obtain
\begin{equation}
    C^{[\alpha]} = - c_0 \mE \delta_{\alpha, 0}. 
\label{eq:C0set}
\end{equation}
Further, subtracting Eqs.~(\ref{eq:rightcondsoln}) and (\ref{eq:leftcondsoln}), and using Eq.~(\ref{eq:C0set}) we obtain
\begin{eqnarray}
    &\left(\wh_j - \mE\right)\ket{J_{1,1}} = \mE \ket{J_{1,1}} \nn \\
    &\implies \wh_j\ket{J_{1,1}} = 2\mE \ket{J_{1,1}}.
\label{eq:hsolnC}
\end{eqnarray}
We have thus shown that a solution to Eqs.~(\ref{eq:rightcond}) and (\ref{eq:leftcond}) for the MPS of Eq.~(\ref{eq:genPottsMPS}) exists, provided the tensor $C$ satisfies Eq.~(\ref{eq:C0set}) and the Hamiltonian term satisfies Eq.~(\ref{eq:hsolnC}).
\section{Families of Perturbed Potts-like Quantum Scarred Hamiltonians}\label{app:Pottsfamily}
Here we show that we can obtain a family of Hamiltonians with perturbed Potts-like quantum scars starting from the MPS of Eq.~(\ref{eq:genPottsMPSarb}).
As shown in Eq.~(\ref{eq:mAcmB}) in App.~\ref{app:examplespotts}, we obtain
\begin{eqnarray}
    &\mA = \{\ket{K_{1,0}}, \ket{K_{2,1}}, \ket{K_{2,0}}, \ket{K_{2,-1}}\} \nn \\
    &\mBt = \{\ket{K_{2,2}}\} \nn \\
    &\implies \mA^c/\mB = \{\ket{K_{2,-2}}, \ket{K_{1,-1}}, \ket{K_{1,1}}, \ket{K_{0,0}}\},    
\end{eqnarray}
where the vectors $\{\ket{K_{m,n}}\}$ have been defined in Eq.~(\ref{eq:Kdefn}). 
Further, defining the subspace $\mC = \{\ket{K_{1,1}}\}$ (see Eq.~(\ref{eq:hsolnC})), we obtain
\begin{equation}
    (\mA/\mBt)/\mC = \{\ket{K_{2,-2}}, \ket{K_{1,-1}}, \ket{K_{0,0}}\}. 
\end{equation}
Thus, the family of Hamiltonians 
\begin{eqnarray}
    &\wH = \sumal{j}{}{\wh_j},\nn \\
    &\wh_j = \mE\ket{K_{2,2}}\bra{K_{2,2}} + 2\mE\ket{K_{1,1}}\bra{K_{1,1}} \nn \\
    &+ \sumal{m, n = 0}{2}{z^{(m, n)}_j \ket{K_{m,-m}}\bra{K_{n,-n}}}
\label{eq:mostgenPottsfamily}
\end{eqnarray}
exhibit a tower of two-site quasiparticle eigenstates starting from the MPS of Eq.~(\ref{eq:genPottsMPSarb}).

\bibliography{mps_scars}

\begin{thebibliography}{75}%
\makeatletter
\providecommand \@ifxundefined [1]{%
 \@ifx{#1\undefined}
}%
\providecommand \@ifnum [1]{%
 \ifnum #1\expandafter \@firstoftwo
 \else \expandafter \@secondoftwo
 \fi
}%
\providecommand \@ifx [1]{%
 \ifx #1\expandafter \@firstoftwo
 \else \expandafter \@secondoftwo
 \fi
}%
\providecommand \natexlab [1]{#1}%
\providecommand \enquote  [1]{``#1''}%
\providecommand \bibnamefont  [1]{#1}%
\providecommand \bibfnamefont [1]{#1}%
\providecommand \citenamefont [1]{#1}%
\providecommand \href@noop [0]{\@secondoftwo}%
\providecommand \href [0]{\begingroup \@sanitize@url \@href}%
\providecommand \@href[1]{\@@startlink{#1}\@@href}%
\providecommand \@@href[1]{\endgroup#1\@@endlink}%
\providecommand \@sanitize@url [0]{\catcode `\\12\catcode `\$12\catcode
  `\&12\catcode `\#12\catcode `\^12\catcode `\_12\catcode `\%12\relax}%
\providecommand \@@startlink[1]{}%
\providecommand \@@endlink[0]{}%
\providecommand \url  [0]{\begingroup\@sanitize@url \@url }%
\providecommand \@url [1]{\endgroup\@href {#1}{\urlprefix }}%
\providecommand \urlprefix  [0]{URL }%
\providecommand \Eprint [0]{\href }%
\providecommand \doibase [0]{http://dx.doi.org/}%
\providecommand \selectlanguage [0]{\@gobble}%
\providecommand \bibinfo  [0]{\@secondoftwo}%
\providecommand \bibfield  [0]{\@secondoftwo}%
\providecommand \translation [1]{[#1]}%
\providecommand \BibitemOpen [0]{}%
\providecommand \bibitemStop [0]{}%
\providecommand \bibitemNoStop [0]{.\EOS\space}%
\providecommand \EOS [0]{\spacefactor3000\relax}%
\providecommand \BibitemShut  [1]{\csname bibitem#1\endcsname}%
\let\auto@bib@innerbib\@empty
\bibitem [{\citenamefont {Deutsch}(1991)}]{deutsch1991quantum}%
  \BibitemOpen
  \bibfield  {author} {\bibinfo {author} {\bibfnamefont {J.~M.}\ \bibnamefont
  {Deutsch}},\ }\href {\doibase 10.1103/PhysRevA.43.2046} {\bibfield  {journal}
  {\bibinfo  {journal} {Phys. Rev. A}\ }\textbf {\bibinfo {volume} {43}},\
  \bibinfo {pages} {2046} (\bibinfo {year} {1991})}\BibitemShut {NoStop}%
\bibitem [{\citenamefont {Srednicki}(1994)}]{srednicki1994chaos}%
  \BibitemOpen
  \bibfield  {author} {\bibinfo {author} {\bibfnamefont {M.}~\bibnamefont
  {Srednicki}},\ }\href {\doibase 10.1103/PhysRevE.50.888} {\bibfield
  {journal} {\bibinfo  {journal} {Phys. Rev. E}\ }\textbf {\bibinfo {volume}
  {50}},\ \bibinfo {pages} {888} (\bibinfo {year} {1994})}\BibitemShut
  {NoStop}%
\bibitem [{\citenamefont {{Nandkishore}}\ and\ \citenamefont
  {{Huse}}(2015)}]{rahul2015review}%
  \BibitemOpen
  \bibfield  {author} {\bibinfo {author} {\bibfnamefont {R.}~\bibnamefont
  {{Nandkishore}}}\ and\ \bibinfo {author} {\bibfnamefont {D.~A.}\ \bibnamefont
  {{Huse}}},\ }\href {\doibase 10.1146/annurev-conmatphys-031214-014726}
  {\bibfield  {journal} {\bibinfo  {journal} {Annual Review of Condensed Matter
  Physics}\ }\textbf {\bibinfo {volume} {6}},\ \bibinfo {pages} {15} (\bibinfo
  {year} {2015})}\BibitemShut {NoStop}%
\bibitem [{\citenamefont {Shiraishi}\ and\ \citenamefont
  {Mori}(2017)}]{mori2017eth}%
  \BibitemOpen
  \bibfield  {author} {\bibinfo {author} {\bibfnamefont {N.}~\bibnamefont
  {Shiraishi}}\ and\ \bibinfo {author} {\bibfnamefont {T.}~\bibnamefont
  {Mori}},\ }\href {\doibase 10.1103/PhysRevLett.119.030601} {\bibfield
  {journal} {\bibinfo  {journal} {Phys. Rev. Lett.}\ }\textbf {\bibinfo
  {volume} {119}},\ \bibinfo {pages} {030601} (\bibinfo {year}
  {2017})}\BibitemShut {NoStop}%
\bibitem [{\citenamefont {Affleck}\ \emph {et~al.}(1988)\citenamefont
  {Affleck}, \citenamefont {Kennedy}, \citenamefont {Lieb},\ and\ \citenamefont
  {Tasaki}}]{Affleck1988}%
  \BibitemOpen
  \bibfield  {author} {\bibinfo {author} {\bibfnamefont {I.}~\bibnamefont
  {Affleck}}, \bibinfo {author} {\bibfnamefont {T.}~\bibnamefont {Kennedy}},
  \bibinfo {author} {\bibfnamefont {E.~H.}\ \bibnamefont {Lieb}}, \ and\
  \bibinfo {author} {\bibfnamefont {H.}~\bibnamefont {Tasaki}},\ }\href
  {https://projecteuclid.org:443/euclid.cmp/1104161001} {\bibfield  {journal}
  {\bibinfo  {journal} {Comm. Math. Phys.}\ }\textbf {\bibinfo {volume}
  {115}},\ \bibinfo {pages} {477} (\bibinfo {year} {1988})}\BibitemShut
  {NoStop}%
\bibitem [{\citenamefont {Moudgalya}\ \emph
  {et~al.}(2018{\natexlab{a}})\citenamefont {Moudgalya}, \citenamefont
  {Rachel}, \citenamefont {Bernevig},\ and\ \citenamefont
  {Regnault}}]{Moudgalya2018a}%
  \BibitemOpen
  \bibfield  {author} {\bibinfo {author} {\bibfnamefont {S.}~\bibnamefont
  {Moudgalya}}, \bibinfo {author} {\bibfnamefont {S.}~\bibnamefont {Rachel}},
  \bibinfo {author} {\bibfnamefont {B.~A.}\ \bibnamefont {Bernevig}}, \ and\
  \bibinfo {author} {\bibfnamefont {N.}~\bibnamefont {Regnault}},\ }\href
  {\doibase 10.1103/PhysRevB.98.235155} {\bibfield  {journal} {\bibinfo
  {journal} {Phys. Rev. B}\ }\textbf {\bibinfo {volume} {98}},\ \bibinfo
  {pages} {235155} (\bibinfo {year} {2018}{\natexlab{a}})}\BibitemShut
  {NoStop}%
\bibitem [{\citenamefont {Moudgalya}\ \emph
  {et~al.}(2018{\natexlab{b}})\citenamefont {Moudgalya}, \citenamefont
  {Regnault},\ and\ \citenamefont {Bernevig}}]{Moudgalya2018b}%
  \BibitemOpen
  \bibfield  {author} {\bibinfo {author} {\bibfnamefont {S.}~\bibnamefont
  {Moudgalya}}, \bibinfo {author} {\bibfnamefont {N.}~\bibnamefont {Regnault}},
  \ and\ \bibinfo {author} {\bibfnamefont {B.~A.}\ \bibnamefont {Bernevig}},\
  }\href {\doibase 10.1103/PhysRevB.98.235156} {\bibfield  {journal} {\bibinfo
  {journal} {Phys. Rev. B}\ }\textbf {\bibinfo {volume} {98}},\ \bibinfo
  {pages} {235156} (\bibinfo {year} {2018}{\natexlab{b}})}\BibitemShut
  {NoStop}%
\bibitem [{\citenamefont {{Bernien}}\ \emph {et~al.}(2017)\citenamefont
  {{Bernien}}, \citenamefont {{Schwartz}}, \citenamefont {{Keesling}},
  \citenamefont {{Levine}}, \citenamefont {{Omran}}, \citenamefont {{Pichler}},
  \citenamefont {{Choi}}, \citenamefont {{Zibrov}}, \citenamefont {{Endres}},
  \citenamefont {{Greiner}}, \citenamefont {{Vuleti{\'c}}},\ and\ \citenamefont
  {{Lukin}}}]{bernien2017probing}%
  \BibitemOpen
  \bibfield  {author} {\bibinfo {author} {\bibfnamefont {H.}~\bibnamefont
  {{Bernien}}}, \bibinfo {author} {\bibfnamefont {S.}~\bibnamefont
  {{Schwartz}}}, \bibinfo {author} {\bibfnamefont {A.}~\bibnamefont
  {{Keesling}}}, \bibinfo {author} {\bibfnamefont {H.}~\bibnamefont
  {{Levine}}}, \bibinfo {author} {\bibfnamefont {A.}~\bibnamefont {{Omran}}},
  \bibinfo {author} {\bibfnamefont {H.}~\bibnamefont {{Pichler}}}, \bibinfo
  {author} {\bibfnamefont {S.}~\bibnamefont {{Choi}}}, \bibinfo {author}
  {\bibfnamefont {A.~S.}\ \bibnamefont {{Zibrov}}}, \bibinfo {author}
  {\bibfnamefont {M.}~\bibnamefont {{Endres}}}, \bibinfo {author}
  {\bibfnamefont {M.}~\bibnamefont {{Greiner}}}, \bibinfo {author}
  {\bibfnamefont {V.}~\bibnamefont {{Vuleti{\'c}}}}, \ and\ \bibinfo {author}
  {\bibfnamefont {M.~D.}\ \bibnamefont {{Lukin}}},\ }\href {\doibase
  10.1038/nature24622} {\bibfield  {journal} {\bibinfo  {journal} {\nat}\
  }\textbf {\bibinfo {volume} {551}},\ \bibinfo {pages} {579} (\bibinfo {year}
  {2017})}\BibitemShut {NoStop}%
\bibitem [{\citenamefont {Fendley}\ \emph {et~al.}(2004)\citenamefont
  {Fendley}, \citenamefont {Sengupta},\ and\ \citenamefont
  {Sachdev}}]{Fendley2004}%
  \BibitemOpen
  \bibfield  {author} {\bibinfo {author} {\bibfnamefont {P.}~\bibnamefont
  {Fendley}}, \bibinfo {author} {\bibfnamefont {K.}~\bibnamefont {Sengupta}}, \
  and\ \bibinfo {author} {\bibfnamefont {S.}~\bibnamefont {Sachdev}},\ }\href
  {\doibase 10.1103/PhysRevB.69.075106} {\bibfield  {journal} {\bibinfo
  {journal} {Phys. Rev. B}\ }\textbf {\bibinfo {volume} {69}},\ \bibinfo
  {pages} {075106} (\bibinfo {year} {2004})}\BibitemShut {NoStop}%
\bibitem [{\citenamefont {Turner}\ \emph
  {et~al.}(2018{\natexlab{a}})\citenamefont {Turner}, \citenamefont
  {Michailidis}, \citenamefont {Abanin}, \citenamefont {Serbyn},\ and\
  \citenamefont {Papic}}]{turner2017quantum}%
  \BibitemOpen
  \bibfield  {author} {\bibinfo {author} {\bibfnamefont {C.}~\bibnamefont
  {Turner}}, \bibinfo {author} {\bibfnamefont {A.}~\bibnamefont {Michailidis}},
  \bibinfo {author} {\bibfnamefont {D.}~\bibnamefont {Abanin}}, \bibinfo
  {author} {\bibfnamefont {M.}~\bibnamefont {Serbyn}}, \ and\ \bibinfo {author}
  {\bibfnamefont {Z.}~\bibnamefont {Papic}},\ }\href {\doibase
  10.1038/s41567-018-0137-5} {\bibfield  {journal} {\bibinfo  {journal} {Nature
  Physics}\ }\textbf {\bibinfo {volume} {14}},\ \bibinfo {pages} {745}
  (\bibinfo {year} {2018}{\natexlab{a}})}\BibitemShut {NoStop}%
\bibitem [{\citenamefont {Turner}\ \emph
  {et~al.}(2018{\natexlab{b}})\citenamefont {Turner}, \citenamefont
  {Michailidis}, \citenamefont {Abanin}, \citenamefont {Serbyn},\ and\
  \citenamefont {Papi\ifmmode~\acute{c}\else \'{c}\fi{}}}]{turner2018quantum}%
  \BibitemOpen
  \bibfield  {author} {\bibinfo {author} {\bibfnamefont {C.~J.}\ \bibnamefont
  {Turner}}, \bibinfo {author} {\bibfnamefont {A.~A.}\ \bibnamefont
  {Michailidis}}, \bibinfo {author} {\bibfnamefont {D.~A.}\ \bibnamefont
  {Abanin}}, \bibinfo {author} {\bibfnamefont {M.}~\bibnamefont {Serbyn}}, \
  and\ \bibinfo {author} {\bibfnamefont {Z.}~\bibnamefont
  {Papi\ifmmode~\acute{c}\else \'{c}\fi{}}},\ }\href {\doibase
  10.1103/PhysRevB.98.155134} {\bibfield  {journal} {\bibinfo  {journal} {Phys.
  Rev. B}\ }\textbf {\bibinfo {volume} {98}},\ \bibinfo {pages} {155134}
  (\bibinfo {year} {2018}{\natexlab{b}})}\BibitemShut {NoStop}%
\bibitem [{\citenamefont {Ho}\ \emph {et~al.}(2019)\citenamefont {Ho},
  \citenamefont {Choi}, \citenamefont {Pichler},\ and\ \citenamefont
  {Lukin}}]{ho2018periodic}%
  \BibitemOpen
  \bibfield  {author} {\bibinfo {author} {\bibfnamefont {W.~W.}\ \bibnamefont
  {Ho}}, \bibinfo {author} {\bibfnamefont {S.}~\bibnamefont {Choi}}, \bibinfo
  {author} {\bibfnamefont {H.}~\bibnamefont {Pichler}}, \ and\ \bibinfo
  {author} {\bibfnamefont {M.~D.}\ \bibnamefont {Lukin}},\ }\href {\doibase
  10.1103/PhysRevLett.122.040603} {\bibfield  {journal} {\bibinfo  {journal}
  {Phys. Rev. Lett.}\ }\textbf {\bibinfo {volume} {122}},\ \bibinfo {pages}
  {040603} (\bibinfo {year} {2019})}\BibitemShut {NoStop}%
\bibitem [{\citenamefont {Michailidis}\ \emph {et~al.}(2020)\citenamefont
  {Michailidis}, \citenamefont {Turner}, \citenamefont
  {Papi\ifmmode~\acute{c}\else \'{c}\fi{}}, \citenamefont {Abanin},\ and\
  \citenamefont {Serbyn}}]{michailidis2019slow}%
  \BibitemOpen
  \bibfield  {author} {\bibinfo {author} {\bibfnamefont {A.~A.}\ \bibnamefont
  {Michailidis}}, \bibinfo {author} {\bibfnamefont {C.~J.}\ \bibnamefont
  {Turner}}, \bibinfo {author} {\bibfnamefont {Z.}~\bibnamefont
  {Papi\ifmmode~\acute{c}\else \'{c}\fi{}}}, \bibinfo {author} {\bibfnamefont
  {D.~A.}\ \bibnamefont {Abanin}}, \ and\ \bibinfo {author} {\bibfnamefont
  {M.}~\bibnamefont {Serbyn}},\ }\href {\doibase 10.1103/PhysRevX.10.011055}
  {\bibfield  {journal} {\bibinfo  {journal} {Phys. Rev. X}\ }\textbf {\bibinfo
  {volume} {10}},\ \bibinfo {pages} {011055} (\bibinfo {year}
  {2020})}\BibitemShut {NoStop}%
\bibitem [{\citenamefont {Khemani}\ \emph {et~al.}(2019)\citenamefont
  {Khemani}, \citenamefont {Laumann},\ and\ \citenamefont
  {Chandran}}]{khemani2019signatures}%
  \BibitemOpen
  \bibfield  {author} {\bibinfo {author} {\bibfnamefont {V.}~\bibnamefont
  {Khemani}}, \bibinfo {author} {\bibfnamefont {C.~R.}\ \bibnamefont
  {Laumann}}, \ and\ \bibinfo {author} {\bibfnamefont {A.}~\bibnamefont
  {Chandran}},\ }\href {\doibase 10.1103/PhysRevB.99.161101} {\bibfield
  {journal} {\bibinfo  {journal} {Phys. Rev. B}\ }\textbf {\bibinfo {volume}
  {99}},\ \bibinfo {pages} {161101} (\bibinfo {year} {2019})}\BibitemShut
  {NoStop}%
\bibitem [{\citenamefont {Lin}\ and\ \citenamefont
  {Motrunich}(2019)}]{lin2019exact}%
  \BibitemOpen
  \bibfield  {author} {\bibinfo {author} {\bibfnamefont {C.-J.}\ \bibnamefont
  {Lin}}\ and\ \bibinfo {author} {\bibfnamefont {O.~I.}\ \bibnamefont
  {Motrunich}},\ }\href {\doibase 10.1103/PhysRevLett.122.173401} {\bibfield
  {journal} {\bibinfo  {journal} {Phys. Rev. Lett.}\ }\textbf {\bibinfo
  {volume} {122}},\ \bibinfo {pages} {173401} (\bibinfo {year}
  {2019})}\BibitemShut {NoStop}%
\bibitem [{\citenamefont {Iadecola}\ \emph {et~al.}(2019)\citenamefont
  {Iadecola}, \citenamefont {Schecter},\ and\ \citenamefont
  {Xu}}]{iadecola2019quantum}%
  \BibitemOpen
  \bibfield  {author} {\bibinfo {author} {\bibfnamefont {T.}~\bibnamefont
  {Iadecola}}, \bibinfo {author} {\bibfnamefont {M.}~\bibnamefont {Schecter}},
  \ and\ \bibinfo {author} {\bibfnamefont {S.}~\bibnamefont {Xu}},\ }\href
  {\doibase 10.1103/PhysRevB.100.184312} {\bibfield  {journal} {\bibinfo
  {journal} {Phys. Rev. B}\ }\textbf {\bibinfo {volume} {100}},\ \bibinfo
  {pages} {184312} (\bibinfo {year} {2019})}\BibitemShut {NoStop}%
\bibitem [{\citenamefont {Choi}\ \emph {et~al.}(2019)\citenamefont {Choi},
  \citenamefont {Turner}, \citenamefont {Pichler}, \citenamefont {Ho},
  \citenamefont {Michailidis}, \citenamefont {Papi\ifmmode~\acute{c}\else
  \'{c}\fi{}}, \citenamefont {Serbyn}, \citenamefont {Lukin},\ and\
  \citenamefont {Abanin}}]{choi2018emergent}%
  \BibitemOpen
  \bibfield  {author} {\bibinfo {author} {\bibfnamefont {S.}~\bibnamefont
  {Choi}}, \bibinfo {author} {\bibfnamefont {C.~J.}\ \bibnamefont {Turner}},
  \bibinfo {author} {\bibfnamefont {H.}~\bibnamefont {Pichler}}, \bibinfo
  {author} {\bibfnamefont {W.~W.}\ \bibnamefont {Ho}}, \bibinfo {author}
  {\bibfnamefont {A.~A.}\ \bibnamefont {Michailidis}}, \bibinfo {author}
  {\bibfnamefont {Z.}~\bibnamefont {Papi\ifmmode~\acute{c}\else \'{c}\fi{}}},
  \bibinfo {author} {\bibfnamefont {M.}~\bibnamefont {Serbyn}}, \bibinfo
  {author} {\bibfnamefont {M.~D.}\ \bibnamefont {Lukin}}, \ and\ \bibinfo
  {author} {\bibfnamefont {D.~A.}\ \bibnamefont {Abanin}},\ }\href {\doibase
  10.1103/PhysRevLett.122.220603} {\bibfield  {journal} {\bibinfo  {journal}
  {Phys. Rev. Lett.}\ }\textbf {\bibinfo {volume} {122}},\ \bibinfo {pages}
  {220603} (\bibinfo {year} {2019})}\BibitemShut {NoStop}%
\bibitem [{\citenamefont {Bull}\ \emph {et~al.}(2020)\citenamefont {Bull},
  \citenamefont {Desaules},\ and\ \citenamefont {Papi\ifmmode~\acute{c}\else
  \'{c}\fi{}}}]{bull2020quantum}%
  \BibitemOpen
  \bibfield  {author} {\bibinfo {author} {\bibfnamefont {K.}~\bibnamefont
  {Bull}}, \bibinfo {author} {\bibfnamefont {J.-Y.}\ \bibnamefont {Desaules}},
  \ and\ \bibinfo {author} {\bibfnamefont {Z.}~\bibnamefont
  {Papi\ifmmode~\acute{c}\else \'{c}\fi{}}},\ }\href {\doibase
  10.1103/PhysRevB.101.165139} {\bibfield  {journal} {\bibinfo  {journal}
  {Phys. Rev. B}\ }\textbf {\bibinfo {volume} {101}},\ \bibinfo {pages}
  {165139} (\bibinfo {year} {2020})}\BibitemShut {NoStop}%
\bibitem [{\citenamefont {Bull}\ \emph {et~al.}(2019)\citenamefont {Bull},
  \citenamefont {Martin},\ and\ \citenamefont {Papi\ifmmode~\acute{c}\else
  \'{c}\fi{}}}]{2bull2019scar}%
  \BibitemOpen
  \bibfield  {author} {\bibinfo {author} {\bibfnamefont {K.}~\bibnamefont
  {Bull}}, \bibinfo {author} {\bibfnamefont {I.}~\bibnamefont {Martin}}, \ and\
  \bibinfo {author} {\bibfnamefont {Z.}~\bibnamefont
  {Papi\ifmmode~\acute{c}\else \'{c}\fi{}}},\ }\href {\doibase
  10.1103/PhysRevLett.123.030601} {\bibfield  {journal} {\bibinfo  {journal}
  {Phys. Rev. Lett.}\ }\textbf {\bibinfo {volume} {123}},\ \bibinfo {pages}
  {030601} (\bibinfo {year} {2019})}\BibitemShut {NoStop}%
\bibitem [{\citenamefont {{Moudgalya}}\ \emph {et~al.}(2019)\citenamefont
  {{Moudgalya}}, \citenamefont {{Bernevig}},\ and\ \citenamefont
  {{Regnault}}}]{moudgalya2019quantum}%
  \BibitemOpen
  \bibfield  {author} {\bibinfo {author} {\bibfnamefont {S.}~\bibnamefont
  {{Moudgalya}}}, \bibinfo {author} {\bibfnamefont {B.~A.}\ \bibnamefont
  {{Bernevig}}}, \ and\ \bibinfo {author} {\bibfnamefont {N.}~\bibnamefont
  {{Regnault}}},\ }\href@noop {} {\bibfield  {journal} {\bibinfo  {journal}
  {arXiv e-prints}\ } (\bibinfo {year} {2019})},\ \Eprint
  {http://arxiv.org/abs/1906.05292} {arXiv:1906.05292 [cond-mat.str-el]}
  \BibitemShut {NoStop}%
\bibitem [{\citenamefont {Hudomal}\ \emph {et~al.}(2020)\citenamefont
  {Hudomal}, \citenamefont {Vasi{\'c}}, \citenamefont {Regnault},\ and\
  \citenamefont {Papi{\'c}}}]{hudomal2019}%
  \BibitemOpen
  \bibfield  {author} {\bibinfo {author} {\bibfnamefont {A.}~\bibnamefont
  {Hudomal}}, \bibinfo {author} {\bibfnamefont {I.}~\bibnamefont {Vasi{\'c}}},
  \bibinfo {author} {\bibfnamefont {N.}~\bibnamefont {Regnault}}, \ and\
  \bibinfo {author} {\bibfnamefont {Z.}~\bibnamefont {Papi{\'c}}},\ }\href
  {\doibase 10.1038/s42005-020-0364-9} {\bibfield  {journal} {\bibinfo
  {journal} {Communications Physics}\ }\textbf {\bibinfo {volume} {3}},\
  \bibinfo {pages} {99} (\bibinfo {year} {2020})}\BibitemShut {NoStop}%
\bibitem [{\citenamefont {\ifmmode \check{Z}\else
  \v{Z}\fi{}nidari\ifmmode~\check{c}\else
  \v{c}\fi{}}(2013)}]{znidaric2019coexistence}%
  \BibitemOpen
  \bibfield  {author} {\bibinfo {author} {\bibfnamefont {M.}~\bibnamefont
  {\ifmmode \check{Z}\else \v{Z}\fi{}nidari\ifmmode~\check{c}\else
  \v{c}\fi{}}},\ }\href {\doibase 10.1103/PhysRevLett.110.070602} {\bibfield
  {journal} {\bibinfo  {journal} {Phys. Rev. Lett.}\ }\textbf {\bibinfo
  {volume} {110}},\ \bibinfo {pages} {070602} (\bibinfo {year}
  {2013})}\BibitemShut {NoStop}%
\bibitem [{\citenamefont {Sala}\ \emph {et~al.}(2020)\citenamefont {Sala},
  \citenamefont {Rakovszky}, \citenamefont {Verresen}, \citenamefont {Knap},\
  and\ \citenamefont {Pollmann}}]{sala2019ergodicity}%
  \BibitemOpen
  \bibfield  {author} {\bibinfo {author} {\bibfnamefont {P.}~\bibnamefont
  {Sala}}, \bibinfo {author} {\bibfnamefont {T.}~\bibnamefont {Rakovszky}},
  \bibinfo {author} {\bibfnamefont {R.}~\bibnamefont {Verresen}}, \bibinfo
  {author} {\bibfnamefont {M.}~\bibnamefont {Knap}}, \ and\ \bibinfo {author}
  {\bibfnamefont {F.}~\bibnamefont {Pollmann}},\ }\href {\doibase
  10.1103/PhysRevX.10.011047} {\bibfield  {journal} {\bibinfo  {journal} {Phys.
  Rev. X}\ }\textbf {\bibinfo {volume} {10}},\ \bibinfo {pages} {011047}
  (\bibinfo {year} {2020})}\BibitemShut {NoStop}%
\bibitem [{\citenamefont {Moudgalya}\ \emph {et~al.}(2019)\citenamefont
  {Moudgalya}, \citenamefont {Prem}, \citenamefont {Nandkishore}, \citenamefont
  {Regnault},\ and\ \citenamefont {Bernevig}}]{moudgalya2019thermalization}%
  \BibitemOpen
  \bibfield  {author} {\bibinfo {author} {\bibfnamefont {S.}~\bibnamefont
  {Moudgalya}}, \bibinfo {author} {\bibfnamefont {A.}~\bibnamefont {Prem}},
  \bibinfo {author} {\bibfnamefont {R.}~\bibnamefont {Nandkishore}}, \bibinfo
  {author} {\bibfnamefont {N.}~\bibnamefont {Regnault}}, \ and\ \bibinfo
  {author} {\bibfnamefont {B.~A.}\ \bibnamefont {Bernevig}},\ }\href@noop {}
  {\bibfield  {journal} {\bibinfo  {journal} {arXiv e-prints}\ } (\bibinfo
  {year} {2019})},\ \Eprint {http://arxiv.org/abs/1910.14048} {arXiv:1910.14048
  [cond-mat.str-el]} \BibitemShut {NoStop}%
\bibitem [{\citenamefont {Pancotti}\ \emph {et~al.}(2020)\citenamefont
  {Pancotti}, \citenamefont {Giudice}, \citenamefont {Cirac}, \citenamefont
  {Garrahan},\ and\ \citenamefont {Ba\~nuls}}]{pancotti2019quantum}%
  \BibitemOpen
  \bibfield  {author} {\bibinfo {author} {\bibfnamefont {N.}~\bibnamefont
  {Pancotti}}, \bibinfo {author} {\bibfnamefont {G.}~\bibnamefont {Giudice}},
  \bibinfo {author} {\bibfnamefont {J.~I.}\ \bibnamefont {Cirac}}, \bibinfo
  {author} {\bibfnamefont {J.~P.}\ \bibnamefont {Garrahan}}, \ and\ \bibinfo
  {author} {\bibfnamefont {M.~C.}\ \bibnamefont {Ba\~nuls}},\ }\href {\doibase
  10.1103/PhysRevX.10.021051} {\bibfield  {journal} {\bibinfo  {journal} {Phys.
  Rev. X}\ }\textbf {\bibinfo {volume} {10}},\ \bibinfo {pages} {021051}
  (\bibinfo {year} {2020})}\BibitemShut {NoStop}%
\bibitem [{\citenamefont {Yang}\ \emph {et~al.}(2020)\citenamefont {Yang},
  \citenamefont {Liu}, \citenamefont {Gorshkov},\ and\ \citenamefont
  {Iadecola}}]{yang2019hilbertspace}%
  \BibitemOpen
  \bibfield  {author} {\bibinfo {author} {\bibfnamefont {Z.-C.}\ \bibnamefont
  {Yang}}, \bibinfo {author} {\bibfnamefont {F.}~\bibnamefont {Liu}}, \bibinfo
  {author} {\bibfnamefont {A.~V.}\ \bibnamefont {Gorshkov}}, \ and\ \bibinfo
  {author} {\bibfnamefont {T.}~\bibnamefont {Iadecola}},\ }\href {\doibase
  10.1103/PhysRevLett.124.207602} {\bibfield  {journal} {\bibinfo  {journal}
  {Phys. Rev. Lett.}\ }\textbf {\bibinfo {volume} {124}},\ \bibinfo {pages}
  {207602} (\bibinfo {year} {2020})}\BibitemShut {NoStop}%
\bibitem [{\citenamefont {Zhao}\ \emph {et~al.}(2020)\citenamefont {Zhao},
  \citenamefont {Vovrosh}, \citenamefont {Mintert},\ and\ \citenamefont
  {Knolle}}]{zhao2020quantum}%
  \BibitemOpen
  \bibfield  {author} {\bibinfo {author} {\bibfnamefont {H.}~\bibnamefont
  {Zhao}}, \bibinfo {author} {\bibfnamefont {J.}~\bibnamefont {Vovrosh}},
  \bibinfo {author} {\bibfnamefont {F.}~\bibnamefont {Mintert}}, \ and\
  \bibinfo {author} {\bibfnamefont {J.}~\bibnamefont {Knolle}},\ }\href
  {\doibase 10.1103/PhysRevLett.124.160604} {\bibfield  {journal} {\bibinfo
  {journal} {Phys. Rev. Lett.}\ }\textbf {\bibinfo {volume} {124}},\ \bibinfo
  {pages} {160604} (\bibinfo {year} {2020})}\BibitemShut {NoStop}%
\bibitem [{\citenamefont {Robinson}\ \emph {et~al.}(2019)\citenamefont
  {Robinson}, \citenamefont {James},\ and\ \citenamefont
  {Konik}}]{robinson2019signatures}%
  \BibitemOpen
  \bibfield  {author} {\bibinfo {author} {\bibfnamefont {N.~J.}\ \bibnamefont
  {Robinson}}, \bibinfo {author} {\bibfnamefont {A.~J.~A.}\ \bibnamefont
  {James}}, \ and\ \bibinfo {author} {\bibfnamefont {R.~M.}\ \bibnamefont
  {Konik}},\ }\href {\doibase 10.1103/PhysRevB.99.195108} {\bibfield  {journal}
  {\bibinfo  {journal} {Phys. Rev. B}\ }\textbf {\bibinfo {volume} {99}},\
  \bibinfo {pages} {195108} (\bibinfo {year} {2019})}\BibitemShut {NoStop}%
\bibitem [{\citenamefont {James}\ \emph {et~al.}(2019)\citenamefont {James},
  \citenamefont {Konik},\ and\ \citenamefont {Robinson}}]{james2019nonthermal}%
  \BibitemOpen
  \bibfield  {author} {\bibinfo {author} {\bibfnamefont {A.~J.~A.}\
  \bibnamefont {James}}, \bibinfo {author} {\bibfnamefont {R.~M.}\ \bibnamefont
  {Konik}}, \ and\ \bibinfo {author} {\bibfnamefont {N.~J.}\ \bibnamefont
  {Robinson}},\ }\href {\doibase 10.1103/PhysRevLett.122.130603} {\bibfield
  {journal} {\bibinfo  {journal} {Phys. Rev. Lett.}\ }\textbf {\bibinfo
  {volume} {122}},\ \bibinfo {pages} {130603} (\bibinfo {year}
  {2019})}\BibitemShut {NoStop}%
\bibitem [{\citenamefont {Lerose}\ \emph {et~al.}(2019)\citenamefont {Lerose},
  \citenamefont {Surace}, \citenamefont {Mazza}, \citenamefont {Perfetto},
  \citenamefont {Collura},\ and\ \citenamefont
  {Gambassi}}]{lerose2019quasilocalized}%
  \BibitemOpen
  \bibfield  {author} {\bibinfo {author} {\bibfnamefont {A.}~\bibnamefont
  {Lerose}}, \bibinfo {author} {\bibfnamefont {F.~M.}\ \bibnamefont {Surace}},
  \bibinfo {author} {\bibfnamefont {P.~P.}\ \bibnamefont {Mazza}}, \bibinfo
  {author} {\bibfnamefont {G.}~\bibnamefont {Perfetto}}, \bibinfo {author}
  {\bibfnamefont {M.}~\bibnamefont {Collura}}, \ and\ \bibinfo {author}
  {\bibfnamefont {A.}~\bibnamefont {Gambassi}},\ }\href@noop {} {\bibfield
  {journal} {\bibinfo  {journal} {arXiv e-prints}\ } (\bibinfo {year}
  {2019})},\ \Eprint {http://arxiv.org/abs/1911.07877} {arXiv:1911.07877
  [cond-mat.stat-mech]} \BibitemShut {NoStop}%
\bibitem [{\citenamefont {Surace}\ \emph {et~al.}(2020)\citenamefont {Surace},
  \citenamefont {Mazza}, \citenamefont {Giudici}, \citenamefont {Lerose},
  \citenamefont {Gambassi},\ and\ \citenamefont
  {Dalmonte}}]{surace2019lattice}%
  \BibitemOpen
  \bibfield  {author} {\bibinfo {author} {\bibfnamefont {F.~M.}\ \bibnamefont
  {Surace}}, \bibinfo {author} {\bibfnamefont {P.~P.}\ \bibnamefont {Mazza}},
  \bibinfo {author} {\bibfnamefont {G.}~\bibnamefont {Giudici}}, \bibinfo
  {author} {\bibfnamefont {A.}~\bibnamefont {Lerose}}, \bibinfo {author}
  {\bibfnamefont {A.}~\bibnamefont {Gambassi}}, \ and\ \bibinfo {author}
  {\bibfnamefont {M.}~\bibnamefont {Dalmonte}},\ }\href {\doibase
  10.1103/PhysRevX.10.021041} {\bibfield  {journal} {\bibinfo  {journal} {Phys.
  Rev. X}\ }\textbf {\bibinfo {volume} {10}},\ \bibinfo {pages} {021041}
  (\bibinfo {year} {2020})}\BibitemShut {NoStop}%
\bibitem [{\citenamefont {Lin}\ \emph {et~al.}(2020)\citenamefont {Lin},
  \citenamefont {Chandran},\ and\ \citenamefont {Motrunich}}]{lin2019pxp}%
  \BibitemOpen
  \bibfield  {author} {\bibinfo {author} {\bibfnamefont {C.-J.}\ \bibnamefont
  {Lin}}, \bibinfo {author} {\bibfnamefont {A.}~\bibnamefont {Chandran}}, \
  and\ \bibinfo {author} {\bibfnamefont {O.~I.}\ \bibnamefont {Motrunich}},\
  }\href {\doibase 10.1103/PhysRevResearch.2.033044} {\bibfield  {journal}
  {\bibinfo  {journal} {Phys. Rev. Research}\ }\textbf {\bibinfo {volume}
  {2}},\ \bibinfo {pages} {033044} (\bibinfo {year} {2020})}\BibitemShut
  {NoStop}%
\bibitem [{\citenamefont {Alhambra}\ \emph {et~al.}(2020)\citenamefont
  {Alhambra}, \citenamefont {Anshu},\ and\ \citenamefont
  {Wilming}}]{alhambra2019revivals}%
  \BibitemOpen
  \bibfield  {author} {\bibinfo {author} {\bibfnamefont {A.~M.}\ \bibnamefont
  {Alhambra}}, \bibinfo {author} {\bibfnamefont {A.}~\bibnamefont {Anshu}}, \
  and\ \bibinfo {author} {\bibfnamefont {H.}~\bibnamefont {Wilming}},\ }\href
  {\doibase 10.1103/PhysRevB.101.205107} {\bibfield  {journal} {\bibinfo
  {journal} {Phys. Rev. B}\ }\textbf {\bibinfo {volume} {101}},\ \bibinfo
  {pages} {205107} (\bibinfo {year} {2020})}\BibitemShut {NoStop}%
\bibitem [{\citenamefont {Pai}\ and\ \citenamefont
  {Pretko}(2019)}]{pai2019robust}%
  \BibitemOpen
  \bibfield  {author} {\bibinfo {author} {\bibfnamefont {S.}~\bibnamefont
  {Pai}}\ and\ \bibinfo {author} {\bibfnamefont {M.}~\bibnamefont {Pretko}},\
  }\href {\doibase 10.1103/PhysRevLett.123.136401} {\bibfield  {journal}
  {\bibinfo  {journal} {Phys. Rev. Lett.}\ }\textbf {\bibinfo {volume} {123}},\
  \bibinfo {pages} {136401} (\bibinfo {year} {2019})}\BibitemShut {NoStop}%
\bibitem [{\citenamefont {Khemani}\ \emph {et~al.}(2020)\citenamefont
  {Khemani}, \citenamefont {Hermele},\ and\ \citenamefont
  {Nandkishore}}]{khemani2019localization}%
  \BibitemOpen
  \bibfield  {author} {\bibinfo {author} {\bibfnamefont {V.}~\bibnamefont
  {Khemani}}, \bibinfo {author} {\bibfnamefont {M.}~\bibnamefont {Hermele}}, \
  and\ \bibinfo {author} {\bibfnamefont {R.}~\bibnamefont {Nandkishore}},\
  }\href {\doibase 10.1103/PhysRevB.101.174204} {\bibfield  {journal} {\bibinfo
   {journal} {Phys. Rev. B}\ }\textbf {\bibinfo {volume} {101}},\ \bibinfo
  {pages} {174204} (\bibinfo {year} {2020})}\BibitemShut {NoStop}%
\bibitem [{\citenamefont {Mukherjee}\ \emph {et~al.}(2020)\citenamefont
  {Mukherjee}, \citenamefont {Nandy}, \citenamefont {Sen}, \citenamefont
  {Sen},\ and\ \citenamefont {Sengupta}}]{mukherjee2019collapse}%
  \BibitemOpen
  \bibfield  {author} {\bibinfo {author} {\bibfnamefont {B.}~\bibnamefont
  {Mukherjee}}, \bibinfo {author} {\bibfnamefont {S.}~\bibnamefont {Nandy}},
  \bibinfo {author} {\bibfnamefont {A.}~\bibnamefont {Sen}}, \bibinfo {author}
  {\bibfnamefont {D.}~\bibnamefont {Sen}}, \ and\ \bibinfo {author}
  {\bibfnamefont {K.}~\bibnamefont {Sengupta}},\ }\href {\doibase
  10.1103/PhysRevB.101.245107} {\bibfield  {journal} {\bibinfo  {journal}
  {Phys. Rev. B}\ }\textbf {\bibinfo {volume} {101}},\ \bibinfo {pages}
  {245107} (\bibinfo {year} {2020})}\BibitemShut {NoStop}%
\bibitem [{\citenamefont {Haldar}\ \emph {et~al.}(2019)\citenamefont {Haldar},
  \citenamefont {Sen}, \citenamefont {Moessner},\ and\ \citenamefont
  {Das}}]{haldar2019scars}%
  \BibitemOpen
  \bibfield  {author} {\bibinfo {author} {\bibfnamefont {A.}~\bibnamefont
  {Haldar}}, \bibinfo {author} {\bibfnamefont {D.}~\bibnamefont {Sen}},
  \bibinfo {author} {\bibfnamefont {R.}~\bibnamefont {Moessner}}, \ and\
  \bibinfo {author} {\bibfnamefont {A.}~\bibnamefont {Das}},\ }\href@noop {}
  {\bibfield  {journal} {\bibinfo  {journal} {arXiv e-prints}\ } (\bibinfo
  {year} {2019})},\ \Eprint {http://arxiv.org/abs/1909.04064} {arXiv:1909.04064
  [cond-mat.other]} \BibitemShut {NoStop}%
\bibitem [{\citenamefont {Ok}\ \emph {et~al.}(2019)\citenamefont {Ok},
  \citenamefont {Choo}, \citenamefont {Mudry}, \citenamefont {Castelnovo},
  \citenamefont {Chamon},\ and\ \citenamefont {Neupert}}]{ok2019topological}%
  \BibitemOpen
  \bibfield  {author} {\bibinfo {author} {\bibfnamefont {S.}~\bibnamefont
  {Ok}}, \bibinfo {author} {\bibfnamefont {K.}~\bibnamefont {Choo}}, \bibinfo
  {author} {\bibfnamefont {C.}~\bibnamefont {Mudry}}, \bibinfo {author}
  {\bibfnamefont {C.}~\bibnamefont {Castelnovo}}, \bibinfo {author}
  {\bibfnamefont {C.}~\bibnamefont {Chamon}}, \ and\ \bibinfo {author}
  {\bibfnamefont {T.}~\bibnamefont {Neupert}},\ }\href {\doibase
  10.1103/PhysRevResearch.1.033144} {\bibfield  {journal} {\bibinfo  {journal}
  {Phys. Rev. Research}\ }\textbf {\bibinfo {volume} {1}},\ \bibinfo {pages}
  {033144} (\bibinfo {year} {2019})}\BibitemShut {NoStop}%
\bibitem [{\citenamefont {Lee}\ \emph {et~al.}(2020)\citenamefont {Lee},
  \citenamefont {Melendrez}, \citenamefont {Pal},\ and\ \citenamefont
  {Changlani}}]{lee2020exact}%
  \BibitemOpen
  \bibfield  {author} {\bibinfo {author} {\bibfnamefont {K.}~\bibnamefont
  {Lee}}, \bibinfo {author} {\bibfnamefont {R.}~\bibnamefont {Melendrez}},
  \bibinfo {author} {\bibfnamefont {A.}~\bibnamefont {Pal}}, \ and\ \bibinfo
  {author} {\bibfnamefont {H.~J.}\ \bibnamefont {Changlani}},\ }\href {\doibase
  10.1103/PhysRevB.101.241111} {\bibfield  {journal} {\bibinfo  {journal}
  {Phys. Rev. B}\ }\textbf {\bibinfo {volume} {101}},\ \bibinfo {pages}
  {241111} (\bibinfo {year} {2020})}\BibitemShut {NoStop}%
\bibitem [{\citenamefont {Schecter}\ and\ \citenamefont
  {Iadecola}(2019)}]{schecter2019weak}%
  \BibitemOpen
  \bibfield  {author} {\bibinfo {author} {\bibfnamefont {M.}~\bibnamefont
  {Schecter}}\ and\ \bibinfo {author} {\bibfnamefont {T.}~\bibnamefont
  {Iadecola}},\ }\href {\doibase 10.1103/PhysRevLett.123.147201} {\bibfield
  {journal} {\bibinfo  {journal} {Phys. Rev. Lett.}\ }\textbf {\bibinfo
  {volume} {123}},\ \bibinfo {pages} {147201} (\bibinfo {year}
  {2019})}\BibitemShut {NoStop}%
\bibitem [{\citenamefont {Chattopadhyay}\ \emph {et~al.}(2020)\citenamefont
  {Chattopadhyay}, \citenamefont {Pichler}, \citenamefont {Lukin},\ and\
  \citenamefont {Ho}}]{chattopadhyay2019quantum}%
  \BibitemOpen
  \bibfield  {author} {\bibinfo {author} {\bibfnamefont {S.}~\bibnamefont
  {Chattopadhyay}}, \bibinfo {author} {\bibfnamefont {H.}~\bibnamefont
  {Pichler}}, \bibinfo {author} {\bibfnamefont {M.~D.}\ \bibnamefont {Lukin}},
  \ and\ \bibinfo {author} {\bibfnamefont {W.~W.}\ \bibnamefont {Ho}},\ }\href
  {\doibase 10.1103/PhysRevB.101.174308} {\bibfield  {journal} {\bibinfo
  {journal} {Phys. Rev. B}\ }\textbf {\bibinfo {volume} {101}},\ \bibinfo
  {pages} {174308} (\bibinfo {year} {2020})}\BibitemShut {NoStop}%
\bibitem [{\citenamefont {Iadecola}\ and\ \citenamefont
  {Schecter}(2020)}]{iadecola2019quantum2}%
  \BibitemOpen
  \bibfield  {author} {\bibinfo {author} {\bibfnamefont {T.}~\bibnamefont
  {Iadecola}}\ and\ \bibinfo {author} {\bibfnamefont {M.}~\bibnamefont
  {Schecter}},\ }\href {\doibase 10.1103/PhysRevB.101.024306} {\bibfield
  {journal} {\bibinfo  {journal} {Phys. Rev. B}\ }\textbf {\bibinfo {volume}
  {101}},\ \bibinfo {pages} {024306} (\bibinfo {year} {2020})}\BibitemShut
  {NoStop}%
\bibitem [{\citenamefont {Mark}\ \emph
  {et~al.}(2020{\natexlab{a}})\citenamefont {Mark}, \citenamefont {Lin},\ and\
  \citenamefont {Motrunich}}]{mark2020unified}%
  \BibitemOpen
  \bibfield  {author} {\bibinfo {author} {\bibfnamefont {D.~K.}\ \bibnamefont
  {Mark}}, \bibinfo {author} {\bibfnamefont {C.-J.}\ \bibnamefont {Lin}}, \
  and\ \bibinfo {author} {\bibfnamefont {O.~I.}\ \bibnamefont {Motrunich}},\
  }\href {\doibase 10.1103/PhysRevB.101.195131} {\bibfield  {journal} {\bibinfo
   {journal} {Phys. Rev. B}\ }\textbf {\bibinfo {volume} {101}},\ \bibinfo
  {pages} {195131} (\bibinfo {year} {2020}{\natexlab{a}})}\BibitemShut
  {NoStop}%
\bibitem [{\citenamefont {Vernier}\ \emph {et~al.}(2019)\citenamefont
  {Vernier}, \citenamefont {O'Brien},\ and\ \citenamefont
  {Fendley}}]{Vernier2019}%
  \BibitemOpen
  \bibfield  {author} {\bibinfo {author} {\bibfnamefont {E.}~\bibnamefont
  {Vernier}}, \bibinfo {author} {\bibfnamefont {E.}~\bibnamefont {O'Brien}}, \
  and\ \bibinfo {author} {\bibfnamefont {P.}~\bibnamefont {Fendley}},\ }\href
  {\doibase 10.1088/1742-5468/ab11c0} {\bibfield  {journal} {\bibinfo
  {journal} {Journal of Statistical Mechanics: Theory and Experiment}\ }\textbf
  {\bibinfo {volume} {2019}},\ \bibinfo {pages} {043107} (\bibinfo {year}
  {2019})}\BibitemShut {NoStop}%
\bibitem [{\citenamefont {Shibata}\ \emph {et~al.}(2020)\citenamefont
  {Shibata}, \citenamefont {Yoshioka},\ and\ \citenamefont
  {Katsura}}]{shibata2019onsagers}%
  \BibitemOpen
  \bibfield  {author} {\bibinfo {author} {\bibfnamefont {N.}~\bibnamefont
  {Shibata}}, \bibinfo {author} {\bibfnamefont {N.}~\bibnamefont {Yoshioka}}, \
  and\ \bibinfo {author} {\bibfnamefont {H.}~\bibnamefont {Katsura}},\ }\href
  {\doibase 10.1103/PhysRevLett.124.180604} {\bibfield  {journal} {\bibinfo
  {journal} {Phys. Rev. Lett.}\ }\textbf {\bibinfo {volume} {124}},\ \bibinfo
  {pages} {180604} (\bibinfo {year} {2020})}\BibitemShut {NoStop}%
\bibitem [{\citenamefont {Yang}(1987)}]{Yang1987}%
  \BibitemOpen
  \bibfield  {author} {\bibinfo {author} {\bibfnamefont {S.-K.}\ \bibnamefont
  {Yang}},\ }\href {\doibase https://doi.org/10.1016/0550-3213(87)90359-2}
  {\bibfield  {journal} {\bibinfo  {journal} {Nuclear Physics B}\ }\textbf
  {\bibinfo {volume} {285}},\ \bibinfo {pages} {639 } (\bibinfo {year}
  {1987})}\BibitemShut {NoStop}%
\bibitem [{\citenamefont {Vafek}\ \emph {et~al.}(2017)\citenamefont {Vafek},
  \citenamefont {Regnault},\ and\ \citenamefont
  {Bernevig}}]{vafek2017entanglement}%
  \BibitemOpen
  \bibfield  {author} {\bibinfo {author} {\bibfnamefont {O.}~\bibnamefont
  {Vafek}}, \bibinfo {author} {\bibfnamefont {N.}~\bibnamefont {Regnault}}, \
  and\ \bibinfo {author} {\bibfnamefont {B.~A.}\ \bibnamefont {Bernevig}},\
  }\href {\doibase 10.21468/SciPostPhys.3.6.043} {\bibfield  {journal}
  {\bibinfo  {journal} {SciPost Phys.}\ }\textbf {\bibinfo {volume} {3}},\
  \bibinfo {pages} {043} (\bibinfo {year} {2017})}\BibitemShut {NoStop}%
\bibitem [{\citenamefont {Yu}\ \emph {et~al.}(2018)\citenamefont {Yu},
  \citenamefont {Luo},\ and\ \citenamefont {Clark}}]{yu2018beyond}%
  \BibitemOpen
  \bibfield  {author} {\bibinfo {author} {\bibfnamefont {X.}~\bibnamefont
  {Yu}}, \bibinfo {author} {\bibfnamefont {D.}~\bibnamefont {Luo}}, \ and\
  \bibinfo {author} {\bibfnamefont {B.~K.}\ \bibnamefont {Clark}},\ }\href
  {\doibase 10.1103/PhysRevB.98.115106} {\bibfield  {journal} {\bibinfo
  {journal} {Phys. Rev. B}\ }\textbf {\bibinfo {volume} {98}},\ \bibinfo
  {pages} {115106} (\bibinfo {year} {2018})}\BibitemShut {NoStop}%
\bibitem [{\citenamefont {Affleck}\ \emph {et~al.}(1987)\citenamefont
  {Affleck}, \citenamefont {Kennedy}, \citenamefont {Lieb},\ and\ \citenamefont
  {Tasaki}}]{Affleck1987}%
  \BibitemOpen
  \bibfield  {author} {\bibinfo {author} {\bibfnamefont {I.}~\bibnamefont
  {Affleck}}, \bibinfo {author} {\bibfnamefont {T.}~\bibnamefont {Kennedy}},
  \bibinfo {author} {\bibfnamefont {E.~H.}\ \bibnamefont {Lieb}}, \ and\
  \bibinfo {author} {\bibfnamefont {H.}~\bibnamefont {Tasaki}},\ }\href
  {\doibase 10.1103/PhysRevLett.59.799} {\bibfield  {journal} {\bibinfo
  {journal} {Phys. Rev. Lett.}\ }\textbf {\bibinfo {volume} {59}},\ \bibinfo
  {pages} {799} (\bibinfo {year} {1987})}\BibitemShut {NoStop}%
\bibitem [{\citenamefont {Kl\"{u}mper}\ \emph {et~al.}(1993)\citenamefont
  {Kl\"{u}mper}, \citenamefont {Schadschneider},\ and\ \citenamefont
  {Zittartz}}]{Klumper1993}%
  \BibitemOpen
  \bibfield  {author} {\bibinfo {author} {\bibfnamefont {A.}~\bibnamefont
  {Kl\"{u}mper}}, \bibinfo {author} {\bibfnamefont {A.}~\bibnamefont
  {Schadschneider}}, \ and\ \bibinfo {author} {\bibfnamefont {J.}~\bibnamefont
  {Zittartz}},\ }\href {\doibase 10.1209/0295-5075/24/4/010} {\bibfield
  {journal} {\bibinfo  {journal} {Europhysics Letters ({EPL})}\ }\textbf
  {\bibinfo {volume} {24}},\ \bibinfo {pages} {293} (\bibinfo {year}
  {1993})}\BibitemShut {NoStop}%
\bibitem [{\citenamefont {Perez-Garcia}\ \emph {et~al.}(2007)\citenamefont
  {Perez-Garcia}, \citenamefont {Verstraete}, \citenamefont {Wolf},\ and\
  \citenamefont {Cirac}}]{perezgarcia2007matrix}%
  \BibitemOpen
  \bibfield  {author} {\bibinfo {author} {\bibfnamefont {D.}~\bibnamefont
  {Perez-Garcia}}, \bibinfo {author} {\bibfnamefont {F.}~\bibnamefont
  {Verstraete}}, \bibinfo {author} {\bibfnamefont {M.~M.}\ \bibnamefont
  {Wolf}}, \ and\ \bibinfo {author} {\bibfnamefont {J.~I.}\ \bibnamefont
  {Cirac}},\ }\href@noop {} {\bibfield  {journal} {\bibinfo  {journal} {Quantum
  Info. Comput.}\ }\textbf {\bibinfo {volume} {7}},\ \bibinfo {pages}
  {401–430} (\bibinfo {year} {2007})}\BibitemShut {NoStop}%
\bibitem [{\citenamefont {Schollw\"{o}ck}(2011)}]{Schollwock2011}%
  \BibitemOpen
  \bibfield  {author} {\bibinfo {author} {\bibfnamefont {U.}~\bibnamefont
  {Schollw\"{o}ck}},\ }\href {\doibase
  https://doi.org/10.1016/j.aop.2010.09.012} {\bibfield  {journal} {\bibinfo
  {journal} {Annals of Physics}\ }\textbf {\bibinfo {volume} {326}},\ \bibinfo
  {pages} {96 } (\bibinfo {year} {2011})},\ \bibinfo {note} {january 2011
  Special Issue}\BibitemShut {NoStop}%
\bibitem [{\citenamefont {Orus}(2014)}]{orus2014practical}%
  \BibitemOpen
  \bibfield  {author} {\bibinfo {author} {\bibfnamefont {R.}~\bibnamefont
  {Orus}},\ }\href {\doibase https://doi.org/10.1016/j.aop.2014.06.013}
  {\bibfield  {journal} {\bibinfo  {journal} {Annals of Physics}\ }\textbf
  {\bibinfo {volume} {349}},\ \bibinfo {pages} {117 } (\bibinfo {year}
  {2014})}\BibitemShut {NoStop}%
\bibitem [{\citenamefont {O'Brien}\ \emph {et~al.}(2020)\citenamefont
  {O'Brien}, \citenamefont {Vernier},\ and\ \citenamefont
  {Fendley}}]{OBrien2019}%
  \BibitemOpen
  \bibfield  {author} {\bibinfo {author} {\bibfnamefont {E.}~\bibnamefont
  {O'Brien}}, \bibinfo {author} {\bibfnamefont {E.}~\bibnamefont {Vernier}}, \
  and\ \bibinfo {author} {\bibfnamefont {P.}~\bibnamefont {Fendley}},\ }\href
  {\doibase 10.1103/PhysRevB.101.235108} {\bibfield  {journal} {\bibinfo
  {journal} {Phys. Rev. B}\ }\textbf {\bibinfo {volume} {101}},\ \bibinfo
  {pages} {235108} (\bibinfo {year} {2020})}\BibitemShut {NoStop}%
\bibitem [{\citenamefont {Totsuka}\ and\ \citenamefont
  {Suzuki}(1995)}]{totsuka1995matrix}%
  \BibitemOpen
  \bibfield  {author} {\bibinfo {author} {\bibfnamefont {K.}~\bibnamefont
  {Totsuka}}\ and\ \bibinfo {author} {\bibfnamefont {M.}~\bibnamefont
  {Suzuki}},\ }\href {\doibase 10.1088/0953-8984/7/8/012} {\bibfield  {journal}
  {\bibinfo  {journal} {Journal of Physics: Condensed Matter}\ }\textbf
  {\bibinfo {volume} {7}},\ \bibinfo {pages} {1639} (\bibinfo {year}
  {1995})}\BibitemShut {NoStop}%
\bibitem [{\citenamefont {Karimipour}\ and\ \citenamefont
  {Memarzadeh}(2008)}]{karimipour2008matrix}%
  \BibitemOpen
  \bibfield  {author} {\bibinfo {author} {\bibfnamefont {V.}~\bibnamefont
  {Karimipour}}\ and\ \bibinfo {author} {\bibfnamefont {L.}~\bibnamefont
  {Memarzadeh}},\ }\href {\doibase 10.1103/PhysRevB.77.094416} {\bibfield
  {journal} {\bibinfo  {journal} {Phys. Rev. B}\ }\textbf {\bibinfo {volume}
  {77}},\ \bibinfo {pages} {094416} (\bibinfo {year} {2008})}\BibitemShut
  {NoStop}%
\bibitem [{\citenamefont {Verstraete}\ and\ \citenamefont
  {Cirac}(2006)}]{verstraete2006matrix}%
  \BibitemOpen
  \bibfield  {author} {\bibinfo {author} {\bibfnamefont {F.}~\bibnamefont
  {Verstraete}}\ and\ \bibinfo {author} {\bibfnamefont {J.~I.}\ \bibnamefont
  {Cirac}},\ }\href {\doibase 10.1103/PhysRevB.73.094423} {\bibfield  {journal}
  {\bibinfo  {journal} {Phys. Rev. B}\ }\textbf {\bibinfo {volume} {73}},\
  \bibinfo {pages} {094423} (\bibinfo {year} {2006})}\BibitemShut {NoStop}%
\bibitem [{\citenamefont {Vidal}(2004)}]{vidal2004efficient}%
  \BibitemOpen
  \bibfield  {author} {\bibinfo {author} {\bibfnamefont {G.}~\bibnamefont
  {Vidal}},\ }\href {\doibase 10.1103/PhysRevLett.93.040502} {\bibfield
  {journal} {\bibinfo  {journal} {Phys. Rev. Lett.}\ }\textbf {\bibinfo
  {volume} {93}},\ \bibinfo {pages} {040502} (\bibinfo {year}
  {2004})}\BibitemShut {NoStop}%
\bibitem [{\citenamefont {Haegeman}\ \emph {et~al.}(2013)\citenamefont
  {Haegeman}, \citenamefont {Osborne},\ and\ \citenamefont
  {Verstraete}}]{haegeman2013post}%
  \BibitemOpen
  \bibfield  {author} {\bibinfo {author} {\bibfnamefont {J.}~\bibnamefont
  {Haegeman}}, \bibinfo {author} {\bibfnamefont {T.~J.}\ \bibnamefont
  {Osborne}}, \ and\ \bibinfo {author} {\bibfnamefont {F.}~\bibnamefont
  {Verstraete}},\ }\href {\doibase 10.1103/PhysRevB.88.075133} {\bibfield
  {journal} {\bibinfo  {journal} {Phys. Rev. B}\ }\textbf {\bibinfo {volume}
  {88}},\ \bibinfo {pages} {075133} (\bibinfo {year} {2013})}\BibitemShut
  {NoStop}%
\bibitem [{\citenamefont {Vanderstraeten}\ \emph {et~al.}(2015)\citenamefont
  {Vanderstraeten}, \citenamefont {Mari\"en}, \citenamefont {Verstraete},\ and\
  \citenamefont {Haegeman}}]{vanderstraeten2015excitations}%
  \BibitemOpen
  \bibfield  {author} {\bibinfo {author} {\bibfnamefont {L.}~\bibnamefont
  {Vanderstraeten}}, \bibinfo {author} {\bibfnamefont {M.}~\bibnamefont
  {Mari\"en}}, \bibinfo {author} {\bibfnamefont {F.}~\bibnamefont
  {Verstraete}}, \ and\ \bibinfo {author} {\bibfnamefont {J.}~\bibnamefont
  {Haegeman}},\ }\href {\doibase 10.1103/PhysRevB.92.201111} {\bibfield
  {journal} {\bibinfo  {journal} {Phys. Rev. B}\ }\textbf {\bibinfo {volume}
  {92}},\ \bibinfo {pages} {201111} (\bibinfo {year} {2015})}\BibitemShut
  {NoStop}%
\bibitem [{\citenamefont {Vanderstraeten}\ \emph {et~al.}(2019)\citenamefont
  {Vanderstraeten}, \citenamefont {Haegeman},\ and\ \citenamefont
  {Verstraete}}]{vanderstraeten2019tangent}%
  \BibitemOpen
  \bibfield  {author} {\bibinfo {author} {\bibfnamefont {L.}~\bibnamefont
  {Vanderstraeten}}, \bibinfo {author} {\bibfnamefont {J.}~\bibnamefont
  {Haegeman}}, \ and\ \bibinfo {author} {\bibfnamefont {F.}~\bibnamefont
  {Verstraete}},\ }\href {\doibase 10.21468/SciPostPhysLectNotes.7} {\bibfield
  {journal} {\bibinfo  {journal} {SciPost Phys. Lect. Notes}\ ,\ \bibinfo
  {pages} {7}} (\bibinfo {year} {2019})}\BibitemShut {NoStop}%
\bibitem [{\citenamefont {den Nijs}\ and\ \citenamefont
  {Rommelse}(1989)}]{denNijs1989string}%
  \BibitemOpen
  \bibfield  {author} {\bibinfo {author} {\bibfnamefont {M.}~\bibnamefont {den
  Nijs}}\ and\ \bibinfo {author} {\bibfnamefont {K.}~\bibnamefont {Rommelse}},\
  }\href {\doibase 10.1103/PhysRevB.40.4709} {\bibfield  {journal} {\bibinfo
  {journal} {Phys. Rev. B}\ }\textbf {\bibinfo {volume} {40}},\ \bibinfo
  {pages} {4709} (\bibinfo {year} {1989})}\BibitemShut {NoStop}%
\bibitem [{\citenamefont {Kennedy}\ and\ \citenamefont
  {Tasaki}(1992)}]{kennedy1992hidden}%
  \BibitemOpen
  \bibfield  {author} {\bibinfo {author} {\bibfnamefont {T.}~\bibnamefont
  {Kennedy}}\ and\ \bibinfo {author} {\bibfnamefont {H.}~\bibnamefont
  {Tasaki}},\ }\href {\doibase 10.1103/PhysRevB.45.304} {\bibfield  {journal}
  {\bibinfo  {journal} {Phys. Rev. B}\ }\textbf {\bibinfo {volume} {45}},\
  \bibinfo {pages} {304} (\bibinfo {year} {1992})}\BibitemShut {NoStop}%
\bibitem [{\citenamefont {Oshikawa}(1992)}]{oshikawa1992hidden}%
  \BibitemOpen
  \bibfield  {author} {\bibinfo {author} {\bibfnamefont {M.}~\bibnamefont
  {Oshikawa}},\ }\href {\doibase 10.1088/0953-8984/4/36/019} {\bibfield
  {journal} {\bibinfo  {journal} {Journal of Physics: Condensed Matter}\
  }\textbf {\bibinfo {volume} {4}},\ \bibinfo {pages} {7469} (\bibinfo {year}
  {1992})}\BibitemShut {NoStop}%
\bibitem [{\citenamefont {P\'erez-Garc\'{\i}a}\ \emph
  {et~al.}(2008)\citenamefont {P\'erez-Garc\'{\i}a}, \citenamefont {Wolf},
  \citenamefont {Sanz}, \citenamefont {Verstraete},\ and\ \citenamefont
  {Cirac}}]{perezgarcia2008string}%
  \BibitemOpen
  \bibfield  {author} {\bibinfo {author} {\bibfnamefont {D.}~\bibnamefont
  {P\'erez-Garc\'{\i}a}}, \bibinfo {author} {\bibfnamefont {M.~M.}\
  \bibnamefont {Wolf}}, \bibinfo {author} {\bibfnamefont {M.}~\bibnamefont
  {Sanz}}, \bibinfo {author} {\bibfnamefont {F.}~\bibnamefont {Verstraete}}, \
  and\ \bibinfo {author} {\bibfnamefont {J.~I.}\ \bibnamefont {Cirac}},\ }\href
  {\doibase 10.1103/PhysRevLett.100.167202} {\bibfield  {journal} {\bibinfo
  {journal} {Phys. Rev. Lett.}\ }\textbf {\bibinfo {volume} {100}},\ \bibinfo
  {pages} {167202} (\bibinfo {year} {2008})}\BibitemShut {NoStop}%
\bibitem [{\citenamefont {Schutz}(1993)}]{Schutz1993}%
  \BibitemOpen
  \bibfield  {author} {\bibinfo {author} {\bibfnamefont {G.}~\bibnamefont
  {Schutz}},\ }\href {http://stacks.iop.org/0305-4470/26/i=18/a=021} {\bibfield
   {journal} {\bibinfo  {journal} {Journal of Physics A: Mathematical and
  General}\ }\textbf {\bibinfo {volume} {26}},\ \bibinfo {pages} {4555}
  (\bibinfo {year} {1993})}\BibitemShut {NoStop}%
\bibitem [{\citenamefont {Parkinson}(1987)}]{Parkinson1987integrability}%
  \BibitemOpen
  \bibfield  {author} {\bibinfo {author} {\bibfnamefont {J.~B.}\ \bibnamefont
  {Parkinson}},\ }\href {\doibase 10.1088/0022-3719/20/36/011} {\bibfield
  {journal} {\bibinfo  {journal} {Journal of Physics C: Solid State Physics}\
  }\textbf {\bibinfo {volume} {20}},\ \bibinfo {pages} {L1029} (\bibinfo {year}
  {1987})}\BibitemShut {NoStop}%
\bibitem [{\citenamefont {Barber}\ and\ \citenamefont
  {Batchelor}(1989)}]{Barber1989}%
  \BibitemOpen
  \bibfield  {author} {\bibinfo {author} {\bibfnamefont {M.~N.}\ \bibnamefont
  {Barber}}\ and\ \bibinfo {author} {\bibfnamefont {M.~T.}\ \bibnamefont
  {Batchelor}},\ }\href {\doibase 10.1103/PhysRevB.40.4621} {\bibfield
  {journal} {\bibinfo  {journal} {Phys. Rev.}\ }\textbf {\bibinfo {volume}
  {B40}},\ \bibinfo {pages} {4621} (\bibinfo {year} {1989})}\BibitemShut
  {NoStop}%
\bibitem [{\citenamefont {Ercolessi}\ \emph {et~al.}(2014)\citenamefont
  {Ercolessi}, \citenamefont {Vodola},\ and\ \citenamefont
  {Silvia}}]{Ercolessi2014Analysis}%
  \BibitemOpen
  \bibfield  {author} {\bibinfo {author} {\bibfnamefont {E.}~\bibnamefont
  {Ercolessi}}, \bibinfo {author} {\bibfnamefont {D.~D.}\ \bibnamefont
  {Vodola}}, \ and\ \bibinfo {author} {\bibfnamefont {F.}~\bibnamefont
  {Silvia}},\ }\href {https://amslaurea.unibo.it/8323/1/ferri_silvia_tesi.pdf}
  {\bibfield  {journal} {\bibinfo  {journal} {``Analysis of the spectrum of the
  spin-1 biquadratic antiferromagnetic chain" (unpublished)}\ } (\bibinfo
  {year} {2014})}\BibitemShut {NoStop}%
\bibitem [{\citenamefont {Parkinson}(1988)}]{Parkinson1988spin}%
  \BibitemOpen
  \bibfield  {author} {\bibinfo {author} {\bibfnamefont {J.~B.}\ \bibnamefont
  {Parkinson}},\ }\href {\doibase 10.1088/0022-3719/21/20/014} {\bibfield
  {journal} {\bibinfo  {journal} {Journal of Physics C: Solid State Physics}\
  }\textbf {\bibinfo {volume} {21}},\ \bibinfo {pages} {3793} (\bibinfo {year}
  {1988})}\BibitemShut {NoStop}%
\bibitem [{\citenamefont {Arovas}(1989)}]{Arovas1989}%
  \BibitemOpen
  \bibfield  {author} {\bibinfo {author} {\bibfnamefont {D.~P.}\ \bibnamefont
  {Arovas}},\ }\href {\doibase https://doi.org/10.1016/0375-9601(89)90921-3}
  {\bibfield  {journal} {\bibinfo  {journal} {Physics Letters A}\ }\textbf
  {\bibinfo {volume} {137}},\ \bibinfo {pages} {431 } (\bibinfo {year}
  {1989})}\BibitemShut {NoStop}%
\bibitem [{\citenamefont {Schuch}\ \emph {et~al.}(2010)\citenamefont {Schuch},
  \citenamefont {Cirac},\ and\ \citenamefont
  {Pérez-García}}]{schuch2010peps}%
  \BibitemOpen
  \bibfield  {author} {\bibinfo {author} {\bibfnamefont {N.}~\bibnamefont
  {Schuch}}, \bibinfo {author} {\bibfnamefont {I.}~\bibnamefont {Cirac}}, \
  and\ \bibinfo {author} {\bibfnamefont {D.}~\bibnamefont {Pérez-García}},\
  }\href {\doibase https://doi.org/10.1016/j.aop.2010.05.008} {\bibfield
  {journal} {\bibinfo  {journal} {Annals of Physics}\ }\textbf {\bibinfo
  {volume} {325}},\ \bibinfo {pages} {2153 } (\bibinfo {year}
  {2010})}\BibitemShut {NoStop}%
\bibitem [{\citenamefont {Shiraishi}(2019)}]{shiraishi2019connection}%
  \BibitemOpen
  \bibfield  {author} {\bibinfo {author} {\bibfnamefont {N.}~\bibnamefont
  {Shiraishi}},\ }\href {\doibase 10.1088/1742-5468/ab342e} {\bibfield
  {journal} {\bibinfo  {journal} {Journal of Statistical Mechanics: Theory and
  Experiment}\ }\textbf {\bibinfo {volume} {2019}},\ \bibinfo {pages} {083103}
  (\bibinfo {year} {2019})}\BibitemShut {NoStop}%
\bibitem [{\citenamefont {Mark}\ \emph
  {et~al.}(2020{\natexlab{b}})\citenamefont {Mark}, \citenamefont {Lin},\ and\
  \citenamefont {Motrunich}}]{mark2019new}%
  \BibitemOpen
  \bibfield  {author} {\bibinfo {author} {\bibfnamefont {D.~K.}\ \bibnamefont
  {Mark}}, \bibinfo {author} {\bibfnamefont {C.-J.}\ \bibnamefont {Lin}}, \
  and\ \bibinfo {author} {\bibfnamefont {O.~I.}\ \bibnamefont {Motrunich}},\
  }\href {\doibase 10.1103/PhysRevB.101.094308} {\bibfield  {journal} {\bibinfo
   {journal} {Phys. Rev. B}\ }\textbf {\bibinfo {volume} {101}},\ \bibinfo
  {pages} {094308} (\bibinfo {year} {2020}{\natexlab{b}})}\BibitemShut
  {NoStop}%
\bibitem [{\citenamefont {Lesanovsky}(2011)}]{lesanovsky2011many}%
  \BibitemOpen
  \bibfield  {author} {\bibinfo {author} {\bibfnamefont {I.}~\bibnamefont
  {Lesanovsky}},\ }\href@noop {} {\bibfield  {journal} {\bibinfo  {journal}
  {Physical Review Letters}\ }\textbf {\bibinfo {volume} {106}},\ \bibinfo
  {pages} {025301} (\bibinfo {year} {2011})}\BibitemShut {NoStop}%
\end{thebibliography}%
\end{document}